\documentclass[structabstract]{aa}  
\usepackage{graphicx}
\usepackage{txfonts}
\usepackage{natbib}
\bibpunct{(}{)}{;}{a}{}{,}
\usepackage{multirow}
\usepackage{longtable,lscape}

\begin{document}
   \title{ARRAKIS: atlas of resonance rings as known in the S$^4$G\thanks{Appendices~A, B, and C are available in electronic form at http://www.aanda.org}\fnmsep\thanks{Tables~A1 and A2 are only available in electronic form at the CDS via anonymous ftp to cdsarc.u-strasbg.fr (130.79.128.5) or via http://cdsweb.u-strasbg.fr/cgi-bin/qcat?J/A+A/}}
   
   \author{S.~Comer\'on\inst{1,2,3}
          \and
          H.~Salo\inst{1}
          \and
          E.~Laurikainen\inst{1,2}
          \and
          J.~H.~Knapen\inst{4,5}
          \and
          R.~J.~Buta\inst{6}
          \and
          M.~Herrera-Endoqui\inst{1}
          \and
          J.~Laine\inst{1}
          \and
          B.~W.~Holwerda\inst{7}
          \and
          K.~Sheth\inst{8}
          \and
          M.~W.~Regan\inst{9}
          \and
          J.~L.~Hinz\inst{10}
          \and
          J.~C.~Mu\~noz-Mateos\inst{11}
          \and
          A.~Gil de Paz\inst{12}
          \and
          K.~Men\'endez-Delmestre\inst{13}
          \and
          M.~Seibert\inst{14}
          \and
          T.~Mizusawa\inst{8,15}
          \and
          T.~Kim\inst{8,11,14,16}
          \and
          S.~Erroz-Ferrer\inst{4,5}
          \and
          D.~A.~Gadotti\inst{10}
          \and
          E.~Athanassoula\inst{17}
          \and
          A.~Bosma\inst{17}
          \and
          L.~C.~Ho\inst{14,18}
          }

   \institute{University of Oulu, Astronomy Division, Department of Physics, P.O.~Box 3000, FIN-90014, Finland\\
              \email{seb.comeron@gmail.com}
         \and
	     Finnish Centre of Astronomy with ESO (FINCA), University of Turku, V\"ais\"al\"antie 20, FI-21500, Piikki\"o, Finland
	 \and
             Korea Astronomy and Space Science Institute, 776, Daedeokdae-ro, Yuseong-gu, Daejeon 305-348, Republic of Korea
         \and
	     Instituto de Astrof\'isica de Canarias, E-38205 La Laguna, Tenerife, Spain
	 \and
	     Departamento de Astrof\'isica, Universidad de La Laguna, E-38200, La Laguna, Tenerife, Spain
	 \and
	     Department of Physics and Astronomy, University of Alabama, Box~870324, Tuscaloosa, AL 35487
	 \and
	     European Space Agency, ESTEC, Keplerlaan 1, 2200 AG, Noorwijk, the Netherlands
	 \and
	     National Radio Astronomy Observatory/NAASC, 520 Edgemont Road, Charlottesville, VA 22903, USA
	 \and
	     Space Telescope Science Institute, 3700 San Antonio Drive, Baltimore, MD 21218, USA
	 \and
	     European Southern Observatory, Casilla 19001, Santiago 19, Chile
	 \and
	     MMTO, University of Arizona, 933 North Cherry Avenue, Tucson, AZ 85721, USA
	 \and
	     Departamento de Astrof\'isica, Universidad Complutense de Madrid, Madrid 28040, Spain
	 \and
	     Universidade Federal do Rio de Janeiro, Observat\'orio do Valongo, Ladeira Pedro Ant\^onio, 43, CEP 20080-090, Rio de Janeiro, Brazil
	 \and
	     The Observatories of the Carnegie Institution for Science, 813, Santa Barbara Street, Pasadena, CA 91101, USA
	 \and
	     Florida Institute of Technology, Melbourne, FL 32901, USA
	 \and
	     Astronomy Program, Department of Physics and Astronomy, Seoul National University, Seoul 151-742, Republic of Korea
	 \and
	     Aix Marseille Universit\'e, CNRS, LAM (Laboratoire d'Astrophysique de Marseille) UMR~7326, 13388, Marseille, France
	 \and
	     Kavli Institute for Astronomy and Astrophysics, Peking University, Beijing 100871, China}

\titlerunning{ARRAKIS}
\authorrunning{Comer\'on, S. et al.}
 
  \abstract
   {Resonance rings and pseudorings (here collectively called rings) are thought to be related to the gathering of material near dynamical resonances caused by non-axisymmetries in galaxy discs. This means that they are the result of secular evolution processes that redistribute material and angular momentum in discs. Studying them may give clues on the formation and growth of bars and other disc non-axisymmetries.}
   {Our aims are to produce a catalogue and an atlas of the rings detected in the Spitzer Survey of Stellar Structure in Galaxies (S$^4$G) and to conduct a statistical study of the data in the catalogue.}
   {We traced the contours of rings previously identified and fitted them with ellipses. We found the orientation of bars by studying the galaxy ellipse fits from the S$^4$G pipeline~4. We used the galaxy orientation data obtained by the S$^4$G pipeline~4 to obtain intrinsic ellipticities and orientations of rings and the bars.}
   {ARRAKIS contains data on      724
\unskip ringed galaxies in the S$^4$G. The frequency of resonance rings in the S$^4$G is of $16\pm1\%$
\unskip and $35\pm1\%$
\unskip for outer and inner features, respectively. Outer rings are mostly found in Hubble stages $-1\leq T\leq4$. Inner rings are found in a broad distribution that covers the range $-1\leq T\leq7$. We confirm that outer rings have two preferred orientations, namely parallel and perpendicular to the bar. We confirm a tendency for inner rings to be oriented parallel to the bar, but we report the existence of a significant fraction (maybe as large as $50\%$) of inner features that have random orientations with respect to the bar. These misaligned inner rings are mostly found in late-type galaxies ($T\geq4$). We find that the fraction of barred galaxies hosting outer (inner) rings is $\sim1.7$ times ($\sim1.3$ times) that in unbarred galaxies.}
   {We confirm several results from previous surveys as well as predictions from simulations of resonant rings and/or from manifold flux tube theory. We report that a significant fraction of inner rings in late-type galaxies have a random orientation with respect to the bar. This may be caused by spiral modes that are decoupled from the bar and dominate the Fourier amplitude spectrum at the radius of the inner ring. The fact that rings are only mildly favoured by bars suggests that those in unbarred galaxies either formed because of weak departures from the axisymmetry of the galactic potential or that they are born because of bars that were destroyed after the ring formation.}


   \maketitle

   
\section{Introduction}

Galaxies are constantly evolving. Their properties change because of fast interactions and mergers \citep{TOOM07} and also because of slow secular evolution \citep[e.g.,][]{KOR04, ATH12A}. Part of the secular evolution in disc galaxies is driven by non-axisymmetries such as bars and oval distortions. Long-lived non-axisymmetries efficiently redistribute material and angular momentum across the discs in a Hubble-Lema\^itre time, which makes understanding them crucial to describe present-day galaxies. This slow process is, among many other consequences, responsible for the creation of pseudobulges \citep{KOR04}, for the radial spread of outer parts of discs \citep{SCH84, ATH12B}, and also for building spectacular rings and pseudorings. In this paper we study rings and pseudorings in a representative sample of nearby galaxies.

Rings are beautiful closed structures made of stars and/or gas. Pseudorings are their open counterparts, sometimes incomplete versions of rings and sometimes formed by spiral arms that almost connect. Here we use the word rings to refer to the set that includes both rings and pseudorings. The set of closed features is referred to as closed rings.

Rings often host intense star formation and/or are made of young blue stars \citep[see, e.g.,][]{BU93, KNA95, CRO96, BU02, KNA05, BU07, CO10, GROU10}. However, ``dead'' purely stellar rings exist as well \citep{BU91, ER99, ER02, CO13}.

The majority of rings seen in normal disc galaxies are likely caused by the influence of dynamical orbital resonances on the motions of gas clouds in the plane of the disc. The presence of a rotating bar or oval sets up a pattern speed and probably drives spiral patterns that can, via action of gravity torques, secularly evolve into more closed, ring-like features \citep{SCH81, SCH84}. The main evidence in support of this idea has come from observations of ring morphologies, intrinsic ring shapes, and intrinsic bar and ring major axis orientations as well as test-particle and $n$-body simulations [see review by \citet{BU96} and \citet{RAU00}].

There are four major and two secondary dynamical resonances that are believed to be important for ring formation  \citep[for detailed reviews of barred galaxy dynamics, see][]{SELL93, ATH12A}. These are defined (in the epicyclic approximation) by the relation between the bar pattern speed, $\Omega_{\rm p}$, the circular angular speed $\Omega$, and the radial epicyclic frequency $\kappa$. The major resonances are the outer Lindblad resonance (OLR, where $\Omega_{\rm p}=\Omega+\kappa/2$), the corotation resonance (CR, where $\Omega_{\rm p}=\Omega$), and the two inner Lindblad resonances (ILRs, where $\Omega_{\rm p}=\Omega-\kappa/2$). The secondary resonances are the inner 4:1 resonance (often called the UHR for ultraharmonic resonance, but which is here referred to as I4R, where $\Omega_{\rm p}=\Omega-\kappa/4$) and the outer 4:1 resonance (O4R, where $\Omega_{\rm p}=\Omega+\kappa/4$). Because $\Omega$ and $\kappa$ depend on the rotation curve of a galaxy, resonance locations will be tied to the gravitational potential of the system. For example, depending on the bar pattern speed and the central mass concentration, one or both ILRs may be absent. Typically, the OLR is located at a radius of roughly twice the bar length, the CR is located slightly outside the end of the bar, and the ILRs are well inside the bar.

Between the outermost ILR and the CR, the main family of orbits, called $x_1$ orbits, is parallel to the bar. In the framework of the epicyclic approximation, each time a main resonance is crossed, the orbits change their orientation by $90^{\rm o}$. Because of this, orbits that are slightly inside a resonance will intersect with those slightly outside it. Graphic depictions of this can be found, for instance, in Figure~11 in \citet{KNA95} and Figure~2 in \citet{ENG97}. Growing bars redistribute the angular momentum in a galaxy. Gas outside CR tends to be moved to the OLR radius \citep{LYN72, SELL81, SCH81}. Gas inside CR tends to move inwards and is collected either close to the I4R and/or close to the ILRs \citep{SCH84}. Thus, orbits near resonances are fed with gas by the bar angular momentum redistribution. Because of the high gas densities reached there and because of the collisions of gas clumps moving in intersecting orbits, star formation starts and resonance rings are created.  Although strictly speaking the epicyclic approximation is no longer valid for strongly perturbed potentials such as those affected by a strong bar \citep[for a good example of that see][]{SA99}, the picture described here remains qualitatively valid.

Rings in barred galaxies are classified according to their size compared to that of the bar. The precise location of a resonance cannot be determined unless the rotation curve of the galaxy and the pattern speed of the bar are known. Thus the resonance interpretations come from a confrontation between theory and observation:
\begin{itemize}
\item Outer rings are found at a radius roughly twice as large as that of the bar and are thought to be typically related to the OLR. Occasionally, some outer rings may be linked to the O4R as in the case of the dimpled ones in \citet{BYRD98}, those in simulations by \citet{RAU00}, and in the modelling of NGC~1433 by \citet{TREU08}. The first outer feature that has been described is that in NGC~1291 \citep{PER22}. 
\item Inner rings are found slightly outside the bar and are thought to be related to the I4R. They have been observed since R.~J.~Mitchell's observations of NGC~4725 in 1858 \citep{PAR80}. They were first described by \citet{CUR18}. 
\item Nuclear rings are found well inside the bar and are generally thought to be related to the ILRs \citep[but see][where nuclear rings are suggested to be caused by the centrifugal barrier encountered by gas migrating to the inner regions of the galaxy]{KIMWT12}. They were first seen in NGC~4321 by \citet{KEEL08} and described in NGC~3351 by \citet{CUR18}. Nuclear rings are especially bright and correlate with the presence of sigma-drops \citep[sigma-drops are nuclear regions in the centre of galaxies where the stellar velocity dispersion is measured to be lower than in the surroundings;][]{CO08b}.
\end{itemize}
We caution about the naming conventions and note that inner rings in barred galaxies are not thought to be related to the inner Lindblad resonances, but to the I4R. 

Based on data from the Third Reference Catalogue of Bright Galaxies \citep[RC3;][]{VAU91}, \citet{BU96} found that roughly 10\% of disc galaxies host outer rings and that 45\% of disc galaxies host inner rings. \citet{CO10} found that the fraction of nuclear rings is roughly 20\% for galaxies with Hubble stages between $T=-3$ and $T=7$.

Rings have been associated to resonances since the pioneering works by \citet{MAR72} and \citet{SCHO76}. Subsequently, this connection has been made in simulations by, among others, \citet{SCH81, SCH84}, \citet{BYRD94}, and \citet{KNA95}. Because of that, outer, inner, and nuclear rings have historically been called resonance rings.

A complementary theory explains the formation of at least some rings. This theory, called manifold theory, shows in a single framework the creation of outer spirals and outer and inner rings. It proposes that they are made of particles trapped in invariant manifolds (tubes of orbits) that start and end in one of the two unstable Lagrangian points at the end of the bar, $L_{1}$ and $L_{2}$ \citep{ROM06, ROM07, ATH09A, ATH09B, ATH10, ATH12B}. This theory predicts the existence of inner rings and that they are mainly oriented along the bar. It also predicts the existence of the outer R$_1$, R$_2$, and R$_1$R$_2$ rings and their orientation with respect to the bar (for a description of outer ring morphologies see Section~\ref{sdescripouter}). It also predicts the locations and axial ratios of rings. In fact, there is no prediction of the resonance ring simulations that the flux tube manifold theory does not predict as well, while the manifold shapes reproduce well that of rings in the response simulations mentioned above. The latter is possible because the values of the radii of the $L_1$, $L_2$, $L_3$, and $L_4$ Lagrangian points in barred galaxies are so close to that of the corotation radius that for observed galaxies they can be considered as equal. In this way it is possible to overcome the approximation of a resonant radius that is only valid in the linear and epicyclic approximations, that is, not in realistic barred galaxies. In the following we write manifold theory instead of flux tube manifold theory, and we use the generic name resonant rings for resonant and flux tube manifold rings.

Rings are also observed in galaxies that do not host a bar \citep[see, e.g., Figures.~3e and 3f in][]{BU95}. \citet{CO10} claimed that the frequency of rings with sizes comparable with those of nuclear rings in unbarred galaxies ($19\pm4\%$) is similar to the frequency of unbarred disc galaxies. One possibility to explain the existence of such rings is that they are remnants left after the bar dissolved \citep[for possible bar dissolution mechanisms, see e.g.,][]{RA91, FR93, BOU02, BOU05} or were destroyed in interactions with other galaxies \citep{ATH96, BER03}. However, the possibility of bar dissolution has to be taken with caution because many simulations showed that bars cannot be easily erased without interactions \citep[see, e.g.,][]{MAR04, DE04, DE06, BER07, ATH13}. Alternatively, these rings that apparently do not fit in the resonance theory might be explained by resonances caused by weak oval distortions, strong spiral patterns \citep{JUNG96, RAU00}, or by the gravitational potential distortions induced during minor mergers and/or interactions. This last possibility has been suggested by \citet{KNA04} for the pseudoring  in NGC~278 and by some numerical experiments \citep{TU06}.

Another kind of interesting morphological feature in disc galaxies are the lenses. Lenses are typically found in early-type disc galaxies, they have flat luminosity profiles and fairly sharp edges. Lenses, like rings, can be classified as outer, inner, or nuclear, based on their sizes. They have been studied in detail in the Near-Infrared atlas of S0-Sa galaxy Survey \citep[NIRS0S;][]{LAU11}. Some features, called ring-lenses and first reported by \citet{KOR79}, have properties between those of rings and lenses; they are lens-like features with a significant enhancement of luminosity close to their edges. In this paper we included ring-lenses together with classical rings.

Rings can be responsible for a significant fraction of the light emitted by a galaxy. For example, the exceptionally bright nuclear ring in NGC~1097 emits $\sim10\%$ of the light of the galaxy at $3.6\,\mu{\rm m}$ \citep{SHETH10}. This can be considered as an estimate of the upper limit of the fraction of light coming from rings in present-day galaxies. However, since rings are often regions of intense star formation with a low mass-to-light ratio even in the mid-infrared, the fraction of the galaxy baryonic mass found in them is smaller than that.

Non-resonant mechanisms also produce rings, but they often have an appearance fairly different from that of resonance rings. A ring can be formed at the largest non-looping orbit whose major axis is parallel to the major axis of the bar. Such a ring is called an x$_{1}$-ring \citep{RE04}. Collisional rings occur when a disc galaxy collides head-on with a satellite \citep{LYNDS76, THEYS77}. Polar rings occur when part of the mass of a satellite is accreted in an orbit perpendicular to the disc plane of the main galaxy \citep{SCH83}. These three types of rings are rarer than resonance rings.

Moreover, some of the smallest nuclear rings \citep[also called ultra-compact nuclear rings;][]{CO08c, CO08a} might be related to the interaction of an AGN with the surrounding interstellar medium \citep{CO10}.

Some rings that are coplanar with the galaxy disc may be related to polar rings. In a few galaxies, in-plane rings are found to be made of counterrotating material postulated to come from a recent minor merger. This could be the case, for instance, for the outer ring in IC~2006 \citep{SCH89, BET01}, the inner ring in NGC~3593 \citep{COR98}, the inner ring in NGC~4138 \citep{JO96, THA97}, and the innermost ring in NGC~7742 \citep{SIL06, MAZ06}.

In this paper we present a catalogue (Appendix~A) and an atlas (Appendix~B) of the rings identified in the Spitzer Survey of Stellar Structure in Galaxies \citep[S$^4$G;][]{SHETH10}. The focus is on resonance rings, hence the title ARRAKIS: atlas of resonance rings as known in the S$^4$G. In the catalogue section, we present data on the projected and intrinsic ring major and minor axes and orientations. We also present the projected and intrinsic highest bar ellipticities, which can be used as a rough estimate of the bar strength, and the bar orientations. The atlas presents images of all the galaxies with rings. We overlay the measured contour of the rings on these images.

The paper is structured as follows: in Section~\ref{sidentif} we briefly describe the S$^4$G and the ring identification process, in Section~\ref{sdescrip} we describe the types of rings detected in the S$^4$G, and in Section~\ref{sprepa} we describe the production process of the catalogue. We compare our measured ring properties with those in previous studies in Section~\ref{scomparison}, we present some statistical results on the ring fraction, their intrinsic axis ratio, their position angle (PA) offset with bars and sizes in Section~\ref{sresults}, and we discuss some of these points in Section~\ref{sdiscuss}. We summarise our results and conclusions in Section~\ref{sconclusions}. The Catalogue is presented in Appendix~A and the atlas of rings is presented in Appendix~B. Appendix~C contains a list of galaxies in the catalogue with a duplicated NGC/IC identification.

\section{Identification of rings and pseudorings in the S$^4$G}

\label{sidentif}

\subsection{The S$^4$G}

The rings in this paper are identified and described using S$^4$G images. The S$^4$G is a mid-infrared survey that has observed a sample of galaxies representative of the local Universe using the InfraRed Array Camera \citep[IRAC;][]{FA04} of the {\it Spitzer} Space Telescope \citep{WER04}. The goal of the S$^4$G is to describe the stellar mass distribution in the local Universe. The mid-infrared is the ideal band to do so because it has little dust absorption. The selection criteria of the sample are the following:

\begin{itemize}
\item{Radio radial heliocentric velocity $v_{\rm radio}<3000\,{\rm km\,s^{-1}}$, which is equivalent to a distance of $D<41$\,Mpc when using a Hubble-Lema\^itre constant of $H_{0}=73\,{\rm km\,s^{-1}\,Mpc^{-1}}$.}
\item{Integrated blue magnitude $m_{B,{\rm corr}}<15.5$\,mag. The magnitude considered here is that corrected for Galactic extinction, inclination, and $K$-correction.}
\item{Angular diameter $D_{25}>1\arcmin$.}
\item{Galactic latitude $|b|>30^{\rm o}$ to avoid the observations to be overly affected by foreground stars and other Galactic emission.}
\end{itemize}

All these parameters were taken from HyperLeda \citep{PA03}. One of the reasons why the sample is representative and not volume-limited is that the galaxies in the sample are limited in diameter and in brightness. Another source of incompleteness is that one of the main inputs of HyperLeda is the RC3, which is only reasonably complete at $m_B<15.5$\,mag (this time the blue magnitude is uncorrected). An additional reason is that some galaxies, especially those of earlier types, did not have a radial velocity in radio in HyperLeda at the time when the sample was defined. The final S$^4$G sample size is 2352 galaxies.

The images were taken to map the galaxies at least up to $1.5\times D_{25}$ and mosaics were produced when needed. The galaxies were observed in the $3.6$ and $4.5\,\mu{\rm m}$ filters of IRAC with a total integration time of four minutes, obtaining images with a surface brightness sensitivity of $\mu({\rm AB})_{3.6\,\mu{\rm m}}\sim27\,{\rm mag\,arcsec^{-2}}$. Such deep images are unprecedented for a large survey of galaxies in the mid-infrared.

Because the distances used here are redshift-independent or Hubble-Lema\^itre flow-corrected, some of the galaxy distances listed in the table in Appendix~A exceed 41\,Mpc, as explained in Section~\ref{sringdiam}.

The S$^4$G data is now public and can be downloaded from the NASA/IPAC Infrared Science Archive (IRSA) website\footnote{http://irsa.ipac.caltech.edu/}.

Some galaxies in the S$^4$G have two identifications in the NGC and IC catalogues \citep{DRE88, DRE95}. To facilitate the comparison of the data presented here with other samples, Appendix~C presents a list of the galaxies with a double identification that have been included in ARRAKIS. This list has been made by using mostly data from The Historically Corrected New General Catalogue \footnote{http://www.ngcicproject.org}, but also from NED and HyperLeda.

\subsection{Morphological classification of S$^4$G galaxies}

Buta et al.~(in preparation) classified most of the galaxies in the S$^4$G using the criteria described in \citet{BU10} and \citet{LAU11}. In brief, the galaxies were classified using the de Vaucouleurs revised Hubble-Sandage system \citep{VAU59} and its revision \citep{VAU63}, which has three dimensions, namely the stage (E, E$^+$, S0$^-$, S0$^{\rm o}$, S0$^+$, S0/a, Sa, Sab, Sb, Sbc, Sc, Scd, Sd, Sdm, Sm, Im), the family (SA, S${\underline{\rm A}}$B, SAB, SA${\underline{\rm B}}$, SB), and the variety (r, ${\underline{\rm r}}$s, rs, r${\underline{\rm s}}$, s). Additional dimensions were added to the classification by indicating which galaxies are highly inclined (spindle or sp), which galaxies are double-staged, which galaxies have shells, lenses, nuclear bars, nuclear discs, triaxial bulges, ans\ae, X-shaped bars, tidal debris, pseudobulges, and other peculiarities. Most important for this work, they also classified which galaxies host outer and nuclear rings. For an extensive review of this matter see \citet{BU13}.

The galaxies were classified using the S$^4$G $3.6\,\mu{\rm m}$ band and interpreting the images as if they were blue light. The classifications in both bands are in general very similar and only differ in galaxies with stages between S0/a and Sc, for which the classification in $3.6\,\mu{\rm m}$ is on average one stage earlier than in the $B$ band \citep{BU10}. This one-stage shift should be taken into account when comparing ARRAKIS results with those in previous studies based on $B$-band imaging.

A problem may appear when, assuming that rings are related to resonances, those in unbarred galaxies are classified as outer, inner, and nuclear rings. In that case, they were classified comparing the ring size with the galaxy radius and/or the location where they appear in relation to spiral arms. This approach may cause some rings to be misclassified, as discussed in \citet{BU95} and \citet{CO10}.

A few of the classified galaxies have two or even three features that fit in the outer, inner, or nuclear categories. For example, NGC~5055 is an SA(rs,rl)bc galaxy and has two inner features. In this case, the feature that appears first in the classification (rs) is the outermost feature. The only exceptions for this rule are R$_1$R$_2^{\prime}$ and R$_1^{\prime}$R$_2^{\prime}$ combinations, for which the order is reversed because of historical reasons.

Because of the limited amount of dust obscuration it suffers, the $3.6\,\mu{\rm m}$ band presents an advantage over optical wavelengths at detecting and describing rings that otherwise could remain hidden. This, combined with the angular resolution of the images in the S$^4$G ($0\farcs75$ pixel size and a full width at half maximum of $\sim2\arcsec$), allowed us to identify rings with diameters down to $\sim10\arcsec$ in low-inclination galaxies. We here considered low-inclination galaxies to be those with $i\leq60^{\rm o}$. This value is based on our ability to measure bar properties as explained in Sections~\ref{sreliability} and \ref{sreliability2}. The ring identification becomes increasingly difficult for higher disc inclinations due to higher optical depth and ring foreshortening \citep[see, e.g., Tables~IV and V in][]{BU96}. A diameter of $10\arcsec$ corresponds to 1.0\,kpc at a distance 20\,Mpc, and to 2.4\,kpc at a distance of 50\,Mpc, below which 98\% of ARRAKIS galaxies are found.

The completeness regarding outer rings is hard to assess; some outer rings are known to be exceedingly faint and some may have surface brightnesses below the sensitivity level of the S$^4$G. Additionally, in some cases, outer features are better seen in blue light: for example, the outer ring in NGC~7743, which is distinguishable in the $B$-band image in \citet{BU07}, is not seen in the mid-infrared. Moreover, in at least one case (NGC~4151) an outer ring has not been detected here because it is larger than the region covered by the S$^4$G frame.

The completeness in the detection of inner rings is very high. When looking at galaxies with $D<20$\,Mpc, we find that only 5\% of inner rings have a diameter smaller than 2.4\,kpc and therefore could not be detected for galaxies at $D=50$\,Mpc. We therefore expect to be missing less than 5\% of the inner rings due to spacial resolution problems.

Regarding nuclear features, we cannot expect to be as complete as we are at finding inner rings: at a distance $D=20$\,Mpc, $5\arcsec$ correspond to a ring radius of $r=500$\,pc, which is similar to the average size of the nuclear rings detected in \citet{CO10}. Therefore, any statistics based on the nuclear rings in ARRAKIS must be considered uncertain and only representative of the behaviour of the largest nuclear rings.

A total of         2347
\unskip S$^4$G galaxies have been classified in Buta et al.~(2013, in preparation). One of these galaxies, NGC~7204 (also named PGC~68061), is a pair composed of NGC~7204A and NGC~7204B, leading the count of S$^4$G classified galaxies to         2348
\unskip. Four S$^4$G galaxies could not be classified because they appear close to very bright saturated stars which makes the galaxy morphology hard to describe. In addition,           70
\unskip galaxies were classified that not belong to the S$^4$G sample, but appear in the S$^4$G frames. Most of these galaxies are satellites of galaxies in the S$^4$G sample or background galaxies. Thus, we have data of a total of         2418
\unskip galaxies for which there is a classification that considers the presence of rings.

The total number of galaxies that appear in S$^4$G frames and have been classified to host rings is      735
\unskip. Of these,      724
\unskip galaxies are included in the original S$^4$G sample. These      724
\unskip galaxies have      295
\unskip outer rings,      609
\unskip inner rings,       47
\unskip nuclear rings,        4
\unskip x$_1$-rings,        4
\unskip collisional rings, and        7
\unskip polar rings.  

To ensure the reliability of the ring statistics we considered only the galaxies in the S$^4$G sample, which, as said before, is representative of the local Universe. However, for the sake of completeness, in the appendices we also included the data and the images corresponding to rings in galaxies not included in the original S$^4$G sample.

\section{Ring morphologies}

\label{sdescrip}

\begin{figure*}
  \includegraphics[width=0.90\textwidth]{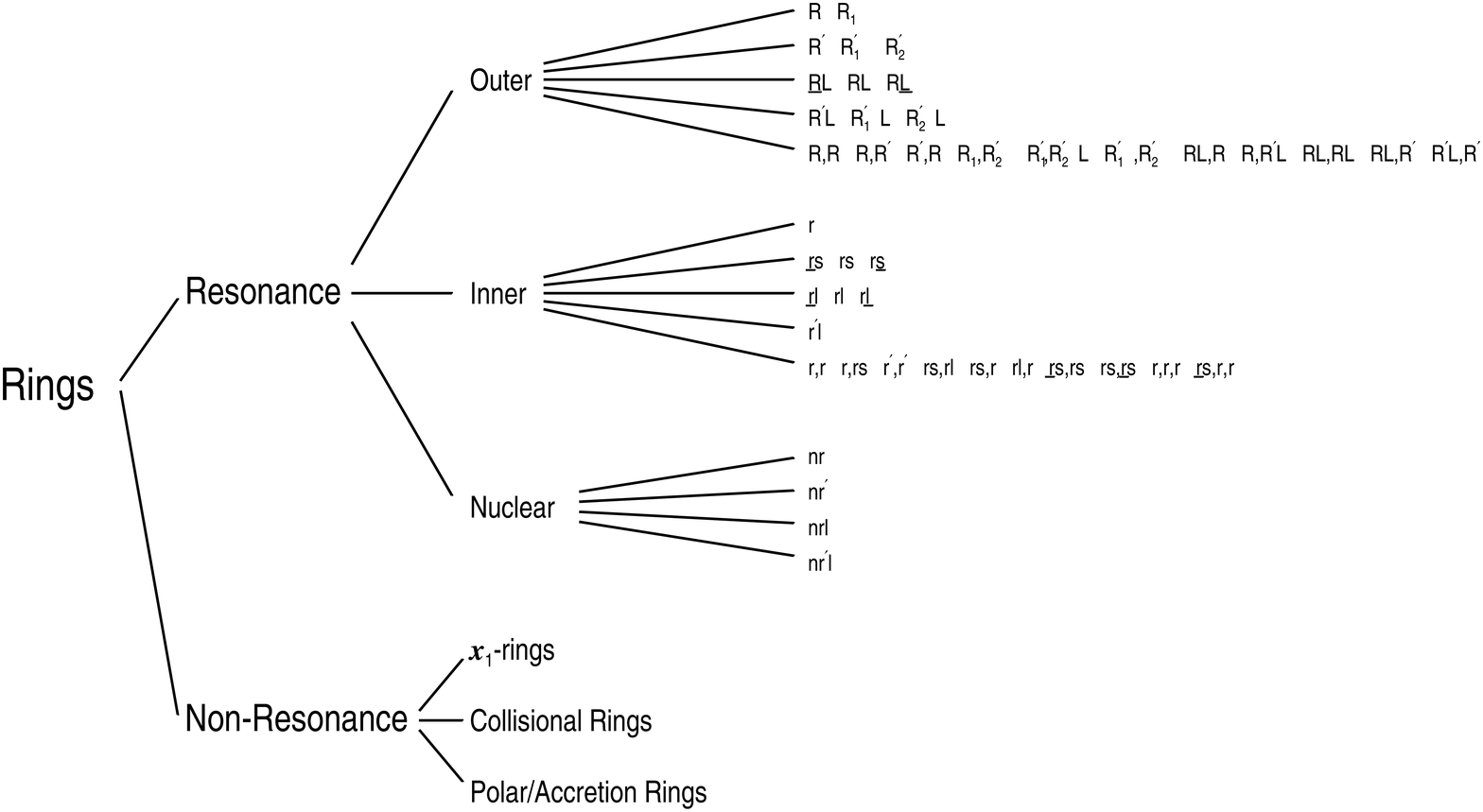}
  \caption{\label{ringstyles} Schematic classification of the ring types found in ARRAKIS. Multiple ring combinations are also included.}  
\end{figure*}

\begin{table}
 \caption{Glossary with some notation used in galaxy classification}
 \label{tringstyles}
 \centering
 \begin{tabular}{l l}
 \hline\hline
 Notation&Feature\\
 \hline
 R&Outer closed ring\\
 R$_1$&Outer closed ring with Type~1 morphology\\
 R$_2$&Outer closed ring with Type~2 morphology\\
 ${\underline{\rm R}}$L, RL, R${\underline{\rm L}}$& Outer closed ring-lenses\\
 L& Outer lens\\
 R$^{\prime}$&Outer pseudoring\\
 R$^{\prime}_1$&Outer pseudoring with Type~1 morphology\\
 R$^{\prime}_2$&Outer pseudoring with Type~2 morphology\\
 ${\underline{\rm R}}^{\prime}$L, R$^{\prime}$L, R$^{\prime}{\underline{\rm L}}$ &Outer pseudoring-lenses\\
 r & Inner closed ring\\
 ${\underline{\rm r}}$l, rl, r${\underline{\rm l}}$& Inner closed ring-lenses\\
 l & Inner lens\\
 r$^{\prime}$,${\underline{\rm r}}$s, rs, r${\underline{\rm s}}$& Inner pseudorings\\
 ${\underline{\rm r}}^{\prime}$l, r$^{\prime}$l, r$^{\prime}{\underline{\rm l}}$ &Inner pseudoring-lenses\\
 nr & Nuclear closed ring\\
 nl & Nuclear lens\\
 nrl & Nuclear closed ring-lens\\
 nr$^{\prime}$ & Nuclear pseudoring\\
 nr$^{\prime}$l & Nuclear pseudoring-lens\\
 x$_1$r & x$_1$-ring\\
 RG & Ring galaxy (collisional ring)\\
 PRG & Polar ring galaxy\\
 \hline
 \end{tabular}
\end{table}

The types of rings studied here are described below. A schematic description of the different ring categories can be found in Figure~\ref{ringstyles}. A glossary of some common morphological features is presented in Table~\ref{tringstyles}. For more information on ring classification and morphology see, for example, \citet{BU07} and \citet{BU13}.

\subsection{Outer rings}
\label{sdescripouter}

\begin{figure*}
\setlength{\tabcolsep}{2.5pt}
  \begin{tabular}{c c c c}
  \includegraphics[width=0.24\textwidth]{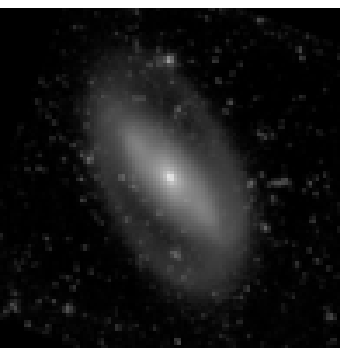}&
  \includegraphics[width=0.24\textwidth]{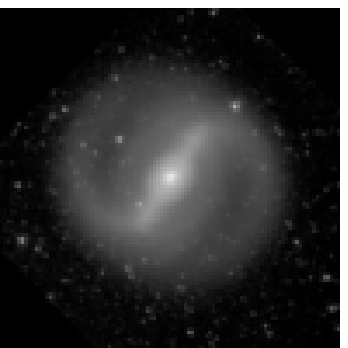}&
  \includegraphics[width=0.24\textwidth]{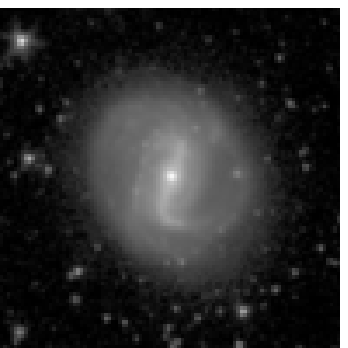}&
  \includegraphics[width=0.24\textwidth]{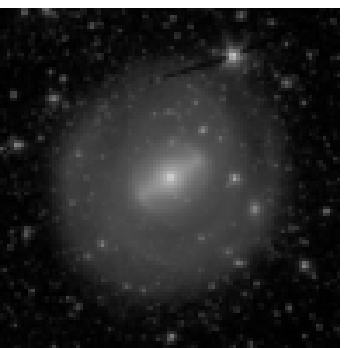}\\
  NGC~5377\hspace{3mm}(R$_1$)SA${\underline{\rm B}}$(s,nl)0/a&
  NGC~4314\hspace{3mm}(R$_1^{\prime}$)SB(rl,nr)a&
  NGC~5757\hspace{3mm}(R$_2^{\prime}$)SB(rs)${\underline{\rm a}}$b&
  NGC~5101\hspace{3mm}(R$_1$R$_2^{\prime}$)SB(${\underline{\rm r}}$s,nl?)0/a\\
  \vspace{-2mm}\\
  \includegraphics[width=0.24\textwidth]{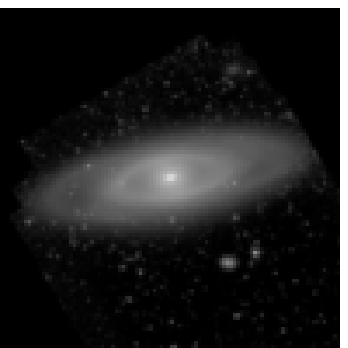}&
  \includegraphics[width=0.24\textwidth]{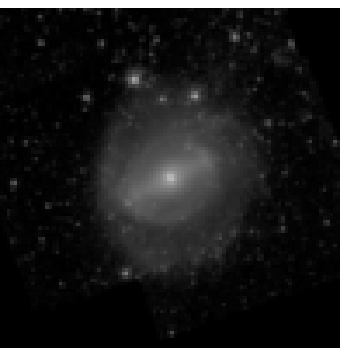}&
  \includegraphics[width=0.24\textwidth]{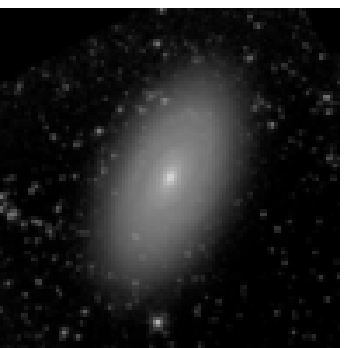}&
  \includegraphics[width=0.24\textwidth]{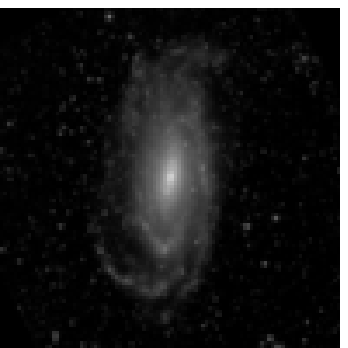}\\
  NGC~4274\hspace{3mm}(R)SB(r,nr)0/a&
  NGC~5850\hspace{3mm}(R$^{\prime}$)SB(r,nr,nb)ab&
  NGC~4380\hspace{3mm}(R)SA(r,l)ab&
  NGC~5033\hspace{3mm}(R$^{\prime}$)SA(rs)c\\
  \vspace{-2mm}\\
  \includegraphics[width=0.24\textwidth]{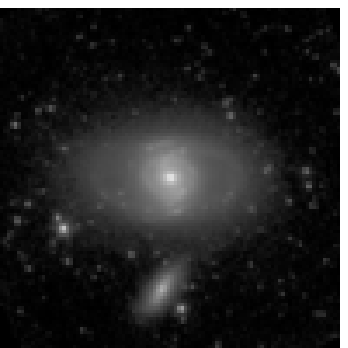}&
  \includegraphics[width=0.24\textwidth]{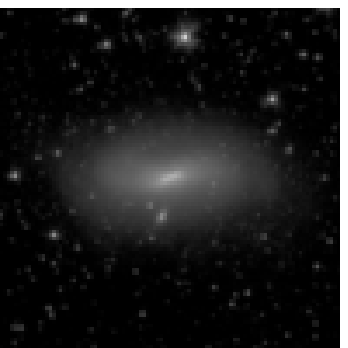}&
  \includegraphics[width=0.24\textwidth]{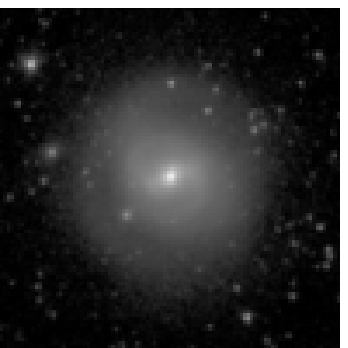}&
  \includegraphics[width=0.24\textwidth]{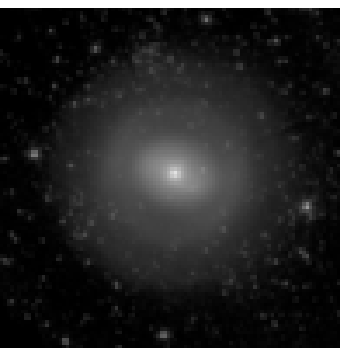}\\
  NGC~4045\hspace{3mm}(R$_1^{\prime}$L)SAB(${\underline{\rm r}}$s,nl)ab&
  NGC~4424\hspace{3mm}(R$_2^{\prime}$L)SB(s)0/a pec&
  NGC~1022\hspace{3mm}(RL)SAB(${\underline{\rm r}}$s)0/a&
  NGC~4457\hspace{3mm}(RR)SAB(l)0$^+$
  \end{tabular}
  \caption{\label{for} Selection of S$^4$G galaxies hosting a variety of outer rings. The images are in the $3.6\,\mu{\rm m}$ band, have a size twice that of the $\mu_{\rm 3.6\,\mu m}=25.5\,{\rm mag\,arcsec^{-2}}$ diameter isophote and are oriented with north up and east left. The top row shows galaxies with outer features compatible with the R$_1$ and R$_2$ ring morphologies seen in simulations by \citet{SCH81, SCH84} and defined by \citet{BU86a}. The second row shows outer rings that cannot be easily recognised to belong to these categories in either barred or unbarred galaxies. The third row has three example of ring-lenses and a rare example of a galaxy with two outer rings. The names of the galaxies and their morphological classification (from Buta et al.~in preparation) can be found below each image.}
\end{figure*}

In barred galaxies, outer rings are those with roughly twice the size of the bar. They are thought to be related to the OLR or, occasionally, to the O4R. Since the work from \citet{VAU59}, outer features have been divided into closed rings (R) and pseudorings (R$^{\prime}$), depending on whether the feature is closed/complete or not. 

Additional subdivisions have been made later based on the morphology of the outer features in the simulations of \citet{SCH81, SCH84} \citep[see Figure~2 in][]{BU86a}. R$_1$ closed rings and R$_1^{\prime}$ pseudorings consist of arms that start at one end of the bar and end after a 180$^{\rm o}$ bend in the other end of the bar. These rings typically have a dimpling in the region of the end of the bar and thus are 8-shaped. R$_2$ closed rings and R$_2^{\prime}$ pseudorings consist of two spiral arms, each starting at one of the ends of the bar and intersecting with each other at a position roughly perpendicular to the main axis of the bar after a 270$^{\rm o}$ bend. ARRAKIS includes some R$_2^{\prime}$ pseudorings, but no R$_2$ rings. According to the simulations from \citet{SCH81, SCH84}, R$_1$ features are expected to be elongated and perpendicular to the major axis of the bar, and R$_2$ features are expected to be elongated and parallel to the major axis of the bar. Sometimes galaxies host an R$_1^{\prime}$R$_2^{\prime}$ or an R$_1$R$_2^{\prime}$ combination. 

When an outer feature could not be unambiguously classified in the R$_1$ and R$_2$ categories (for example in unbarred galaxies), it was classified as R in the case of closed rings and as R$^{\prime}$ in the case of pseudorings.

Some outer features have properties intermediate between rings and lenses and were classified as outer ring-lenses. They are indicated by adding an L to the morphological classifications, for instance, RL and R$_1^{\prime}$L. Subtler degrees of ringness of outer ring-lenses are sometimes described by underlines, forming the sequence R, ${\underline{\rm R}}$L, RL, R${\underline{\rm L}}$, L.

A selection of galaxies showing a variety of outer rings observed in the S$^4$G is presented in Figure~\ref{for}.

\subsection{Inner rings}

\begin{figure*}
\setlength{\tabcolsep}{2.5pt}
  \begin{tabular}{c c c c}
  \includegraphics[width=0.24\textwidth]{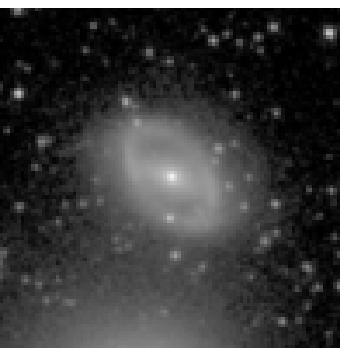}&
  \includegraphics[width=0.24\textwidth]{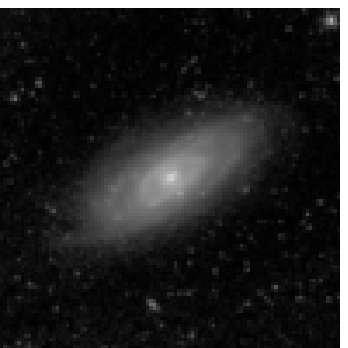}&
  \includegraphics[width=0.24\textwidth]{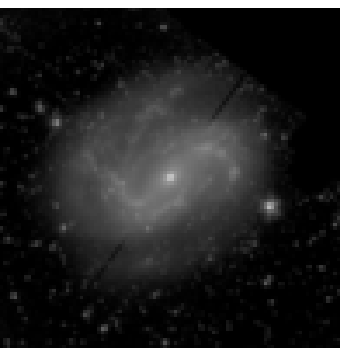}&
  \includegraphics[width=0.24\textwidth]{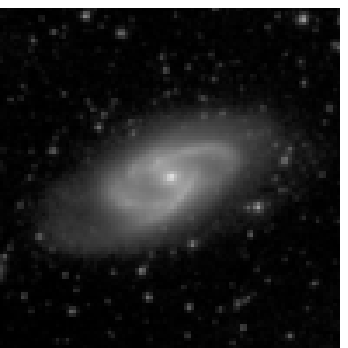}\\
  NGC~5636\hspace{3mm}SAB(r)0/a&
  NGC~3705\hspace{3mm}(R$^{\prime}$)SAB(${\underline{\rm r}}$s)b&
  NGC~4051\hspace{3mm}SAB(rs)b&
  NGC~150\hspace{3mm}SAB(r${\underline{\rm s}}$)ab\\
  \vspace{-2mm}\\
  \includegraphics[width=0.24\textwidth]{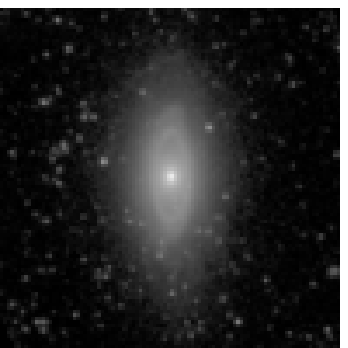}&
  \includegraphics[width=0.24\textwidth]{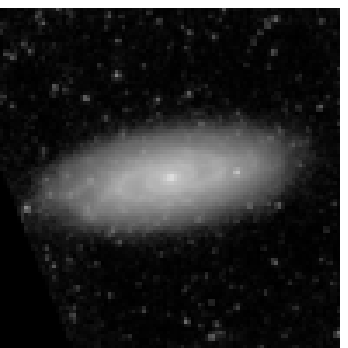}&
  \includegraphics[width=0.24\textwidth]{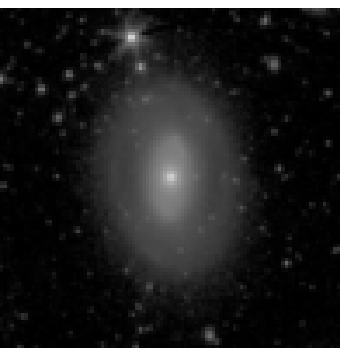}&
  \includegraphics[width=0.24\textwidth]{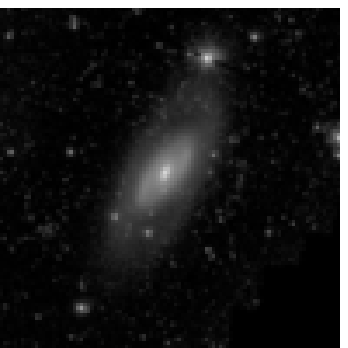}\\
  NGC~3900\hspace{3mm}SA(r)0/a&
  NGC~4062\hspace{3mm}SA(rs)b&
  NGC~2962\hspace{3mm}(R)S${\underline{\rm A}}$B$_{\rm a}$(rl)0$^+$&
  NGC~1415\hspace{3mm}(RL)SAB$_{\rm a}$(r$^{\prime}$l,nr)0$^{+}$\\
  \end{tabular}
  \caption{\label{fir} Selection of S$^4$G galaxies hosting a variety of inner rings presented in the same way as those in Figure~\ref{for}. The top row presents a succession of increasingly more spiral-like inner features in a series of SAB or moderately barred galaxies. The second line shows example of inner rings in unbarred galaxies and two examples of ring-lenses in early-type disc galaxies.}
\end{figure*}

In barred galaxies, inner rings are those that are roughly the size of the bar or slightly larger than the bar. They are thought to be related to the I4R. They are classified in a sequence of openness of the spiral arms that curve to form the ring. This sequence ranges from completely closed inner rings to spirals in the following order, r, ${\underline{\rm r}}$s, rs, r${\underline{\rm s}}$, s \citep{VAU63}. Underlines indicate transitions between the r, rs, and s varieties. Intermediate varieties between r and s may also indicate rings that are not complete. Features classified as inner pseudorings (${\underline{\rm r}}$s, rs, r${\underline{\rm s}}$) are sometime denoted as r$^{\prime}$ in the literature.

Inner features can also be ring-lenses. Closed inner ring-lenses are indicated as rl and inner pseudoring-lenses are indicated as r$^{\prime}$l. Subtler degrees of ringness of inner ring-lenses are sometimes described by underlines, forming the sequence r, ${\underline{\rm r}}$l, rl, r${\underline{\rm l}}$, l.

A selection of galaxies showing a variety of inner rings observed in the S$^4$G is presented in Figure~\ref{fir}.

\subsection{Nuclear rings}

\begin{figure*}
\setlength{\tabcolsep}{2.5pt}
  \begin{tabular}{c c c c}
  \includegraphics[width=0.24\textwidth]{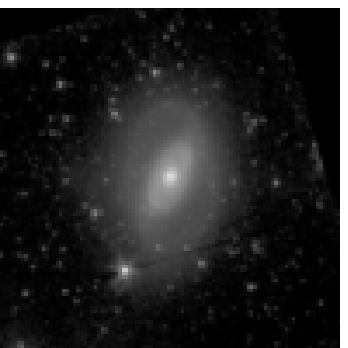}&
  \includegraphics[width=0.24\textwidth]{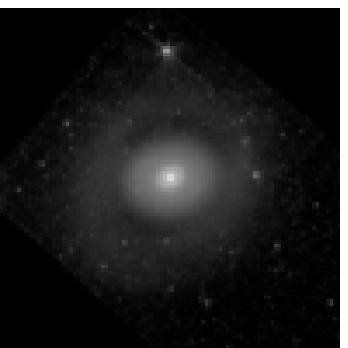}&
  \includegraphics[width=0.24\textwidth]{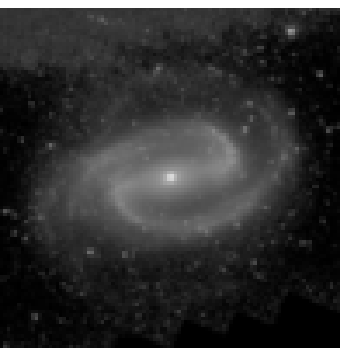}&
  \includegraphics[width=0.24\textwidth]{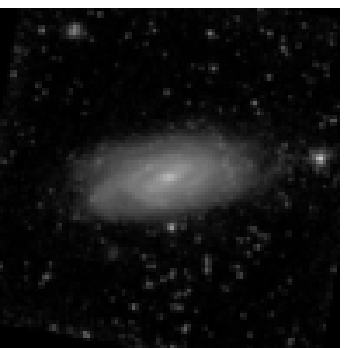}\\
  \includegraphics[width=0.24\textwidth]{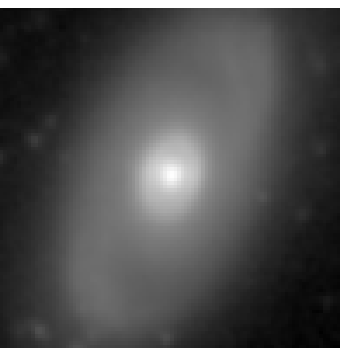}&
  \includegraphics[width=0.24\textwidth]{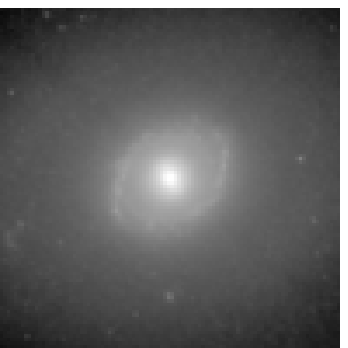}&
  \includegraphics[width=0.24\textwidth]{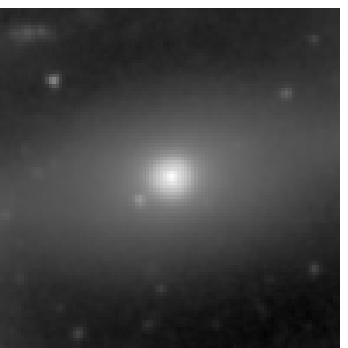}&
  \includegraphics[width=0.24\textwidth]{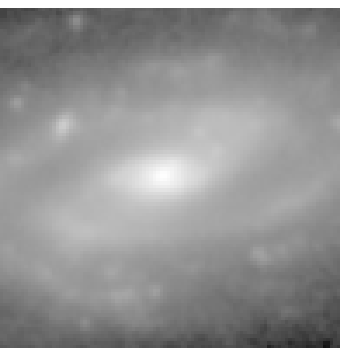}\\  
  IC~4214\hspace{3mm}(R$_1$)S${\underline{\rm A}}$B$_{\rm a}$(r$^{\prime}$l,nr,nb)0/a&
  NGC~4736\hspace{3mm}(R)S${\underline{\rm A}}$B(rl,nr$^{\prime}$,nl,nb)a&
  NGC~1300\hspace{3mm}(R$^{\prime}$)SB(s,nrl)b&
  NGC~1090\hspace{3mm}SAB(s,nr$^{\prime}$l)bc\\
  \end{tabular}
  \caption{\label{fnr} Selection of S$^4$G galaxies hosting a variety of nuclear features. The images in the top row are presented in the same way as those in Figure~\ref{for}. The bottom panels are zoomed by a factor of five compared with those in the top row. The frames show galaxies hosting a ring, a pseudoring, a ring-lens, and a pseudoring-lens, respectively.}
\end{figure*}

Although generally small, several nuclear features are visible in the S$^4$G images. In barred galaxies, nuclear rings are found inside bars. They are thought to be related to the ILRs \citep[but see][]{KIMWT12}. They are expected to be located between the outer and the inner ILR when both are present and inside the ILR when the galaxy has only one of them \citep[see, e.g.,][]{SHLO99, SHETH00}. Nuclear rings have been studied in detail in the atlas of images of nuclear rings \citep[AINUR;][]{CO10}.

Nuclear features are classified into the closed nuclear ring (nr) and nuclear pseudoring (nr$^{\prime}$) subtypes depending on whether the feature appears completely closed or not. Closed nuclear ring-lenses and pseudoring-lenses are denoted as nrl and nr$^{\prime}$l, respectively.

A selection of galaxies showing a variety of nuclear rings observed in the S$^4$G is presented in Figure~\ref{fnr}.

\subsection{x$_{1}$, collisional, polar, and accretion rings}

\begin{figure*}
\setlength{\tabcolsep}{2.5pt}
  \begin{tabular}{c c c c}
  \includegraphics[width=0.24\textwidth]{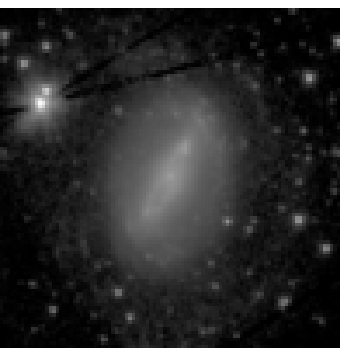}&
  \includegraphics[width=0.24\textwidth]{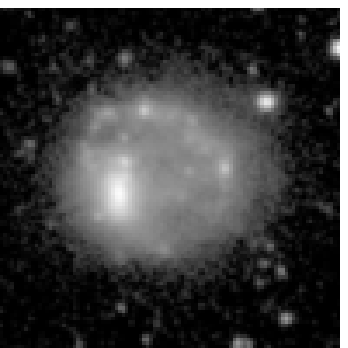}&
  \includegraphics[width=0.24\textwidth]{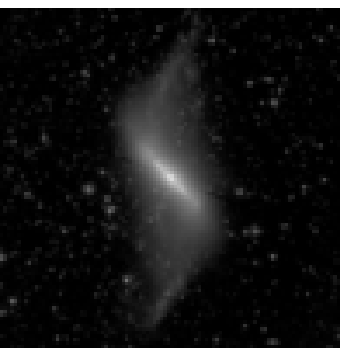}&
  \includegraphics[width=0.24\textwidth]{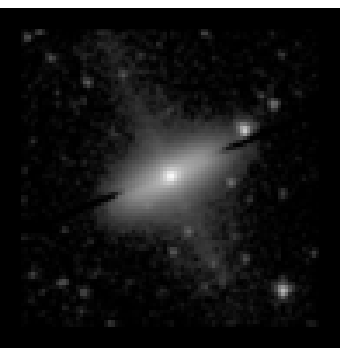}\\
  NGC~6012\hspace{3mm}(R$^{\prime}$)SB(r,x$_1$r)ab&
  NGC~2793\hspace{3mm}RG pec&
  NGC~660\hspace{3mm}PRG&
  NGC~5122\hspace{3mm}S0$^{-}$ sp + PRG sp\\
  \end{tabular}
  \caption{\label{fweird} Selection of S$^4$G galaxies hosting a variety of non-resonance rings presented in the same way as those in Figure~\ref{for}. The first frame shows a galaxy with an x$_1$-ring, the second one a galaxy with a collisional one, and the last two show galaxies with polar rings.}
\end{figure*}

Although the rings discussed in this subsection have a non-resonant origin, they have been included in ARRAKIS for the sake of completeness. 

x$_1$-rings, indicated as x$_1$r in the morphological classification, are very elongated and parallel to bars. Collisional rings, indicated as RG, are in general as large as outer rings, but their centres are typically significantly offset from the galaxy nucleus. They often appear to be intrinsically elliptical \citep[see, e.g.,][]{AP96}. Finally, polar rings, indicated as PRG, are typically seen as needle-like features roughly perpendicular to the major axis of the galaxy, although they can be at other angles, like in the case of NGC~660, which is shown in Figure~\ref{fweird} \citep{WHIT90, DRIEL95}.

Accretion rings are thought to be similar to polar rings, but in their case the material has been accreted at a small angle compared to the galaxy main plane. Since accretion rings are indistinguishable from resonance rings unless kinematic data are available, we included them in the statistics together with resonance rings. To our knowledge, the only rings in this paper that belong to this category are the inner ring in NGC~3593 \citep{COR98}, the inner ring in NGC~4138 \citep{JO96, THA97}, and the innermost ring in NGC~7742 \citep{SIL06, MAZ06}.

A selection of images showing galaxies with x$_1$, collisional, and polar rings observed in the S$^4$G is presented in Figure~\ref{fweird}.

\section{Preparation of the catalogue}

\label{sprepa}

\subsection{Description of the projected shapes of rings}

\begin{figure*}
  \includegraphics[width=1.\textwidth]{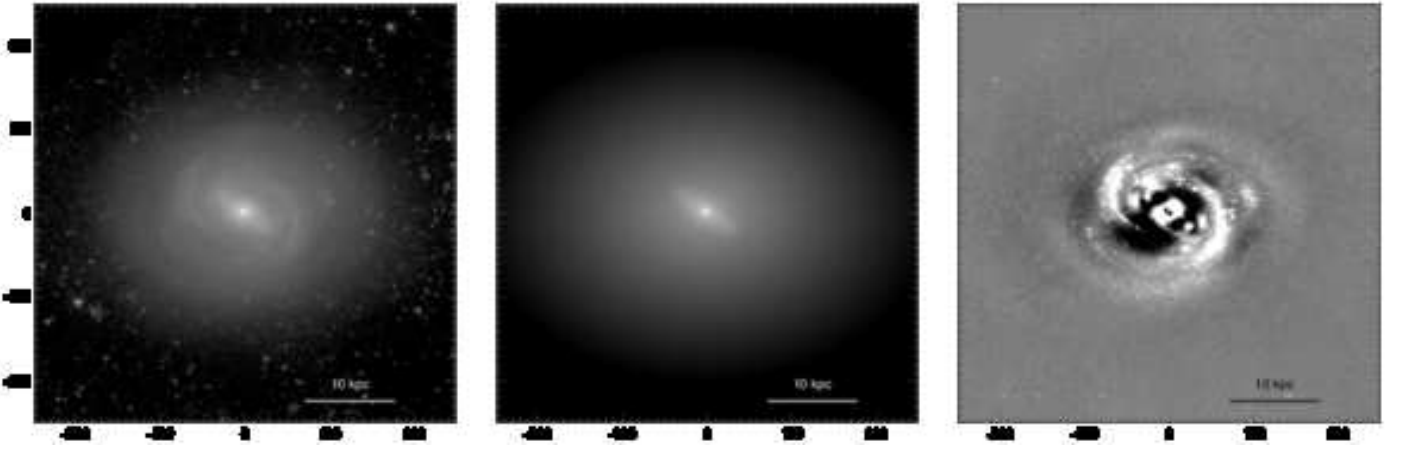}
  \caption{\label{fsubtr} Original image, P4 model, and model-subtracted image of NGC~4579. The model has three components, namely a bulge, a bar, and a disc. NGC~4579 is a galaxy classified as (RL,R$^{\prime}$)SB(r${\underline{\rm s}}$)a by Buta et al.~(in preparation). NGC~4579 has an exceedingly diffuse outer ring as seen in the first frame. The ring becomes more obvious in the model-subtracted image. The axis units are arcseconds.}
\end{figure*}

\begin{figure*}
  \includegraphics[width=1.\textwidth]{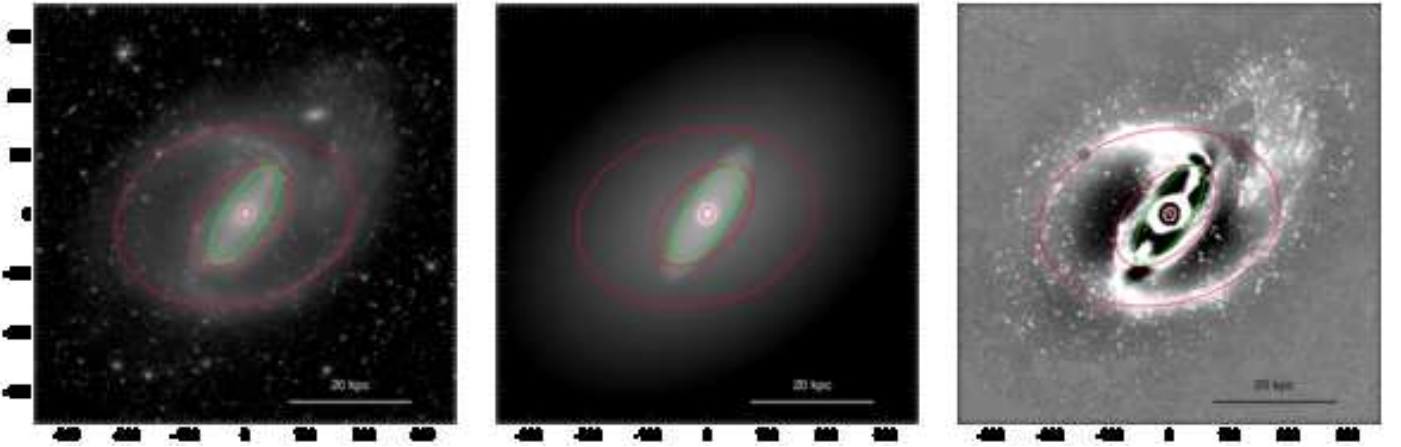}
  \caption{\label{fellipr} Original image, P4 model, and model-subtracted image of NGC~1097, which is classified as an (R$^{\prime}$)SB(rs,nr)a${\underline{\rm b}}$ pec galaxy in Buta et al.~(in preparation). The red ellipses indicate in an outside-in order the average of the two ellipse fits made to the outer, the inner, and the nuclear rings. The green ellipses are the result of an ellipse fit at the radius of the highest bar ellipticity. The red crosses indicate the centre of the galaxy. The axis units are arcseconds.}
\end{figure*}

To describe the shapes of rings, we assumed that they are intrinsically circular or elliptical, as was done in other statistical ring studies \citep{VAU80, BU86a, BU93, BU95, KNA02, CO10, GROU10, LAU11}. Because an ellipse in projection is also an ellipse, the geometrical description of a projected ring in these studies was done by giving the sizes of its major and minor axes ($D_{\rm r}$ and $d_{\rm r}$) or by giving its major axis and its axial ratio ($q_{\rm r}=d_{\rm r}/D_{\rm r}$). To fully characterize the ring geometrical properties, its projected major axis position angle, ${\rm PA_{r}}$, is sometimes given as well.

Although the original S$^4$G frames have been used for the identification of the rings in Buta et al.~(in preparation), working with these images is not always optimal when describing their shape because the contrast between rings and the rest of the galaxy is sometimes very low. A way to overcome this problem is to model the emission of the main components of the galaxy (bulge, bar, disc, etc.) and subtract this model from the original image. This results in the so-called residual image where the disc substructure, such as rings and spiral arms, appears contrast-enhanced. This procedure was preferred to other methods to enhance rings such as unsharp masking because the galaxies were modelled by the S$^4$G pipeline~4 (P4; Salo et al.~in preparation). In short, P4 models were made by fitting the S$^4$G images with up to four components (nucleus, bulges, discs, and/or bars) using version 3.0 of {\sc Galfit} \citep{PENG02, PENG10}. When fitting the model, the disc ellipticity ($\epsilon_{\rm d}=1-q_{\rm d}$) and position angle (PA$_{\rm d}$) were fixed to the values corresponding to deep outer-disc isophotes obtained with an ellipse-fitting routine. A model galaxy image and a residual image were obtained as an output of P4. For the few ARRAKIS galaxies not included in the S$^4$G sample, additional ellipse fits and decompositions were made using the P4 software.

We examined the original and residual images of the galaxies and decided for each ring in which image, original or residual, it was better defined, and thus in which image its shape was easier to measure. Although a priori this should always be the residual image (see, e.g., the example in Figure~\ref{fsubtr}), because of the galaxies with complicated morphologies and/or because P4 models were intentionally kept simple, some residual images have artefacts that hinder defining the outline of a ring. This occurs more often when looking at the inner features in galaxies with strong bars. 

After selecting the image in which the rings were better defined, we used IRAF's\footnote{IRAF is distributed by the National Optical Astronomy Observatories, which are operated by the Association of Universities for Research in Astronomy, Inc., under cooperative agreement with the National Science Foundation.} \citep{TO86, TO93} TVMARK task to describe them. When used in interactive mode, TVMARK allows marking points in an image displayed in DS9 \citep{JO03}. This was used to characterise the shape of rings by marking by hand bright star-forming patches belonging to the feature or stellar emission that probably belonged to the ring, as was done in \citet{GROU10}. All the ring features were measured by the same person (S.~Comer\'on). This procedure was repeated twice for each ring, typically with several days to a few months between the two measurements.

The points marking the outline of the rings for each of the two sets of measurements were fitted with an ellipse using a least-squares fitting algorithm. The output of the fits was the ring diameter in the major axis direction $D_{\rm r}$, the diameter along the minor axis direction $d_{\rm r}$, and the major axis PA (PA$_{\rm r}$). The data used for the statistics in this paper are presented in Appendix~A and were obtained by averaging the $D_{\rm r}$, $d_{\rm r}$, and PA$_{\rm r}$ values obtained from the fits to the two sets of measurements for each ring. Ellipses made from these averaged values are overlayed on the atlas images and in the example shown in Figure~\ref{fellipr}.

\subsection{Accuracy of the measured projected ring properties}

\label{saccuracy}

\begin{figure*}
  \begin{tabular}{c c}
  \includegraphics[width=0.45\textwidth]{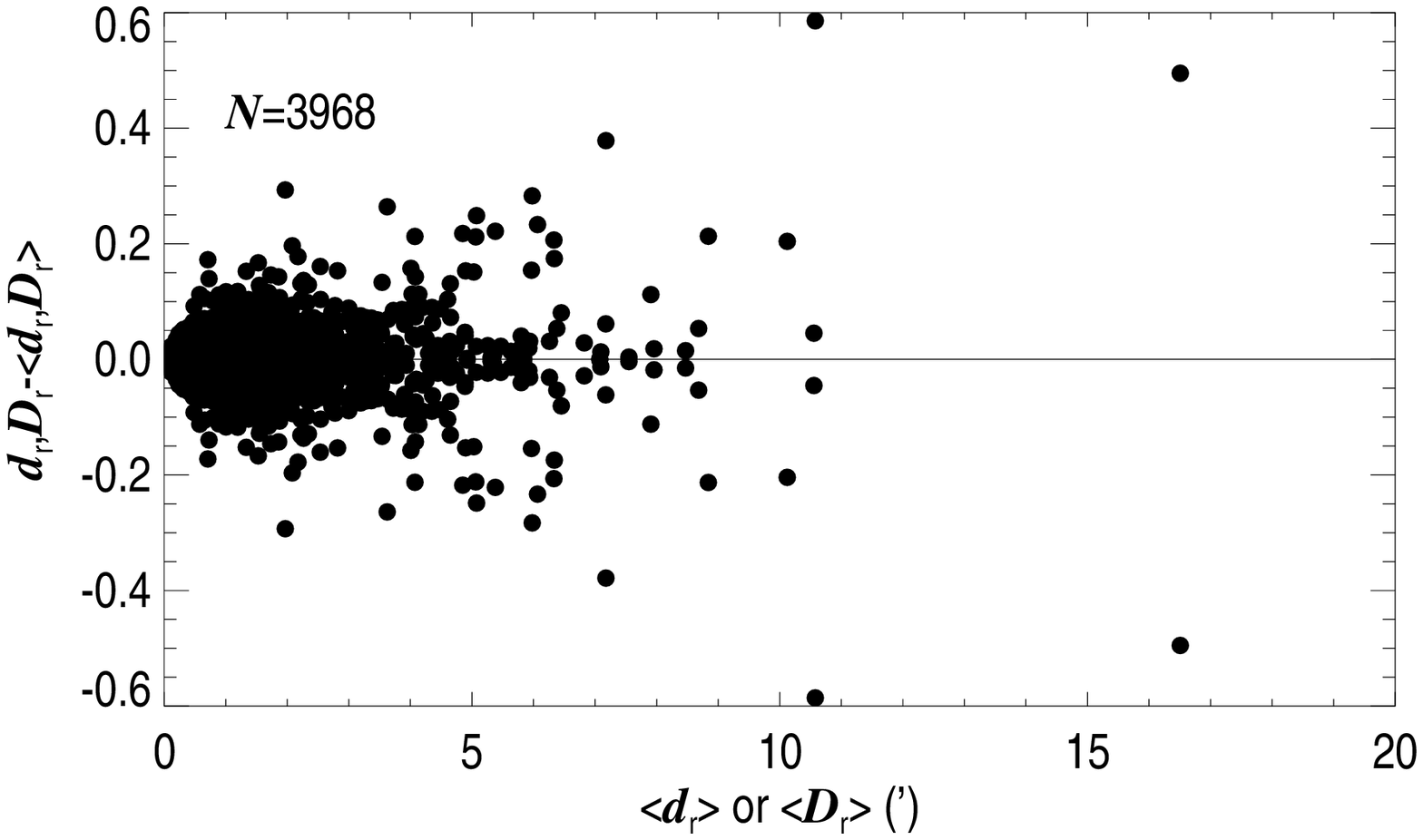}&
  \includegraphics[width=0.45\textwidth]{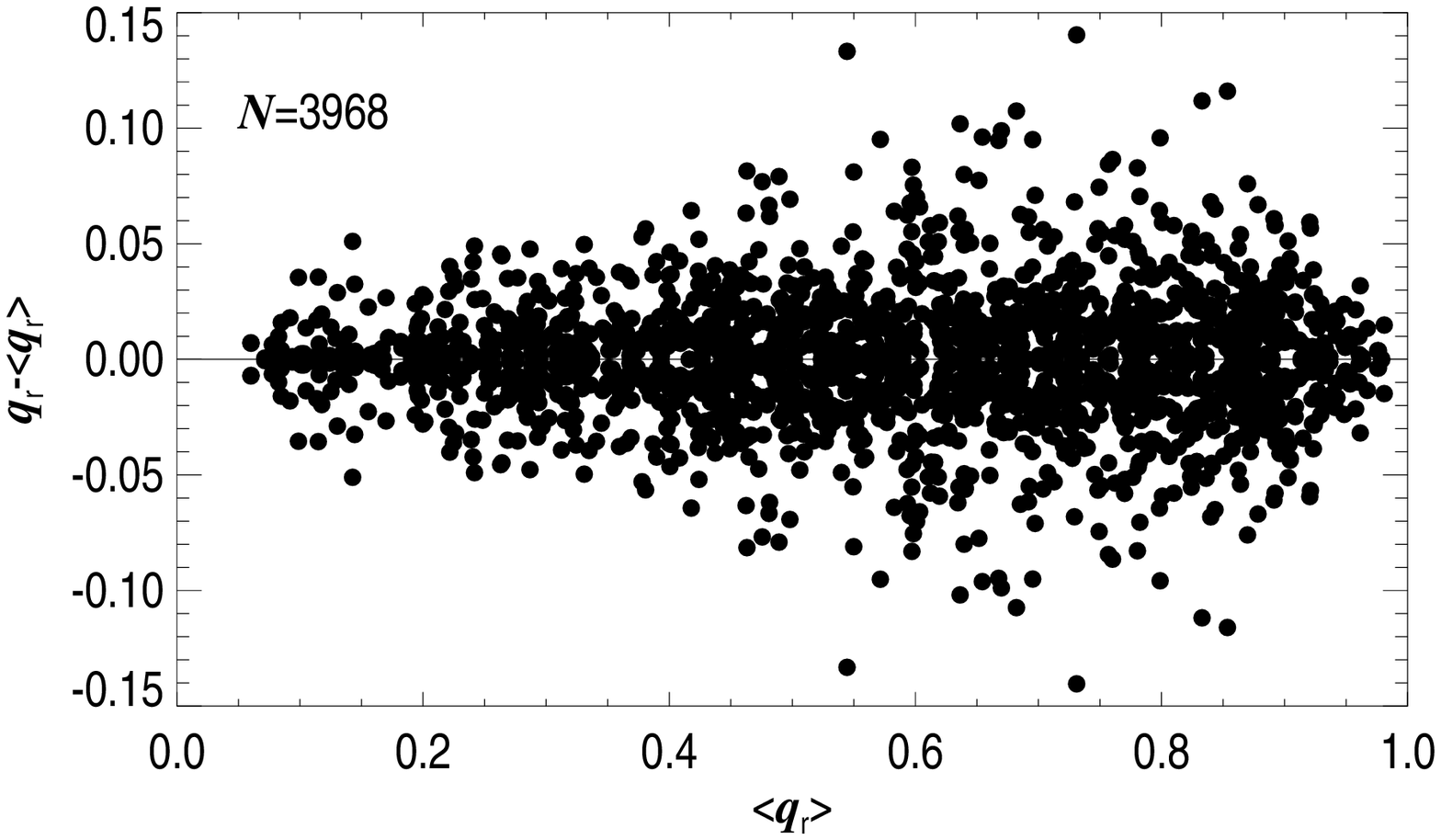}\\
  \includegraphics[width=0.45\textwidth]{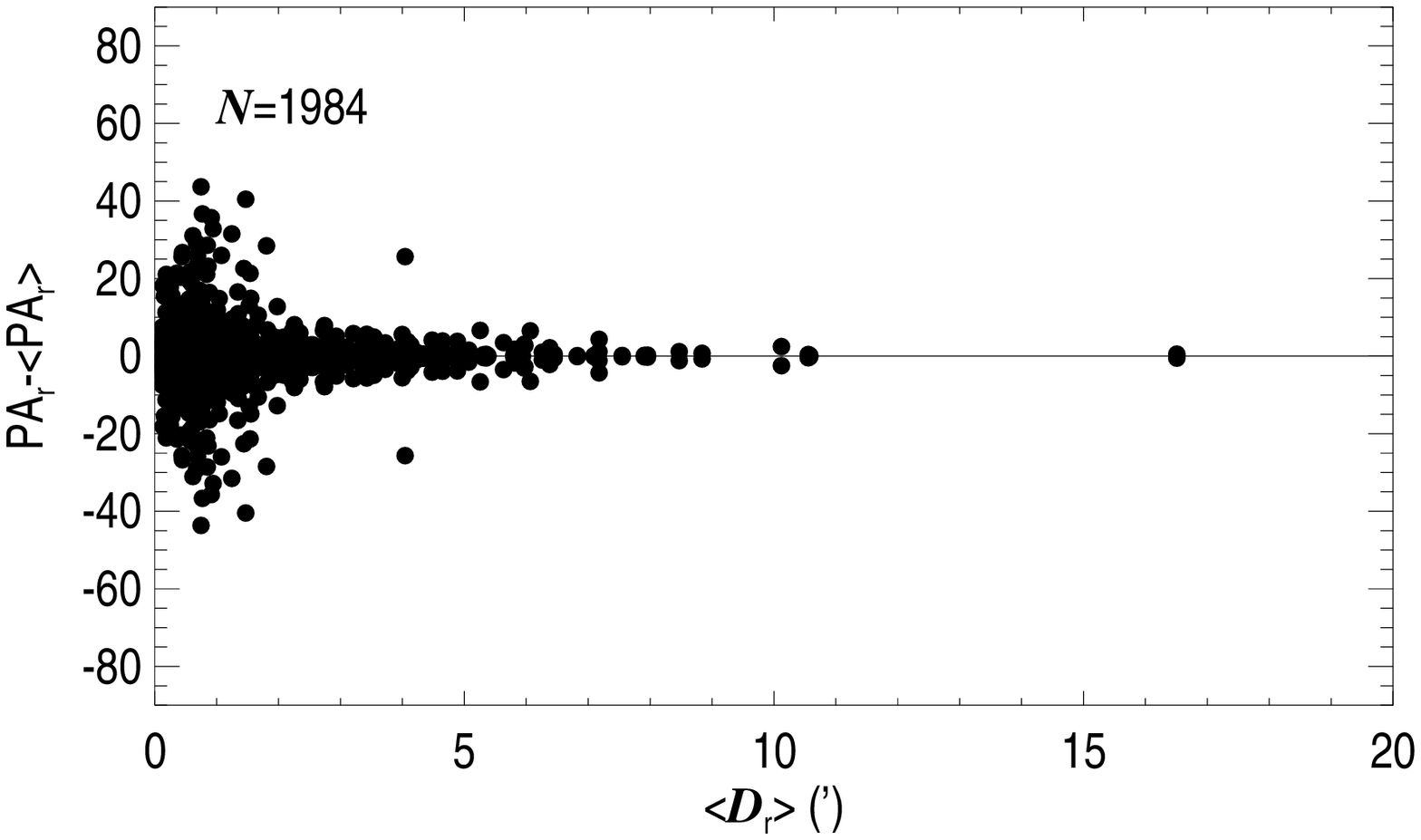}&
  \includegraphics[width=0.45\textwidth]{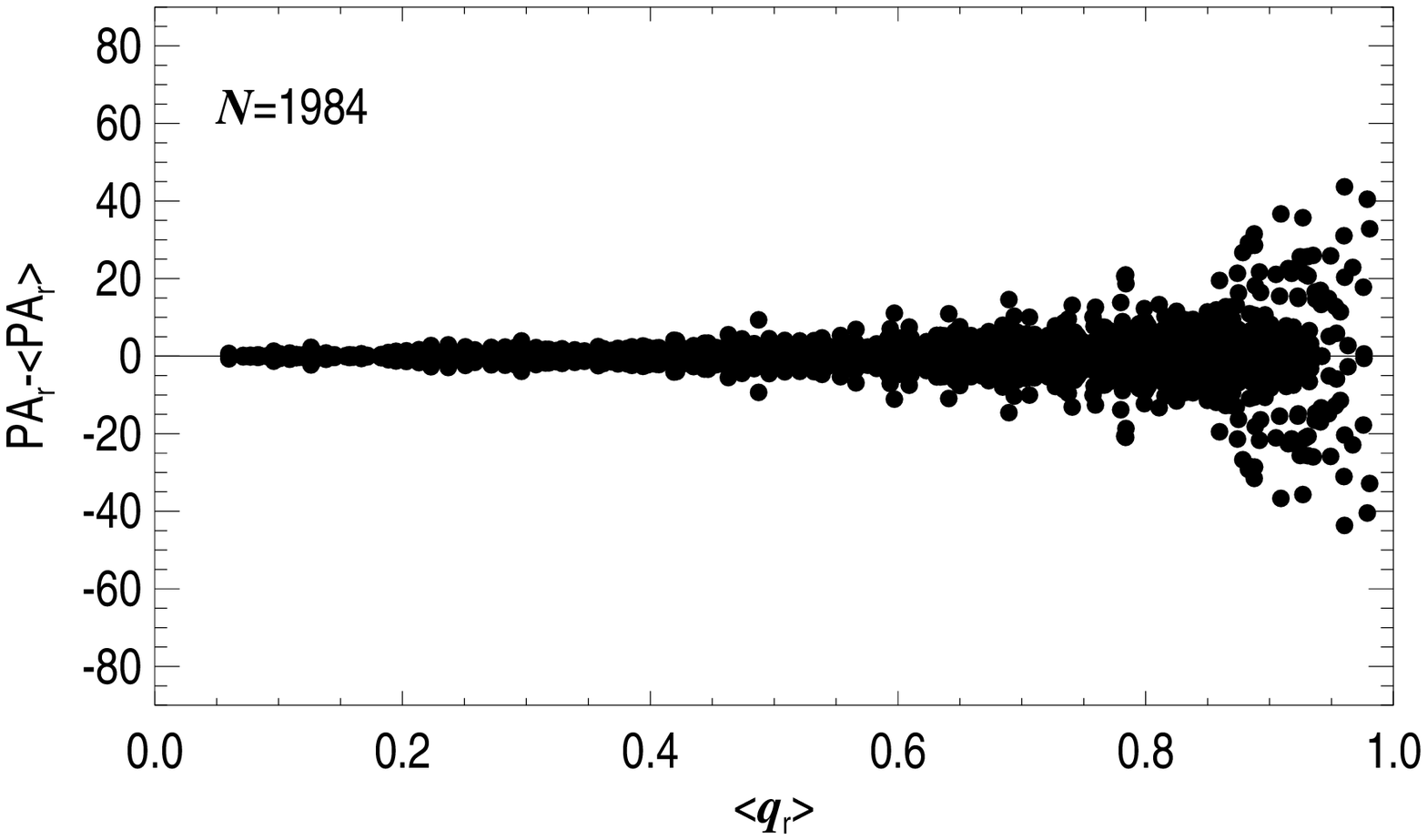}\\
  \end{tabular}
  \caption{\label{ferr} Top-left panel: residuals of the ring diameters along the major and minor axes as a function of the averaged diameters, $\left<d_{\rm r},D_{\rm r}\right>$. Top-right panel: residuals of the ring axis ratios, $q_{\rm r}=d_{\rm r}/D_{\rm r}$, as a function of the mean axis ratios $\left<q_{\rm r}\right>=\left<d_{\rm r}\right>/\left<D_{\rm r}\right>$. Bottom-left panel: residuals of the ring position angles as a function of the averaged ring major axis $\left<D_{\rm r}\right>$. Bottom-right panel: residuals of the ring position angles as a function of the mean ring axis ratios. $N$ refers to the number of points in each plot.}
\end{figure*}

\begin{figure}
  \begin{tabular}{l}
  \includegraphics[width=0.45\textwidth]{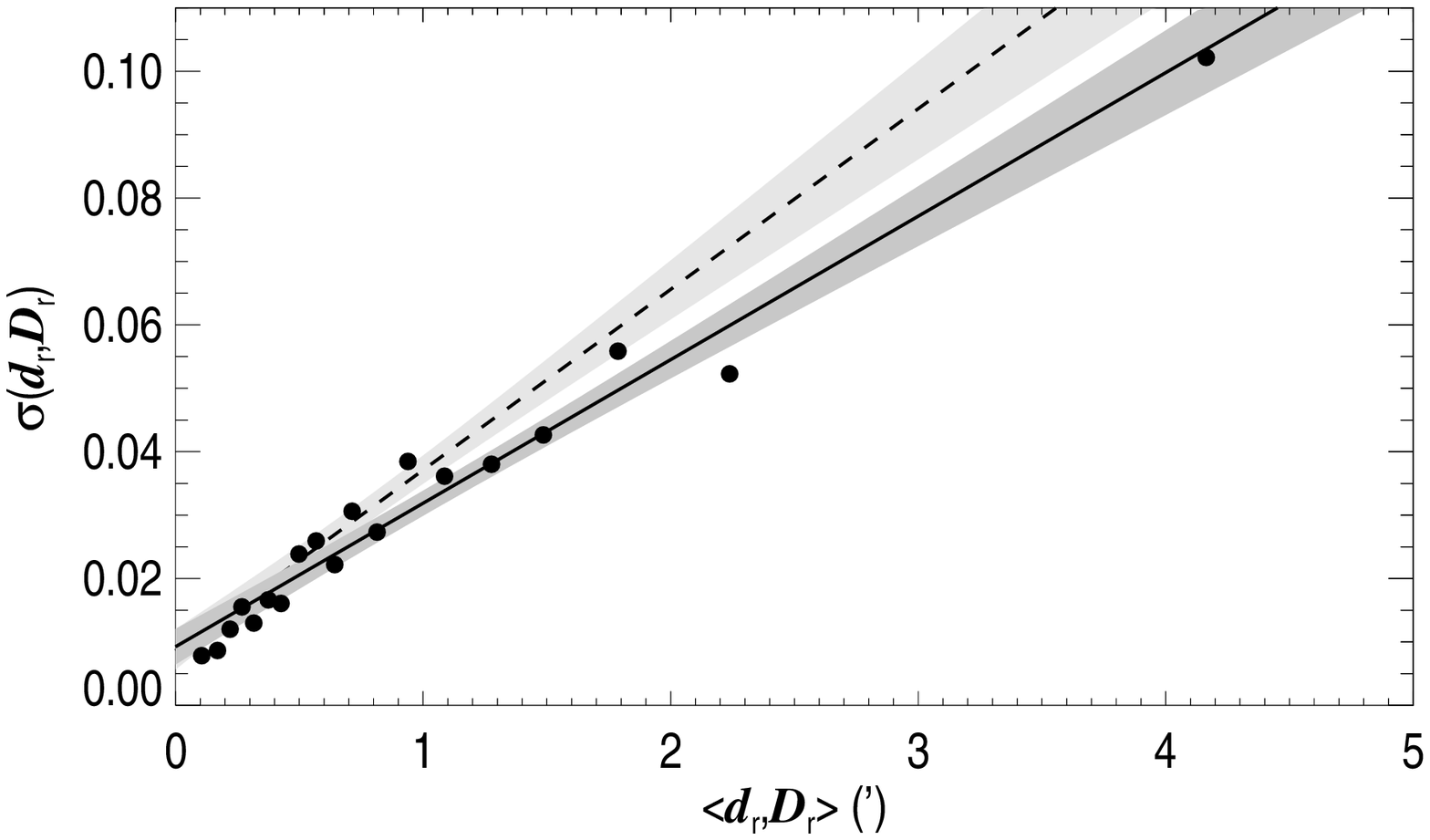}\\
  \includegraphics[width=0.45\textwidth]{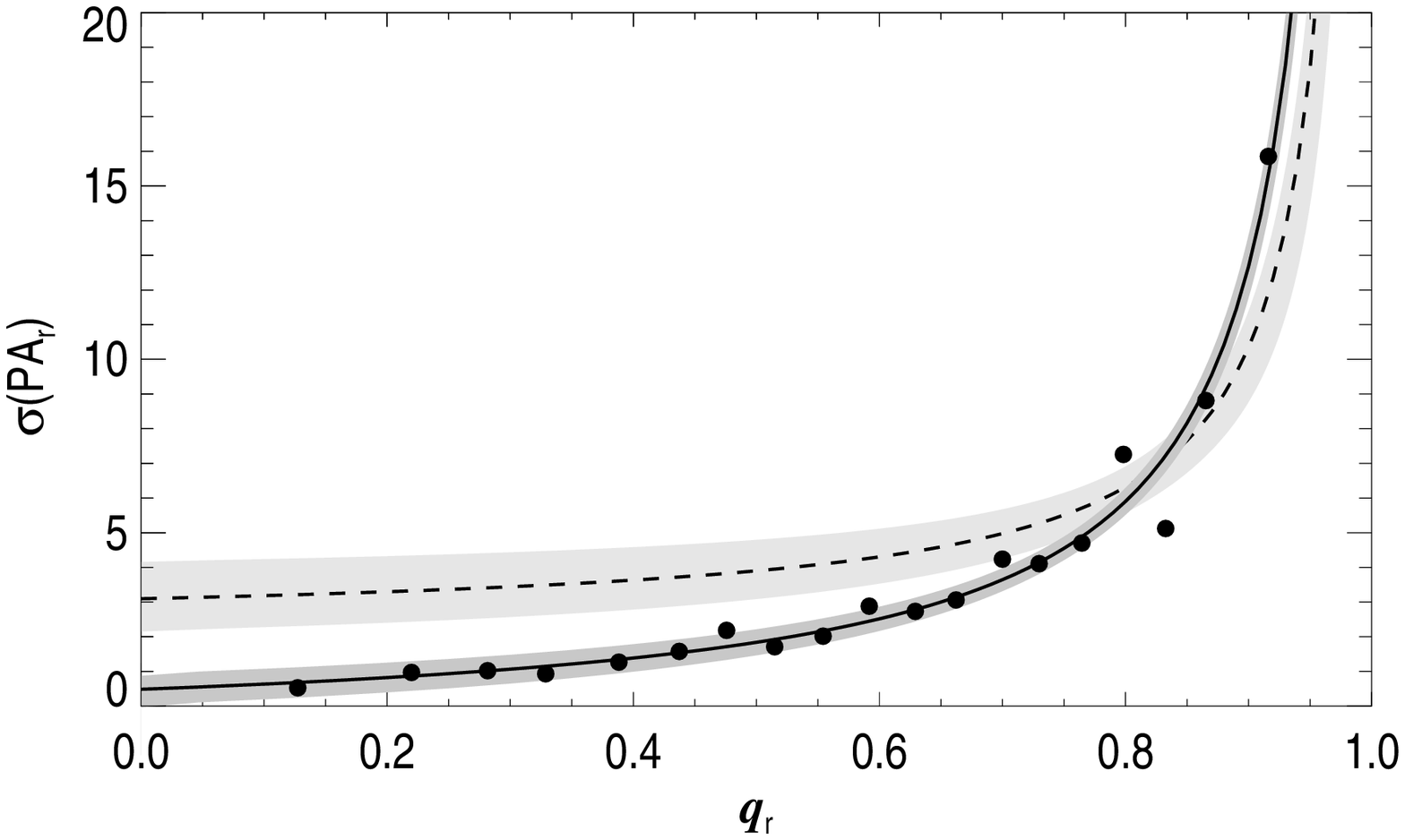}\\
  \end{tabular}
  \caption{\label{ferr2} Top panel: internal errors of $d_{\rm r},D_{\rm r}$ as a function of $\left<d_{\rm r},D_{\rm r}\right>$. Bottom panel: internal errors of PA$_{\rm r}$ as a function of $q_{\rm r}$. The solid lines represent the fits described in Equations~\ref{e1} and \ref{e2} and the dashed lines correspond to the same expressions as presented in the CSRG. Darker grey areas indicate the 95\% confidence interval of the fit for ARRAKIS and lighter grey indicates the same for the CSRG.}
\end{figure}

To study the accuracy of the $D_{\rm r}$, $d_{\rm r}$, and PA$_{\rm r}$ measurements, we examined the residuals between the values of these parameters obtained for each of the two fits and the averaged values, that is, $d_{\rm r},D_{\rm r}-\left<d_{\rm r},D_{\rm r}\right>$ and PA$_{\rm r}-\left<\rm{PA}_{\rm r}\right>$. These residuals can be considered as a rough estimate of internal errors caused by the observer's judgement (Figure~\ref{ferr}). However, we warn that these error estimates do not include those associated with the use of a given method when measuring ring properties. A different ring measurement method may yield values differing from those presented here by many times the error estimate that we discuss.

The top-left panel in Figure~\ref{ferr} shows how the error in the major and minor diameters tends to grow with ring radii: for small diameters, the $d_{\rm r},D_{\rm r}-\left<d_{\rm r},D_{\rm r}\right>$ never exceeds $0\farcm2$, but for larger diameters, it can increase up to $0\farcm6$. The four points with larger $d_{\rm r},D_{\rm r}-\left<d_{\rm r},D_{\rm r}\right>$ correspond, from left to right, to the major axis diameters of the outer features of NGC~4736 and NGC~4258. The outer ring of NGC~4736 appears to be double in its south-eastern side. Our choice in selecting the outline of the ring changed between the two measurements, which explains the large error. NGC~4258 is a galaxy for which the S$^4$G frame does not fully cover the disc, thus a significant fraction of the outer ring lies outside the frame and cannot be studied, causing a large uncertainty in the $D_{\rm r}$ measurement. The range of $d_{\rm r},D_{\rm r}-\left<d_{\rm r},D_{\rm r}\right>$ values is similar to that found in the catalog of southern ringed galaxies \citep[CSRG; Figure~5a in][]{BU95}. 

The errors in the measured ring axis ratios are presented in the top-right panel of Figure~\ref{ferr}. We found that the average error is $|\Delta q_{\rm r}|=0.02$.

The PA$_{\rm r}$ uncertainty is larger for smaller and/or rounder rings, as seen in the bottom panels of Figure~\ref{ferr}; this is natural since for a round ring PA$_{\rm r}$ is undefined. The range of PA$_{\rm r}-\left<{\rm PA}_{\rm r}\right>$ in our study and its behaviour as a function of $D$ and $\left<q_{\rm r}\right>=1-\epsilon_{\rm r}=\left<d_{\rm r}\right>/\left<D_{\rm r}\right>$ is qualitatively similar to that in Figures~5c,d of the CSRG.

Because the CSRG is the largest ring catalogue with an internal error estimate, it can be compared with our results. The CSRG data come from analogically measuring diameters and PAs in photographic plates, but here we present data measured by digital means, which may yield significantly different accuracies. 

The CSRG presents two internal error functions that we reproduce here with S$^4$G data. The internal error function for the diameter residuals was obtained by sorting the residuals according to their mean diameter and then calculating the standard deviation for groups of 200 residuals, $\sigma(d_{\rm r},D_{\rm r})$ (top panel in Figure~\ref{ferr2}). As in the CSRG, we linearly fitted the points and obtained
\begin{equation}
 \label{e1}
 \sigma(d_{\rm r},D_{\rm r})=0\farcm009+0.023\left<d_{\rm r},D_{\rm r}\right>
\unskip.
\end{equation}
Since in ARRAKIS the error grows slower with increasing $d_{\rm r},D_{\rm r}$, our accuracy is slightly better than that obtained in the CSRG ($\sigma(d_{\rm r},D_{\rm r})=0\farcm009+0.029\left<d_{\rm r},D_{\rm r}\right>$). The 95\% confidence bands in Figure~\ref{ferr2} were calculated with the assumption that the errors are normally distributed. The statistics for the confidence bands corresponding to the CSRG were obtained by measuring the position of the data points in a magnified version of Figure~5e in the CSRG.

The internal errors for the ring orientations were calculated by sorting the PA$_{\rm r}$ residuals according to their $q_{\rm r}$ and then calculating the standard deviation for groups of 100 residuals (bottom panel in Figure~\ref{ferr2}). As in the CSRG, we produced a fit that scales inversely with the fitted ellipticity
\begin{equation}
 \label{e2}
 \sigma({\rm PA}_{\rm r})=-0\fdg87+1\fdg35/(1-q_{\rm r})
\unskip, 
\end{equation}
and thus our PA$_{\rm r}$ determinations are better than those in the CSRG for $q_{\rm r}<0.83
\unskip$ [in the CSRG $\sigma({\rm PA}_{\rm r})=2\fdg29+0\fdg81/(1-q_{\rm r})$]. Our PA$_{\rm r}$ measurements are at least twice as precise than those in the CSRG for $q_{\rm r}<0.53
\unskip$. However, perhaps surprisingly, the manual PA$_{\rm r}$ determination in the CSRG seems to have fewer internal errors than ARRAKIS for rings that are nearly round ($q_{\rm r}>0.83
\unskip$). This is no problem because ${\rm PA_{r}}$ becomes meaningless for almost round rings. We calculated the confidence bands for this fit in the same way as for the top panel. For the CSRG fit, we made the statistics based on the information on CSRG's Figure~5f.

We can thus conclude that the internal errors in ARRAKIS are the smallest for such a large dataset of rings so far and that this can be attributed to the use of digital deep images.

\subsection{Obtaining the projected bar parameters}

Simulations by \citet{SCH81, SCH84} and the flux tube manifold orbital calculations \citep{ATH09A, ATH09B} predict R$_1$ rings to be elongated and perpendicular to bars, R$_2$ rings to be elongated and parallel to bars, and inner rings to be elongated and parallel to bars. It is thus interesting to know the bar PA in galaxies with rings. We also measured the highest bar ellipticity $\epsilon_{\rm b}=1-q_{\rm b}=1-d_{\rm b}/D_{\rm b}$ ($D_{\rm b}$ and $d_{\rm b}$ are the bar major and minor diameters and $q_{\rm b}$ is the bar axis ratio). $\epsilon_{\rm d}$ roughly scales with more sophisticated bar strength indicators such as the bar torque \citep{BLOCK01, LAU02}.

The bar data were obtained by studying the ellipticity fits from S$^4$G's P4. We searched for the peak ellipticity within the radius of the bar, regardless of how big the drop in the ellipticity after the maximum was or how the PA varied with radius. We considered the PA at that radius to be representative of the bar orientation. Moreover, the maximum ellipticity radius is typically a lower estimate of the bar length \citep{ER05}, except in galaxies in which the bar merges with rings and/or spiral arms. In the catalogue of Appendix~A, we present the diameters along the major and minor axes of the ellipse fits at the position of the bar maximum ellipticity ($D_{\rm b}$ and $d_{\rm b}$) and the bar PA at the same position, (PA$_{\rm b}$). We checked that each individual PA$_{\rm b}$ roughly corresponds to what would be expected from visual inspection. If it did not, we labelled it as unsuitable for bar analysis, as discussed at the end of Section~\ref{sreliability}.

\subsection{Obtaining the intrinsic ring and bar shape and orientation}

\label{sdeproj}

ARRAKIS is the first large ring catalogue that provides intrinsic ring shapes and orientations. The reason is that deprojecting rings requires knowing, within a few percent, the disc orientation parameters, that is, the ellipticity, $\epsilon_{d}=1-q_{\rm d}$, and the position angle, PA$_{\rm d}$. Under the assumption of discs being intrinsically circular, one needs very deep images to measure $\epsilon_{d}$ and PA$_{\rm d}$ in the outer disc, a region where the effect of perturbing non-axisymmetries such as bars is most likely weak. This was not possible for the CSRG because it was produced using photographic plates, but was performed in \citet{KNA02}, \citet{CO10}, and \citet{GROU10}.

The S$^4$G P4 makes use of deprojection parameters obtained from isophote fits at low surface brightness (for more details see Salo et al.~in preparation). These are the deprojection parameters that we used here for obtaining intrinsic ring and bar shapes ($q_{\rm r,0}$ and $q_{\rm b,0}$) except for the few ARRAKIS galaxies not in the S$^4$G sample (but appearing in S$^4$G frames) and NGC~4698. For this last galaxy our interpretation of the deprojection parameters is very different from that in P4. P4 deprojection parameters were also used to calculate the counter-clockwise angular distance between the line of nodes and the ring and bar major axis ($\theta_{\rm r}$ and $\theta_{\rm b}$), under the assumptions that the outer parts of discs are circular and that rings and bars are roughly elliptical. The equations that we used for deprojection can be found in Appendix~A of \citet{GA07}. The deprojected $D$ and $d$ values for both rings and bars are presented in the catalogue of Appendix~A. The angle difference between the deprojected major axis and the line of nodes of the deprojection is also indicated. The only rings for which we have not included deprojected parameters are the polar rings, because they are known not to lie in the galaxy disc plane.

\subsection{Reliability of the deprojected bar parameters}

\label{sreliability}

\begin{table}
 \caption{Parameters of the model galaxies used to study the reliability of the bar deprojected parameters}
 \label{tmodels}
 \centering
 \begin{tabular}{c c c c c c c}
 \hline\hline
 Model & $\rho_{\rm bulge,0}$ & $a$ & $b$ & $c$ & Bulge/ & Bar/ \\
       &                      &     &     &     & Total  & Total\\
 \hline
 Model~1 & 5  & 30 & 10 & 3 & 11\% & 3\% \\
 Model~2 & 10 & 30 & 10 & 3 & 20\% & 2\% \\
 Model~3 & 5  & 50 & 15 & 4 & 11\% & 8\%\\
 Model~4 & 10 & 50 & 15 & 4 & 19\% & 7\%\\
 \hline
 \end{tabular}
  \tablefoot{Bulge/Total stands for the bulge-to-total mass ratio and Bar/Total stands for the bar-to-total mass ratio.}
\end{table}

\begin{figure*}
  \includegraphics[width=0.9\textwidth]{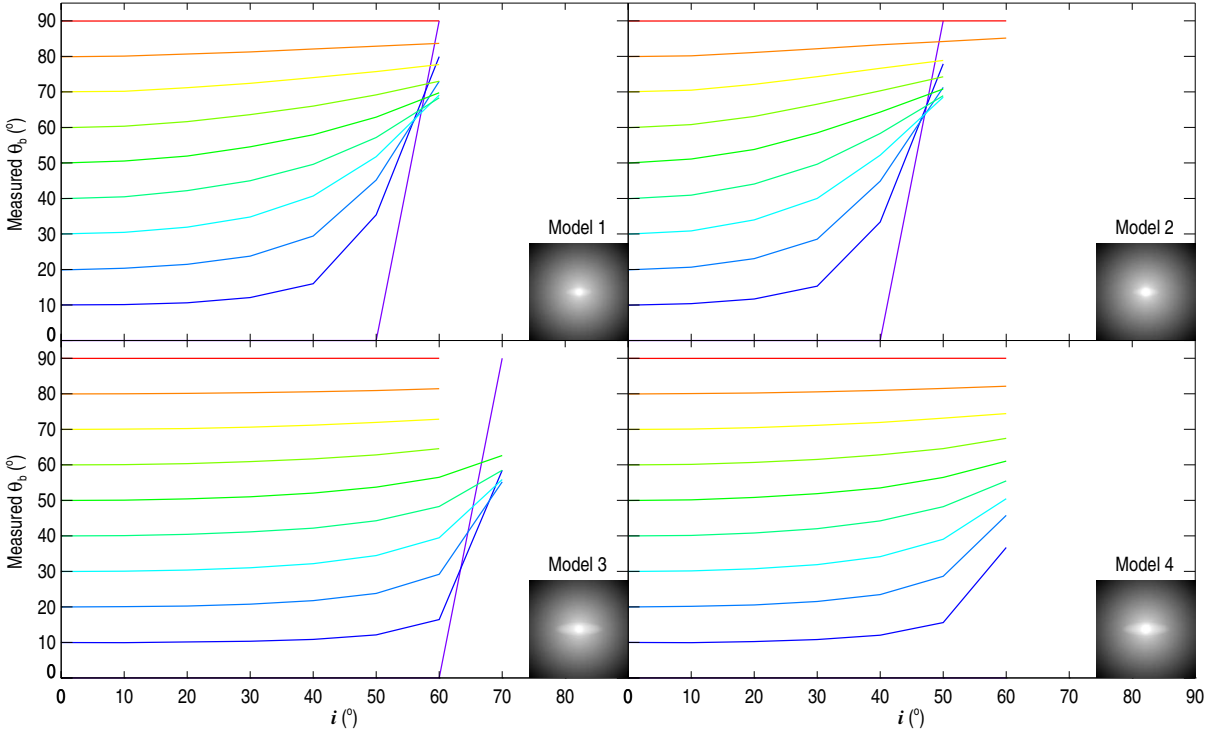}
  \caption{\label{fbpa} Measured orientation of the bar with respect to the line of nodes of the galaxy ($\theta_{\rm b}$) as a function of the galaxy inclination ($i$) for each of our four galaxy models described in the text and Table~\ref{tmodels}. Each coloured line represents a single real $\theta_{\rm b}$ value, which is the one corresponding to the position at which the line crosses the $i=0^{\rm o}$ axis, that is, from purple to red (bottom to top), $\theta_{\rm b}=0^{\rm o}$, $\theta_{\rm b}=10^{\rm o}$, $\theta_{\rm b}=20^{\rm o}$, $\theta_{\rm b}=30^{\rm o}$, $\theta_{\rm b}=40^{\rm o}$, $\theta_{\rm b}=50^{\rm o}$, $\theta_{\rm b}=60^{\rm o}$, $\theta_{\rm b}=70^{\rm o}$, $\theta_{\rm b}=80^{\rm o}$, and $\theta_{\rm b}=90^{\rm o}$ . The images at the bottom-right corner of each panel represent the corresponding galaxy when $\theta_{\rm b}=0^{\rm o}$ and $i=0^{\rm o}$.}
\end{figure*}

\begin{figure*}
  \includegraphics[width=0.9\textwidth]{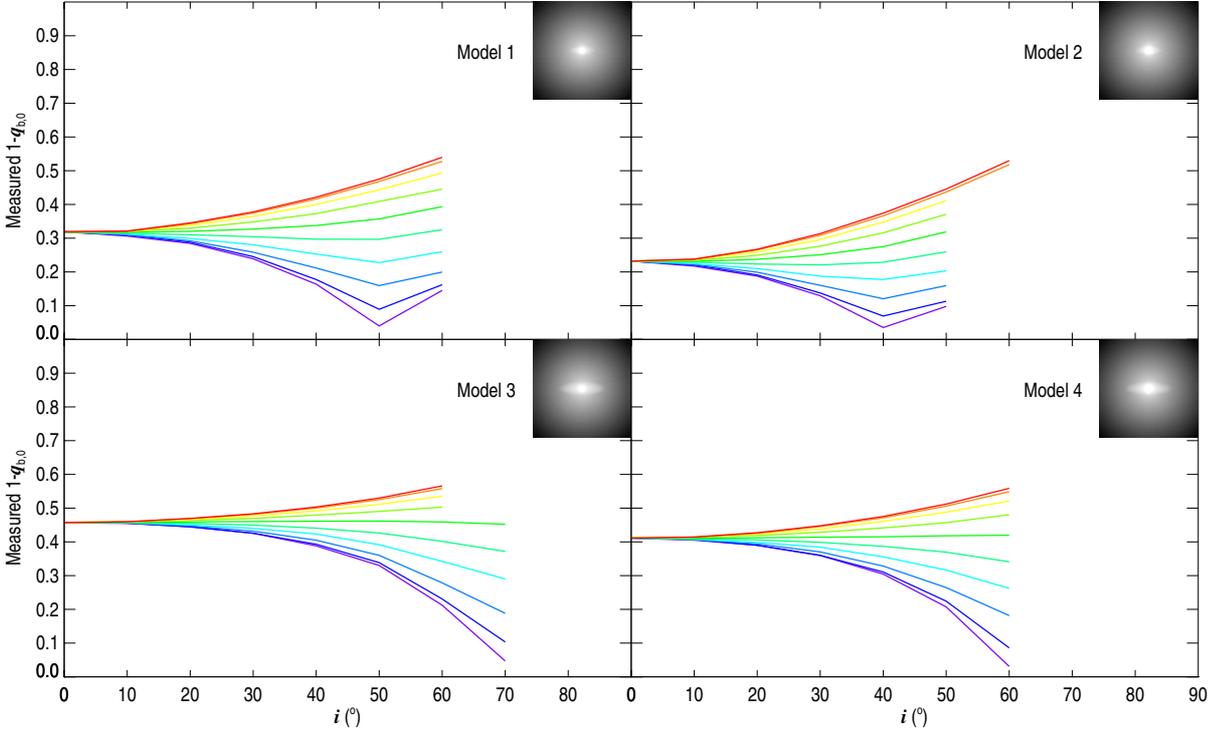}
  \caption{\label{fbellip} Measured bar ellipticity (measured $1-q_{\rm b,0}$) as a function the galaxy inclination ($i$) for each of our four galaxy models. Each coloured line represents a single $\theta_{\rm b}$ value colour-coded in the same way as in Figure~\ref{fbpa} (also ordered here from bottom to top). The intrinsic $1-q_{\rm b,0}$ value is that corresponding to $i=0^{\rm o}$. The images at the top-right corner of each panel represent the corresponding galaxy when $\theta_{\rm b}=0^{\rm o}$ and $i=0^{\rm o}$.}
\end{figure*}

The presence of a bar can be recognised in very inclined galaxies and sometimes also in edge-on galaxies \citep[see, e.g., how box/peanut bulges are related to bars in][]{KUIJ95}. However, obtaining their orientation and ellipticity from ellipse fits can be very difficult. Indeed, the ellipse deprojection would give precise bar parameters for isolated infinitely thin bars. But real bars have some amount of vertical thickening and are embedded in galaxies with discs and bulges. As a natural consequence of that, the more inclined a galaxy, the harder it is to obtain accurate bar intrinsic parameters.

To estimate how precise our bar measurements are, we prepared a set of four toy model galaxies with an exponential disc, a classical spherical bulge, and a Ferrers bar \citep{FER77}. Assuming that the centre of the galaxy is located at $x=0$, $y=0$, and $z=0$ and that the disc lies in the $x-y$ plane, this simplified disc can be described as
\begin{equation}
 \rho_{\rm d}(R,z)=\rho_{\rm d,0}\,e^{-R/h_{\rm R}}\,e^{-|z|/h_{\rm z}},
\end{equation}
where $R$ is the radius in the plane of the galaxy ($R=\sqrt{x^2+y^2}$), $\rho_{\rm d,0}$ stands for the disc central mass density, $h_{\rm R}$ for the disc scale-length, and $h_{\rm z}$ for the disc scale-height.

To describe the bulge we used a Jaffe profile \citep{JAF83} which, when projected onto a plane, results approximately in a de Vaucouleurs profile \citep{VAU48}:
\begin{equation}
 \rho_{\rm bulge}(r)=\rho_{\rm bulge,0}\,\left(\frac{r}{r_{0}}\right)^{-2}\left(1+\frac{r}{r_{0}}\right)^{-2},
\end{equation}
where $r$ is the 3D radius ($r=\sqrt{x^2+y^2+z^2}$), $\rho_{\rm bulge,0}$ controls the bulge mass, and $r_{0}$ controls its size. 

Finally, the bar was described as
\begin{equation}
 \rho_{\rm bar}(m)=\rho_{\rm b,0}\left(1-m^2\right)\,\,\,{\rm for}\,\,\,m\leq1,\,\rho_{\rm b}=0\,\,\,{\rm for}\,\,\,m>1,
\end{equation}
where $\rho_{\rm bar,0}$ is the central bar mass density and
\begin{equation}
 m^2(x,y,z)=\frac{x^2}{a^2}+\frac{y^2}{b^2}+\frac{z^2}{c^2}.
\end{equation}
Here, $a$ is the major axis, $b$ is the minor axis, and $c$ is the vertical axis of the bar.

The four model galaxies had (in arbitrary units) $\rho_{\rm d,0}=1$, $\rho_{\rm bar,0}=1$, $h_{\rm R}=30$, $h_{\rm z}=5$, and $r_0=5$. The other parameters were model-dependent and are listed in Table~\ref{tmodels}. Table~\ref{tmodels} also presents the fraction of mass in the bulge and in the bar for each model. As seen there, the fraction of mass in bars is rather low and thus our fit to models is a worst-case study.

We decided that the $x$ axis would be the line of nodes of our rotation. We first rotated the bar by a given angle $\theta_{\rm b}$ with respect to the line of nodes and then inclined the galaxy by an angle $i$. We defined $\theta_{\rm b}$ to vary from $\theta_{\rm b}=0^{\rm o}$ to $\theta_{\rm b}=90^{\rm o}$ in steps of $10^{\rm o}$. We explored the space of galaxy inclinations from $i=0^{\rm o}$ to $i=80^{\rm o}$ in steps of $10^{\rm o}$. The rotated and inclined galaxies were then projected onto 2D fits images. We used these images to measure the bar properties by fitting ellipticity profiles. We considered that a bar ellipticity and PA could be measured when a peak in ellipticity with prominence larger than 0.01 occurred within the bar radius. We used the ellipticity fit at the maximum ellipticity peak to obtain the deprojected bar parameters, as was done for observed bars in Section~\ref{sdeproj}. The results on the recovery of the original bar PA with respect to the line of nodes, $\theta_{\rm b}$, and the original bar ellipticity, $\epsilon_{\rm b,0}=1-q_{\rm b,0}$, are presented in Figures~\ref{fbpa} and \ref{fbellip}. In several cases in highly inclined galaxies, no clear maximum in ellipticity was found, therefore no deprojection of the bar properties was obtained.

The angle $\theta_{\rm b}$ is typically well recovered for galaxies with $i\leq40^{\rm o}$ when the bar is small compared to the bulge size (models~1 and 2 in Figure~\ref{fbpa}). For galaxies with longer bars (models~3 and 4), $\theta_{\rm b}$ is well recovered for $i\leq60^{\rm o}$. The reason for this is that the bulge, which is intrinsically spherical, and to some extent also the disc, soften the isophotes, which makes them less elliptical. In some cases, for low real $\theta_{\rm b}$ values and high disc inclinations, this effect is so strong that the measured projected ellipticity becomes lower than the ellipticity of the projected disc, causing the measured $\theta_{\rm b}$ to shift by some $90^{\rm o}$ from the real value.

The measured intrinsic bar ellipticity $\epsilon_{\rm b,0}=1-q_{\rm b,0}$ is highly dependent on the real $\theta_{\rm b}$. When the bar is parallel to the line of nodes (real $\theta_{\rm b}\sim0^{\rm o}$), $1-q_{\rm b,0}$ becomes increasingly underestimated for increasing $i$. For bars oriented perpendicular to the line of nodes (real $\theta_{\rm b}\sim90^{\rm o}$), the effect is reversed.

Because the S$^4$G sample is under-abundant in early-type galaxies, bulges are in general not expected to play a significant role when producing ellipticity profiles. Therefore, we considered the parameters of the bars in galaxies with $i\leq60^{\rm o}$ (corresponding to a disc ellipticity $\epsilon_{\rm d}=1-q_{\rm d}\leq0.5$) to be reliably measured. The properties of bars in such discs almost always appear in the catalogue in Appendix~A and the ellipse fit to the highest bar ellipticity is overlayed on the images of the atlas in Appendix~B.

However, even for galaxies with $i\leq60^{\rm o}$ six bar fits were found to be unreliable and were excluded from the appendices. The reason for this is that the bar fit is obviously wrong in galaxies with bars with a low contrast (NGC~2552) or that are influenced by surrounding bright regions such as a ring (NGC~2967, NGC~4351, NGC~5595, NGC~5744, and NGC~6923). For five additional galaxies with $\epsilon_{\rm d}\leq0.5$ classified as barred by Buta et al.~(in preparation), ESO~443-80, NGC~2460, NGC~5633, NGC~6278, and UGC~5814, we were unable to distinguish a peak corresponding to a bar in the ellipticity profiles. Finally, the centre of NGC~4108B is affected by an image artefact that made impossible to measure the bar properties. For galaxies with $i>60^{\rm o}$, we did not include the bar properties in the appendices even for those that are obviously barred.

Moreover, even though inner rings are defined to be surrounding the bar, for           73
\unskip galaxies the maximum in ellipticity that we fitted as an estimate for the bar radius is slightly larger than the inner ring radius in projection and/or in deprojection. This is caused by our very simple approach at describing bars. In most of these cases the bar merges with the ring and the junction points have strong star-forming regions, ans\ae, and/or the beginning of spiral arms, which affect the ellipticity fits in such a way as to move the ellipticity maximum outwards. In other cases, the reason that the bar is larger than the fitted ring diameter may be that some rings slightly deviate from an elliptical shape in such a way that the bar fits within them. These deviations are not captured by our simple ring-fitting approach. In none of the cases considered for the statistics in this paper would this change the bar orientation by more than a few degrees from what would be measured visually.

\subsection{Reliability of the deprojected ellipse parameters in discs with a finite thickness}

\label{sreliability2}

\begin{figure}
  \begin{tabular}{l}
  \includegraphics[width=0.45\textwidth]{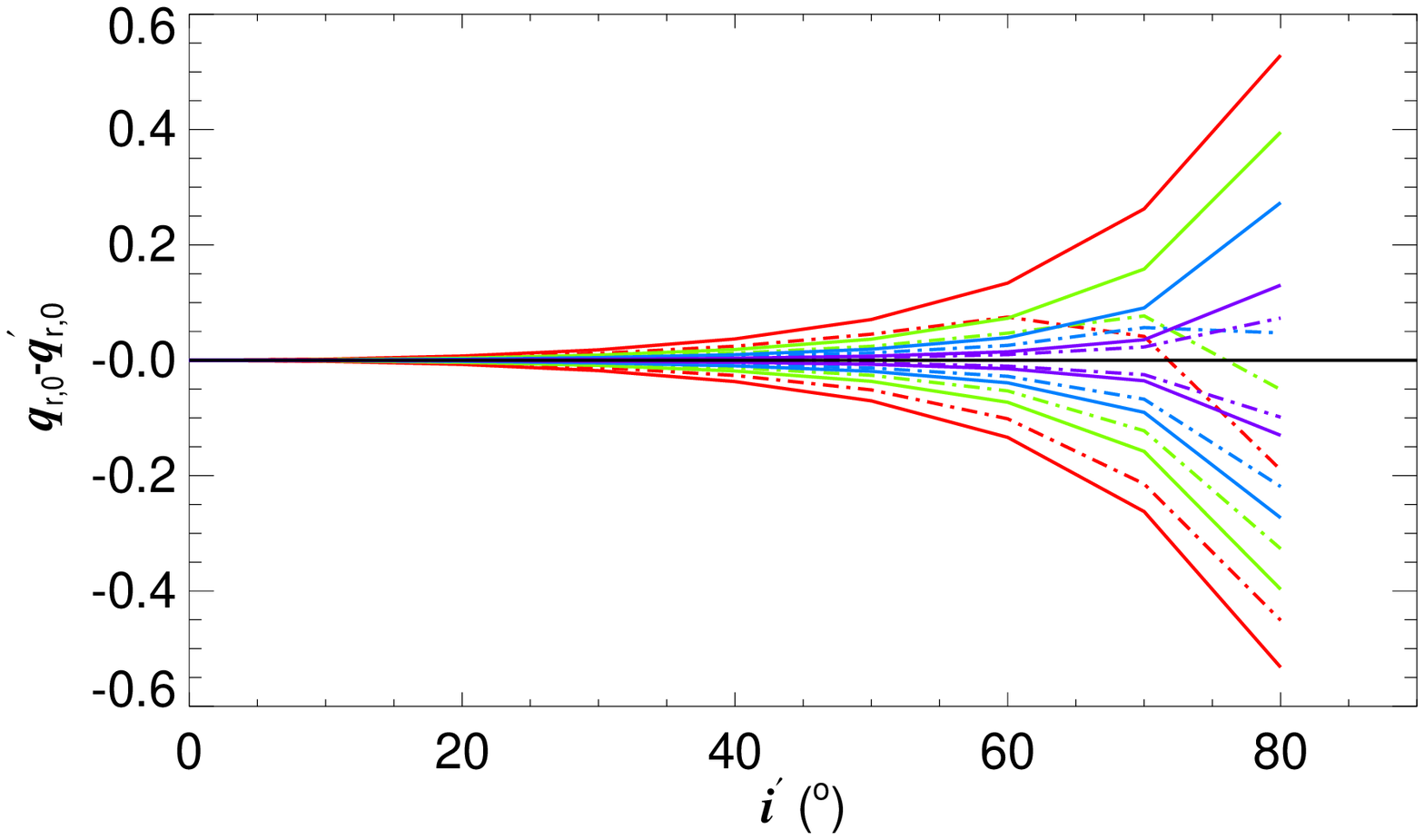}\\
  \includegraphics[width=0.45\textwidth]{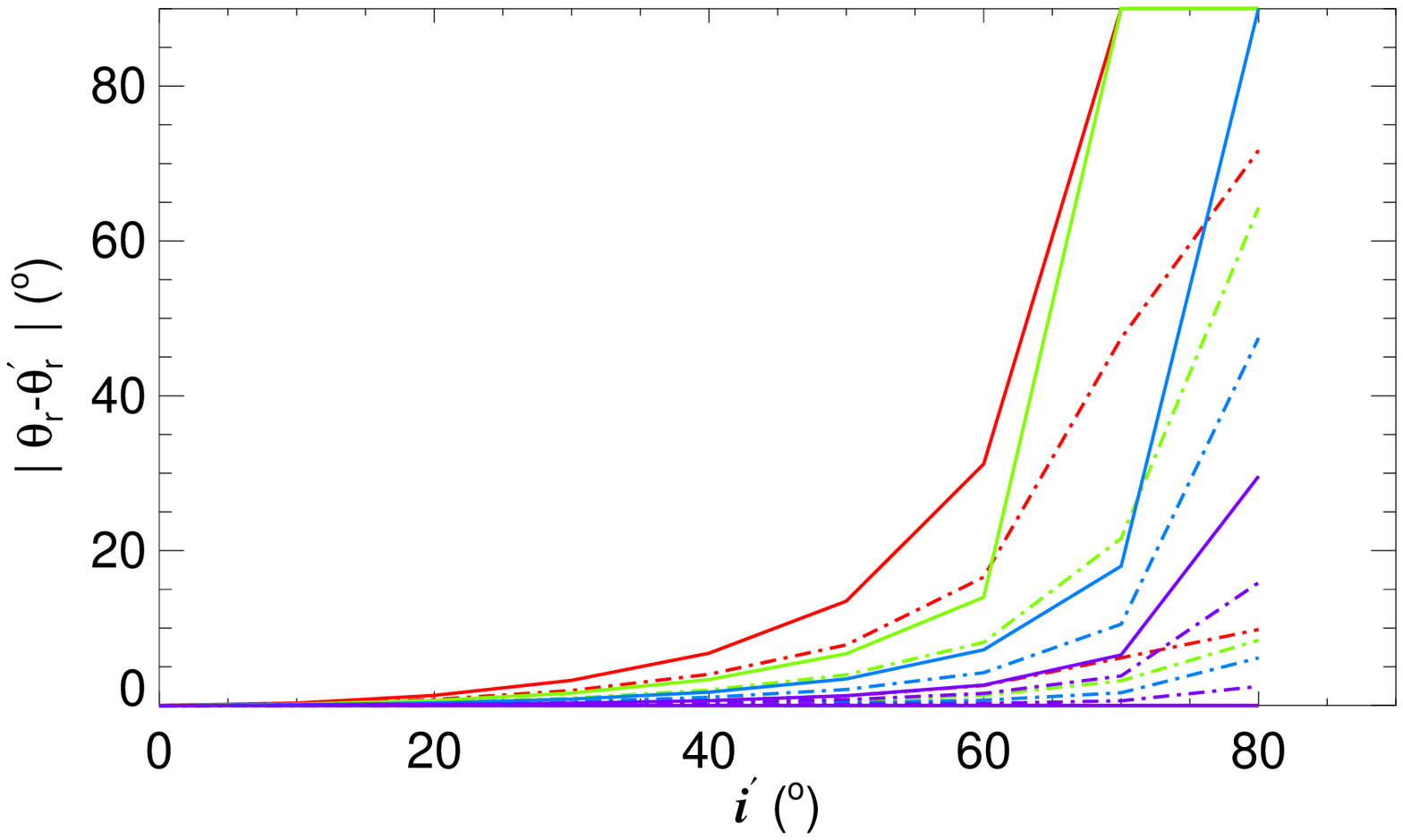}\\
  \end{tabular}
  \caption{\label{frq} Top panel: difference between the measured intrinsic axis ratio of a ring, $q_{\rm r0}$, and the real one, $q^{\prime}_{\rm r,0}$, as a function of the real galaxy inclination, $i^{\prime}$, for four disc scale-height to scale-length ratios. The colour code corresponds in an outside-in order to $h_{\rm z}/h_{\rm R}=1/3$ (red), $h_{\rm z}/h_{\rm R}=1/5$ (green), $h_{\rm z}/h_{\rm R}=1/7$ (blue), and $h_{\rm z}/h_{\rm R}=1/10$ (purple). Bottom panel: difference of the measured angle difference between the orientation of the major axis of a ring with respect to the line of nodes, $\theta_{\rm r}$, with respect to the real one, $\theta^{\prime}_{\rm r}$, as a function of the real galaxy inclination. In both panels, the solid lines indicate the maximum and the minimum $q_{\rm r,0}-q^{\prime}_{\rm r,0}$ ($|\theta_{\rm r}-\theta^{\prime}_{\rm r}|$) for a given $i^{\prime}$ colour-coded as in the top panel. The dashed lines indicate the limits of the region enclosing 68.2\% of $q_{\rm r,0}-q^{\prime}_{\rm r,0}$ ($|\theta_{\rm r}-\theta^{\prime}_{\rm r}|$) values at a given $i^{\prime}$.}
\end{figure}

The reliability of the deprojected ring parameters depends on the quality of the P4 measured disc deprojection parameters. One source of error when deprojecting is the assumption that discs are infinitely thin. Indeed, the thicker a galaxy disc, the more the measured galaxy inclination, $i$, will differ from the real inclination, $i^{\prime}$.

The thickness of discs has been studied by \citet{GRIJS98}. He found that the ratio between the disc scale-height and the disc scale-length ($h_{\rm z}$ and $h_{\rm R}$, respectively) decrease monotonically with Hubble stage. The average $h_{\rm z}/h_{\rm R}$ is around 1/3 for S0 galaxies and decreases to 1/9 for Sd galaxies. However, for a given stage the scatter in $h_{\rm z}/h_{\rm R}$ is large, so the relationship described by \citet{GRIJS98} cannot be used for correcting the finite thickness effects in deprojected parameters without adding a substantial error. Instead, in this section we adopt the approach of quantifying possible biases in the deprojected parameters.

\citet{HUB26} described the formalism for finding inclination of an oblate ellipsoid,
\begin{equation}
 {\rm cos}^2i^{\prime}=\frac{q_{\rm d}^2-q_{\rm z}^2}{1-q_{\rm z}^2},
\end{equation}
where $q_{\rm d}$ is the observed galaxy disc axis ratio and $q_{\rm z}$ is the galaxy flattening, that is, $h_{\rm z}/h_{\rm R}$. From this, and knowing that ${\rm cos}\,i=q_{\rm d}$, we can derive the formula for calculating the $i$ corresponding to a given $i^{\prime}$:
\begin{equation}
\label{eqdeproj}
 i={\rm arccos}\left(\sqrt{q_{\rm z}^2+(1+q_{\rm z}^2)\,{\rm cos}^2i^{\prime}}\right).
\end{equation}
From this expression we can deduce that the maximum bias at finding a disc inclination comes from the galaxies with thicker discs, namely S0s with $h_{\rm z}/h_{\rm R}\sim1/3$. Most of our results in Section~\ref{sresults} are based on galaxies with $i\leq60^{\rm o}$. For an S0 galaxy with $i^{\prime}=60^{\rm o}$ $\Delta i=i^{\prime}-i\sim5^{\rm o}$. Accordingly, for most of the galaxies with $i\leq60^{\rm o}$, $i\sim i^{\prime}$.

To examine how the biased $i$ values calculated from Equation~\ref{eqdeproj} would affect our ring deprojection values, we created four model galaxies with $h_{\rm z}/h_{\rm R}=1/3$, $h_{\rm z}/h_{\rm R}=1/5$, $h_{\rm z}/h_{\rm R}=1/7$, and $h_{\rm z}/h_{\rm R}=1/10$. We then inclined each of them with angles in the range $i^{\prime}=0^{\rm o}$ to $i^{\prime}=80^{\rm o}$ with $10^{\rm o}$ intervals.

We checked how the underestimated inclination measurements would affect the measured intrinsic ring ellipticities, $q_{\rm r,0}$, and the angle between their major axes and the lines of nodes of the deprojection, $\theta_{\rm r}$. To do this, we considered a population of rings with a Gaussian distribution of intrinsic axis ratios, $q^{\prime}_{\rm r,0}$, with the centre of the Gaussian at $q^{\prime}_{\rm r,0}=0.8$ and a dispersion of $\sigma^{\prime}=0.13$. These values are representative of the measured axis ratio distribution of inner rings as seen in Section~\ref{sshape}. For each galaxy model with a given $h_{\rm z}/h_{\rm R}$ and for each of the studied inclinations, we obtained 1000 random ring axis ratios. The intrinsic axis ratio of each of these rings was recovered after projecting it using the real galaxy inclination, $i^{\prime}$, and deprojecting it using the measured galaxy inclination, $i$. This was repeated for 19 angle differences between the ring major axis and the line of nodes used for inclining the model galaxy. These angles ranged from $\theta^\prime_{\rm r}=0^{\rm o}$ to $\theta^\prime_{\rm r}=90^{\rm o}$ in steps of $5^{\rm o}$.

In the top panel in Figure~\ref{frq} we study the accuracy of the measured ring axis ratios. If, as in our results section, we focus on galaxies with $i^\prime\leq60^{\rm o}$, the errors at measuring $q_{\rm r,0}$ are always below $|\Delta q_{\rm r,0}|=0.05$ except for galaxies with $h_{\rm z}/h_{\rm R}=1/3$ and a few cases with $h_{\rm z}/h_{\rm R}=1/5$. Except in exceptional cases for $h_{\rm z}/h_{\rm R}=1/3$, $|\Delta q_{\rm r,0}|\leq0.10$. For $i^\prime\leq40^{\rm o}$, the uncertainties in $q_{\rm 0}$ due to the finite thickness of discs are always rather small and in the order of the uncertainty in the measurement of $q_{\rm r}$ ($|\Delta q_{\rm r}|=0.02$) as measured in Section~\ref{saccuracy}.

In the bottom panel in Figure~\ref{frq} we study the accuracy of the measured ring intrinsic orientation. We find that for galaxies with $i^\prime\leq60^{\rm o}$, the maximum $|\theta_{\rm r}-\theta^\prime_{\rm r}|$ is always below $10^{\rm o}$ except, again, for galaxies with $h_{\rm z}/h_{\rm R}=1/3$ and in a few cases in those with $h_{\rm z}/h_{\rm R}=1/5$.

We conclude this section by saying that as found for bars in Section~\ref{sreliability}, results from deprojected parameters seem in general reliable for galaxies with $i\leq60^{\rm o}$. The exception are the moderately inclined galaxies ($i=40-60^{\rm o}$) with a larger disc thickness relative to the scale-length (S0 galaxies), whose deprojected ring parameters are in some cases not very accurate.

\section{Comparison of ring properties in ARRAKIS with the data in the literature}
\label{scomparison}

In this section we compare the measured properties of rings with those reported in a set of significant papers in the literature that have samples in common with ours.

\begin{figure}
  \includegraphics[width=0.45\textwidth]{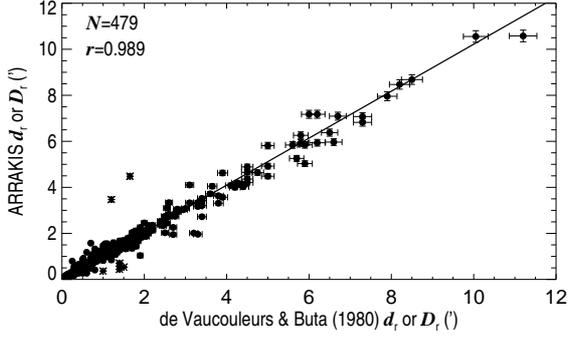}
  \caption{\label{fVAU80} Comparison of the angular diameters of ARRAKIS rings with those in VB80. The line corresponds to a linear fit to the data points in the plot. The outliers due to the wrong identification of the resonance to which a ring is linked in VB80 are marked with asterisks and are excluded from the fit. The numbers in the top-left corner indicate the number of points considered in the fit and their correlation coefficient.}
\end{figure}

In several cases, significant discrepancies have been found between the ring diameters in ARRAKIS and those in the literature. In most of them, we found that this is because the features are classified in a different way in ARRAKIS and in the literature. For example, the feature with an ARRAKIS major diameter of $1\farcm48$ in NGC~4736 was classified as an inner ring by \citet{VAU80}. Here, this feature is considered as a nuclear ring because it is found inside a very broad bar/oval that was unnoticed by \citet{VAU80}. The broad bar in NGC~4736 is here recognized as being surrounded by a subtle ring-lens $4\farcm48$ in major diameter that was considered to be an outer lens in \citet{VAU80}. Therefore, when comparing the inner ring in \citet{VAU80} with that in ARRAKIS, we are actually comparing two different features. Such features with a wrong classification appear as outliers in Figures~\ref{fVAU80}$-$\ref{fNIRS0S} and have not been considered for the statistics in this section.

\subsection{ARRAKIS and \citet{VAU80}}

De Vaucouleurs \& Buta (1980; hereafter VB80) compiled a catalogue that includes the diameter of outer and inner rings in 532 bright galaxies. The S$^4$G and the VB80 samples have          303
\unskip galaxies in common.

We compared the sizes of the rings listed both in ARRAKIS and VB80. For this, we performed a linear least-squares fit of the major and minor axis diameter measurements in the two surveys. The fit takes into account the errors both in the ARRAKIS and the VB80 measurements. The ARRAKIS diameter errors were considered to follow the internal error function in Equation~\ref{e1}. VB80 did not quantify their errors. Since their data are collected from photographic plates, we considered that their errors are probably similar to those in the CSRG, and therefore we used the CSRG internal error function for their diameters.

The fit we obtained is
\begin{eqnarray}
 d_{\rm r},D_{\rm r}({\rm ARRAKIS})= \nonumber\\
 0\farcm013\pm0\farcm006+(1.021\pm0.003)(d_{\rm r},D_{\rm r}) (\rm VB80)
\unskip
\end{eqnarray}
with a correlation factor $r=0.989$
\unskip. This means that the agreement between the measurements in ARRAKIS and in VB80 is very good.

\subsection{ARRAKIS and \citet{BU93}}

\begin{figure}
  \includegraphics[width=0.45\textwidth]{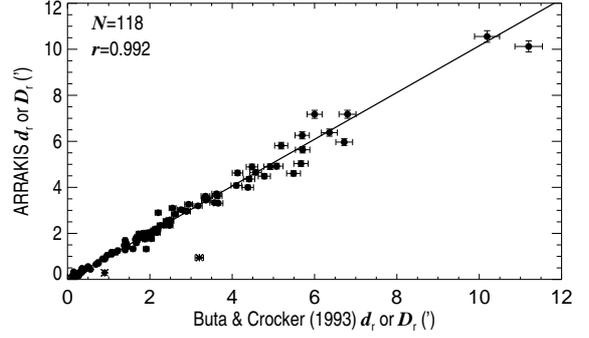}
  \caption{\label{fBU93}\bf As in Figure~\ref{fVAU80}, but with data from BC93.}
\end{figure}

\citet[][hereafter BC93]{BU93} studied galaxies with nuclear components, mostly nuclear rings, but also lenses and spirals. In BC93, galaxies with nuclear features also had their inner and outer features classified. The BC3 catalogue lists 64 galaxies,           37
\unskip of which are included in the S$^4$G. Figure~\ref{fBU93} compares the sizes of the features measured in ARRAKIS with those in BC93.

As with the VB80 data, we performed a least-squares fit to compare the BC93 data with ours. Again, the CSRG internal error function was used to describe the errors in the diameters in BC3. The fit gives
\begin{eqnarray}
 d_{\rm r},D_{\rm r}({\rm ARRAKIS})= \nonumber\\
 -0\farcm008\pm0\farcm016+(1.016\pm0.006)(d_{\rm r},D_{\rm r}) (\rm BC93)
\unskip
\end{eqnarray}
with a correlation factor $r=0.992$
\unskip.

\subsection{ARRAKIS and the CSRG}

\begin{figure*}
  \begin{tabular}{c c}
  \includegraphics[width=0.45\textwidth]{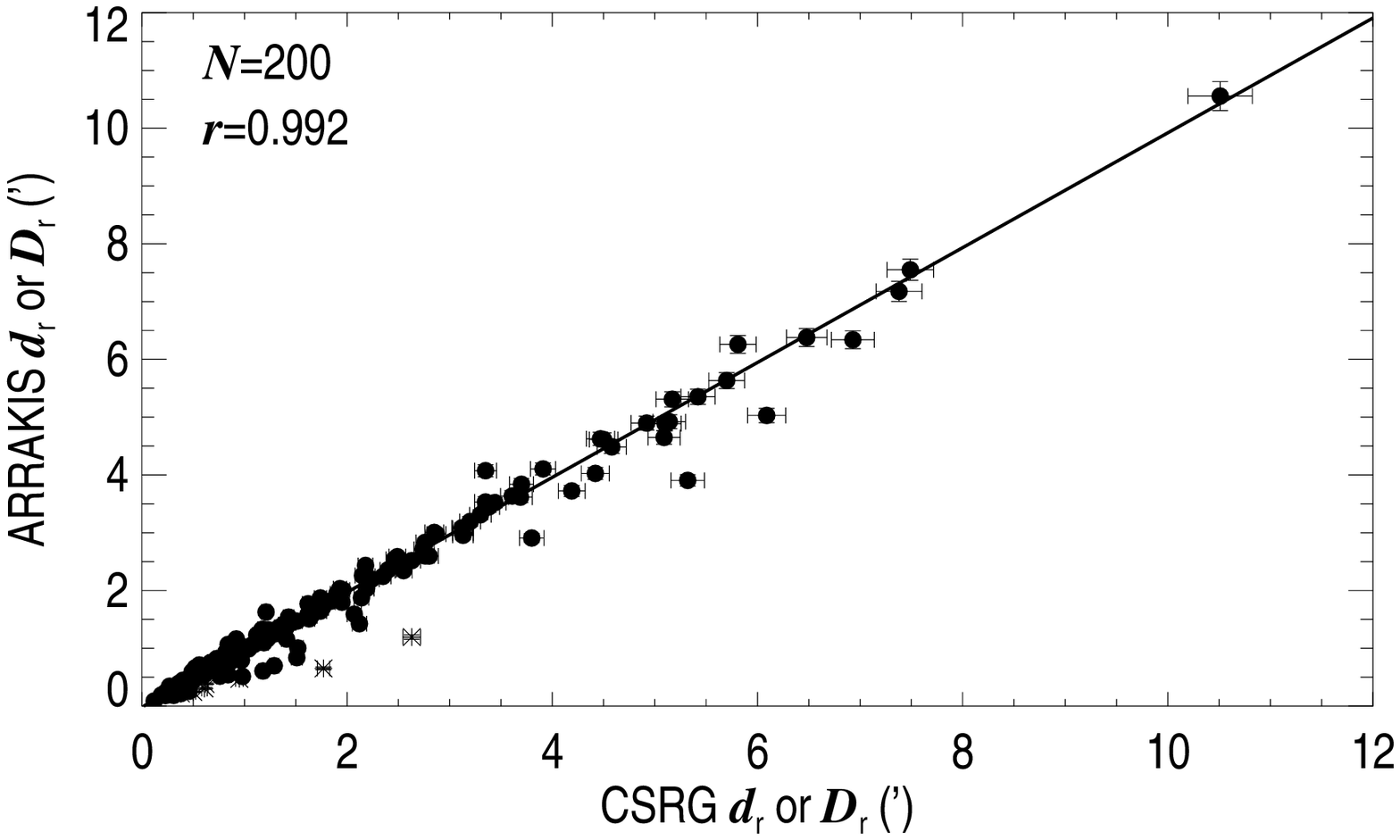}&
  \includegraphics[width=0.45\textwidth]{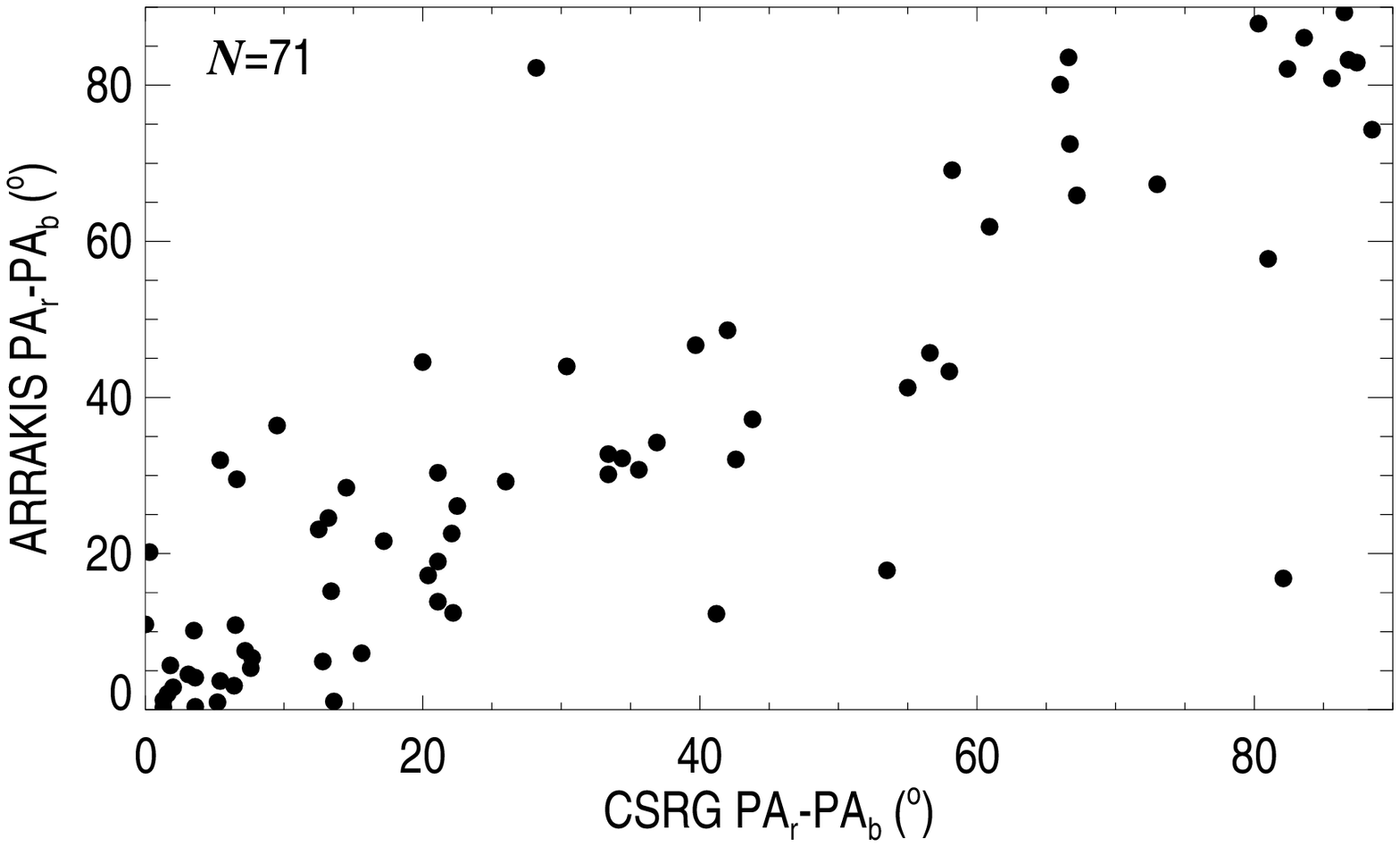}\\
  \includegraphics[width=0.45\textwidth]{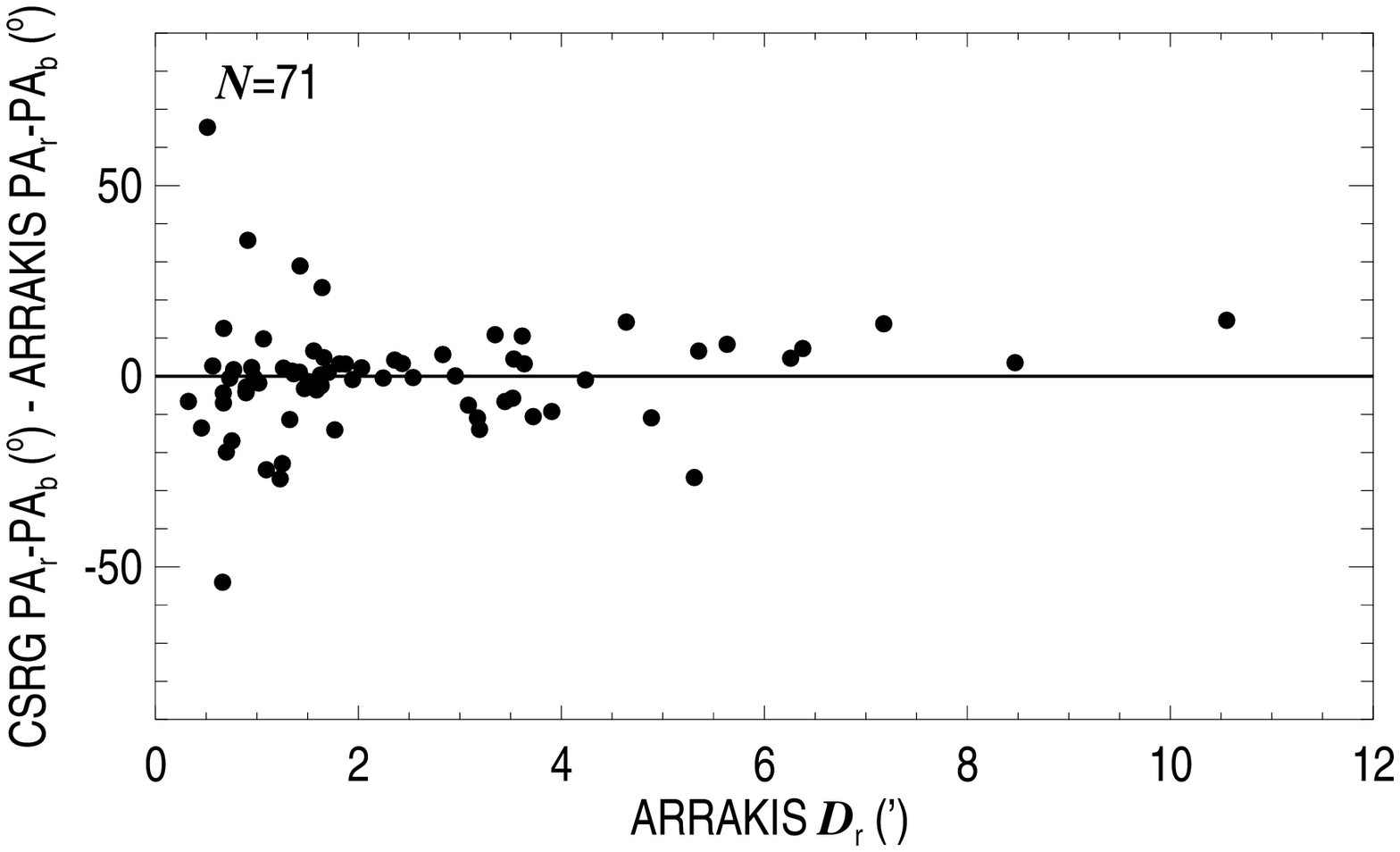}&
  \includegraphics[width=0.45\textwidth]{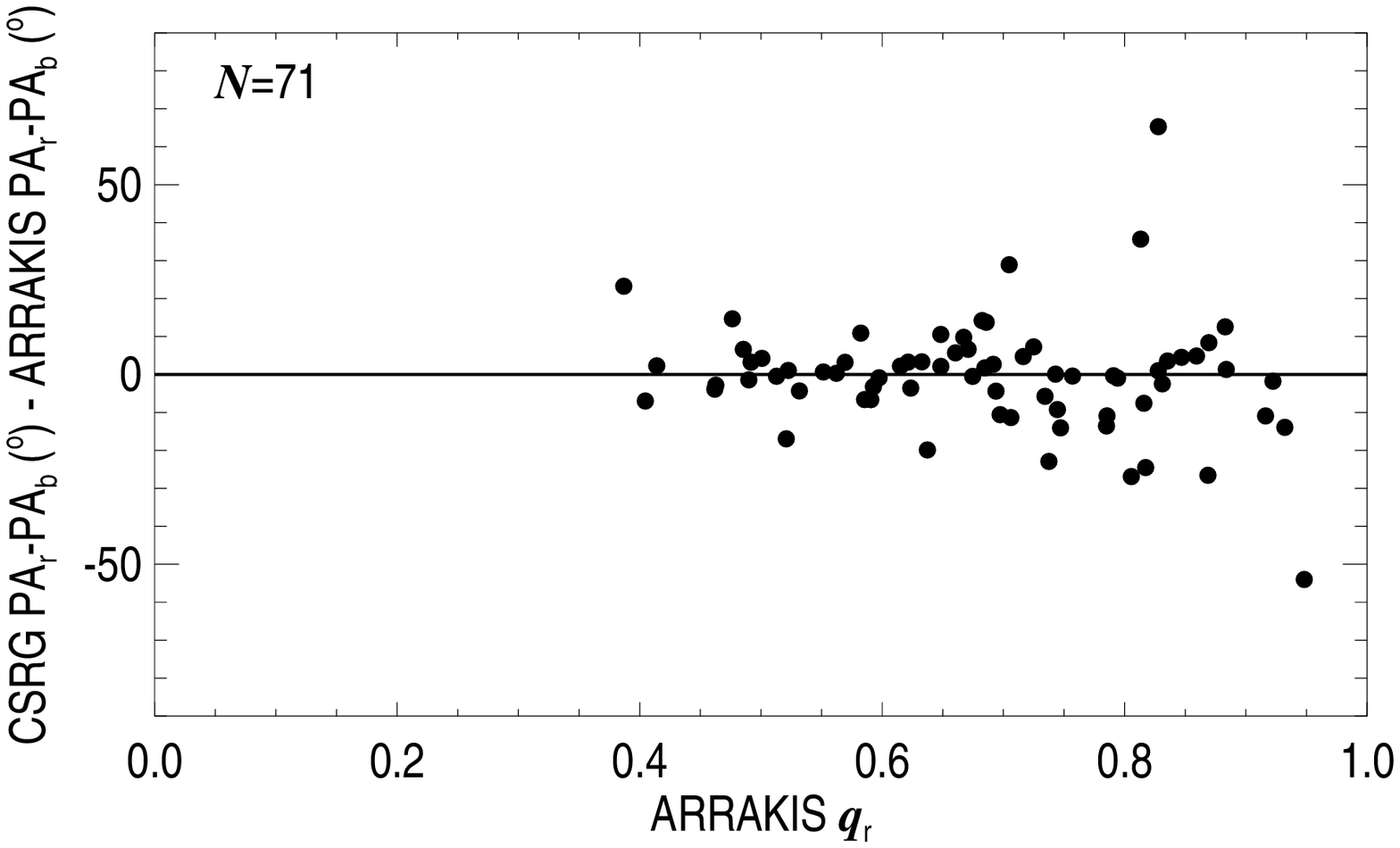}\\
  \end{tabular}
  \caption{\label{fCSRG} Top-left panel: As in Figure~\ref{fVAU80}, but with data from the CSRG. Top-right panel: comparison of the PA offset between the major axis of bars and that of rings in the CSRG and ARRAKIS. Bottom panels: difference in the PA offsets between the bar and the ring major axis as a function of the ring size and of the ring axis ratio. In the three last panels, only points for barred galaxies with $\epsilon_{\rm d}\leq0.5$ are presented.}
\end{figure*}

The CSRG includes 3692 galaxies that host outer and inner rings and lenses south of declination $\delta=-17^{\rm o}$. The CSRG has          120
\unskip galaxies in common with the S$^4$G.

In the top-left panel of Figure~\ref{fCSRG} we compare the size of a feature when it is identified in both the CSRG and ARRAKIS. A linear fit to the data accounting for the ARRAKIS and CSRG interal error functions at measuring diameters yields 
\begin{eqnarray}
 d_{\rm r},D_{\rm r}({\rm ARRAKIS})= \nonumber\\
 -0\farcm018\pm0\farcm011+(0.994\pm0.005)(d_{\rm r},D_{\rm r}) (\rm CSRG)
\unskip,
\end{eqnarray}
and a correlation factor $r=0.992$
\unskip, again indicating excellent agreement.

The CSRG describes the projected ring orientation as the PA offset between the bar and the ring major axis. ARRAKIS has such data only for galaxies with $q_{\rm d}>0.5$ because we considered more inclined galaxies to have unreliable bar axis ratio determinations (Section~\ref{sreliability}). The correlation between the ring orientation values for the CSRG and ARRAKIS is not as tight as it is for the angular sizes (top-right panel in Figure~\ref{fCSRG}). This is probably due to the analogue procedure used for building the CSRG. Indeed, it is conceivable to obtain accurate results on ring diameters using a magnifying device, but the determination of the major axis of bars and especially rings is doomed to be much more subjective. 

The difference between the CSRG ${\rm PA}_{\rm r}-{\rm PA}_{\rm b}$ values and ours is plotted in the bottom-left and bottom-right panels of Figure~\ref{fCSRG} as a function of the ring size and the ring axis ratio. The differences in ${\rm PA}_{\rm r}-{\rm PA}_{\rm b}$ are larger than our internal error for $PA_{\rm r}$, which is generally below $10^{\rm o}$ for rings with $D_{\rm r}>3\arcmin$ and/or $q_{\rm r}<0.7$ (Figure~\ref{ferr}).

\subsection{ARRAKIS and AINUR}

\begin{figure}
  \includegraphics[width=0.45\textwidth]{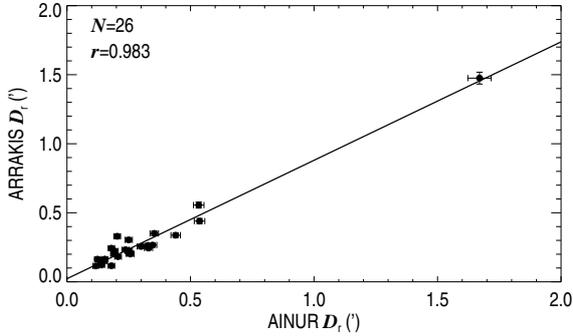}
  \caption{\label{fAINUR}\bf As in Figure~\ref{fVAU80}, but with data from AINUR.}
\end{figure}

AINUR contains the most complete catalogue of nuclear rings to date, mostly relying on {\it Hubble} Space Telescope images that in some cases have an angular resolution as high as $\sim0\farcs1$. Because of that, it contains detailed information of angularly small rings that cannot be expected to be described this thoroughly in ARRAKIS. Sixty out of 107 galaxies in AINUR are in the S$^4$G sample. Of these           60
\unskip galaxies           26
\unskip are reported here to have a nuclear ring.

When comparing the major axis diameter of features classified as nuclear rings both in AINUR and ARRAKIS, we obtain
\begin{eqnarray}
 d_{\rm r},D_{\rm r}({\rm ARRAKIS})= \nonumber\\
 0\farcm022\pm0\farcm010+(0.858\pm0.029)(d_{\rm r},D_{\rm r}) (\rm AINUR)
\unskip,
\end{eqnarray}
with a correlation factor $r=0.983$
\unskip. Here, the internal error functions for the diameters are those in Equation~\ref{e1} both for AINUR and ARRAKIS. We chose this error function for AINUR because, as for ARRAKIS, the ring measurements were made on digital images. This fit is much poorer than those found for VAU80, BU93, CSRG, and NIRS0S. This is probably because of the difficulty found in measuring the properties of features that are often barely resolved in the S$^4$G. The difference may also reflect that the procedures used to measure the ring properties in the two works are different.

\subsection{ARRAKIS and NIRS0S}

\begin{figure*}
  \begin{tabular}{c c}
  \includegraphics[width=0.45\textwidth]{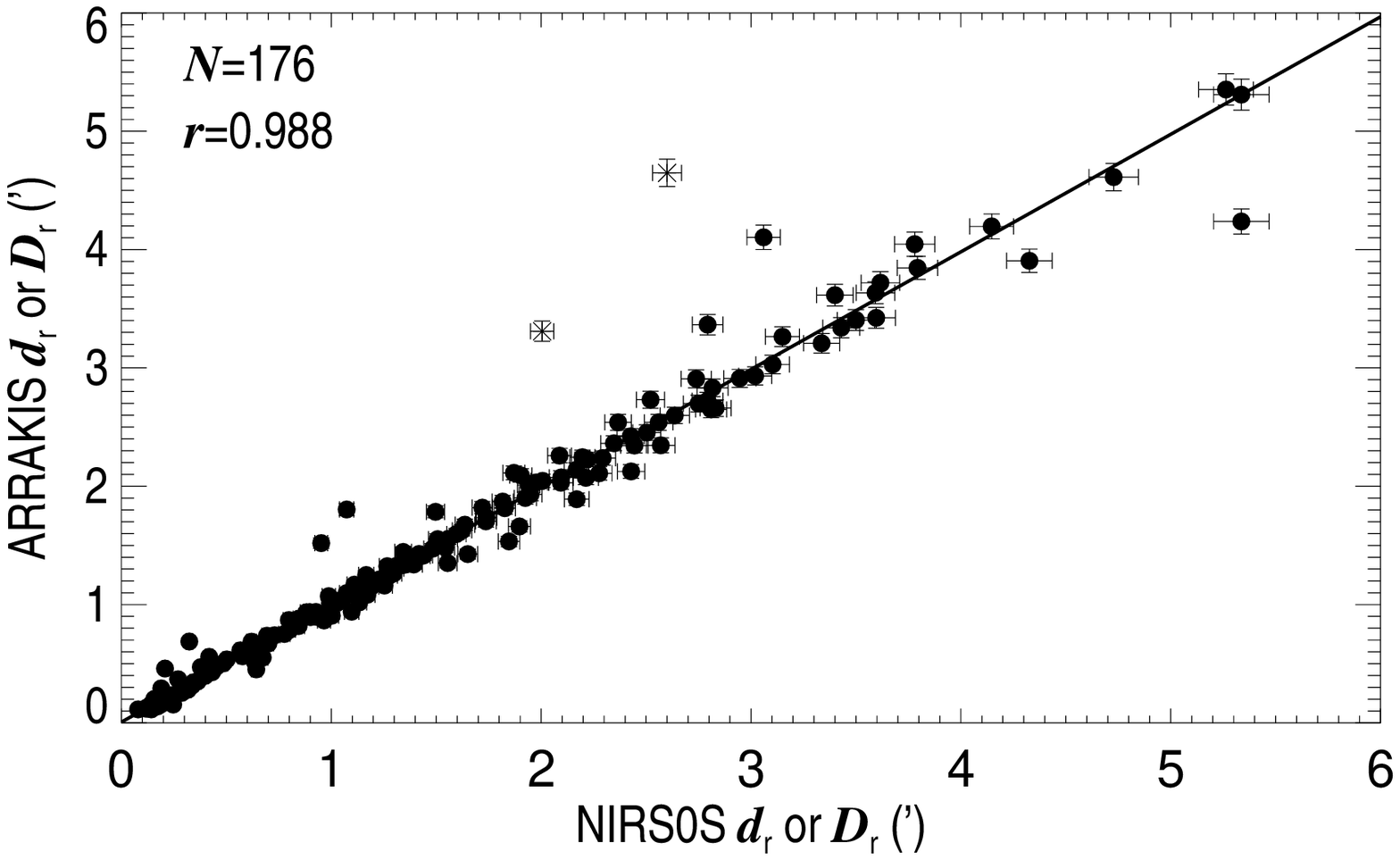}&
  \includegraphics[width=0.45\textwidth]{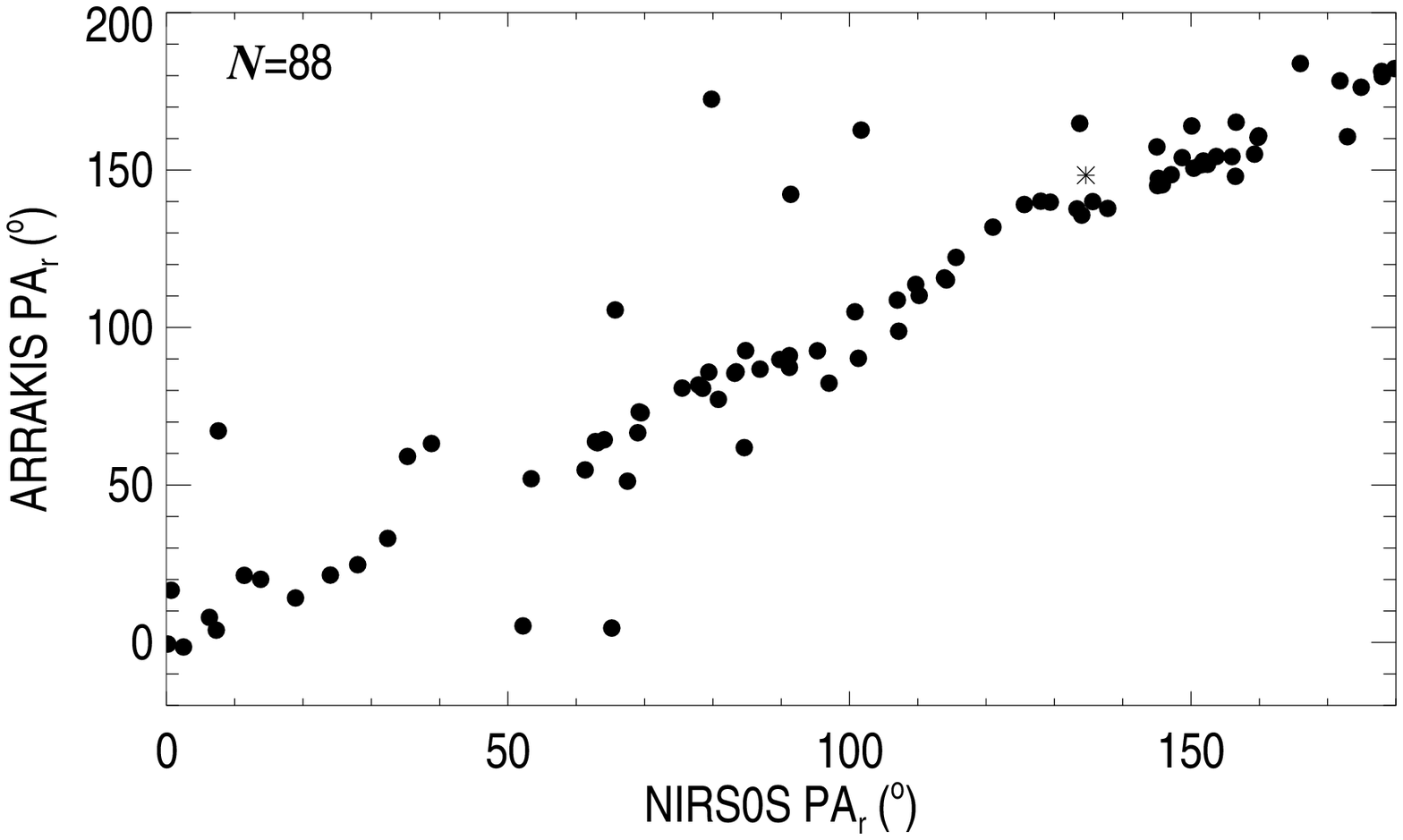}\\
  \includegraphics[width=0.45\textwidth]{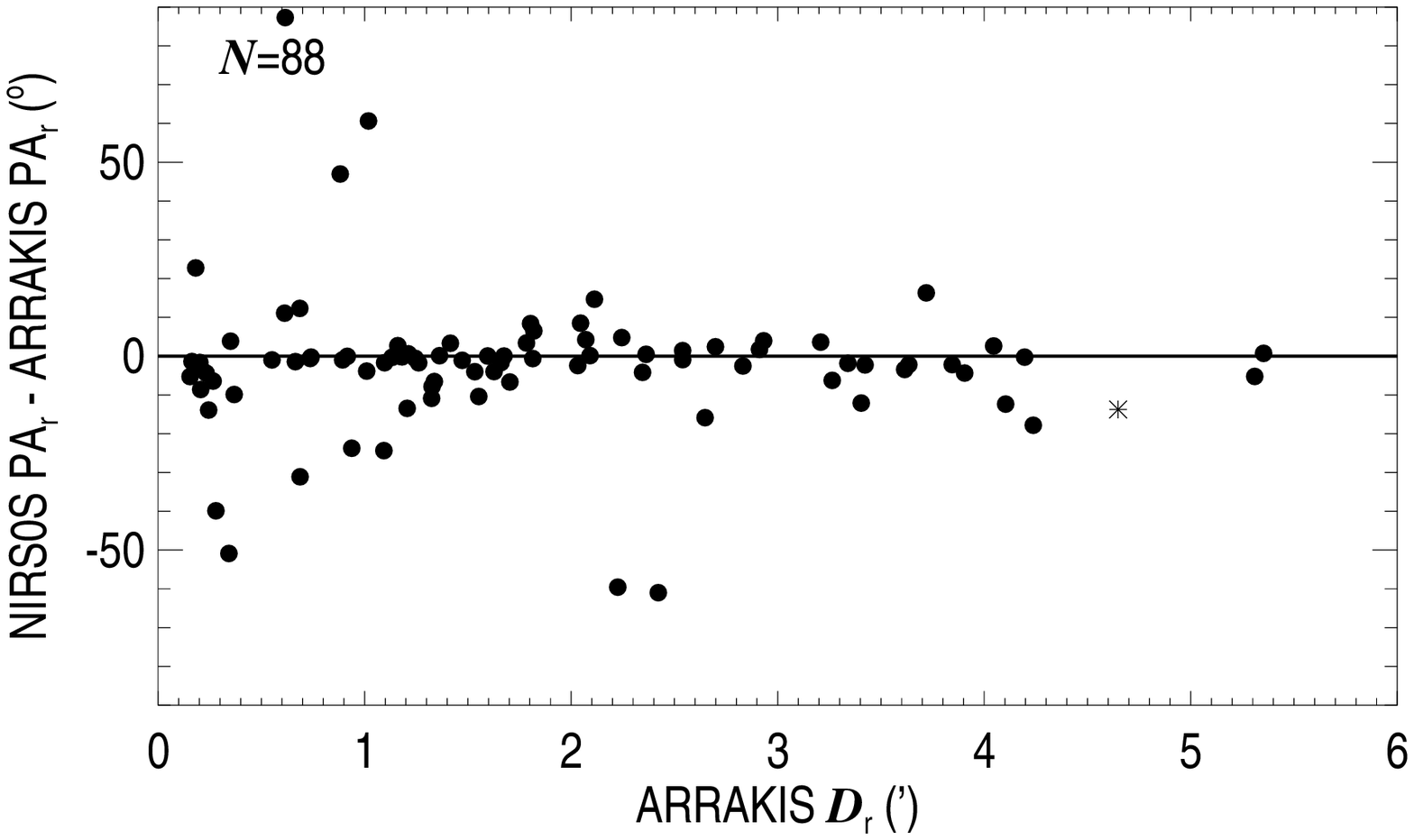}&
  \includegraphics[width=0.45\textwidth]{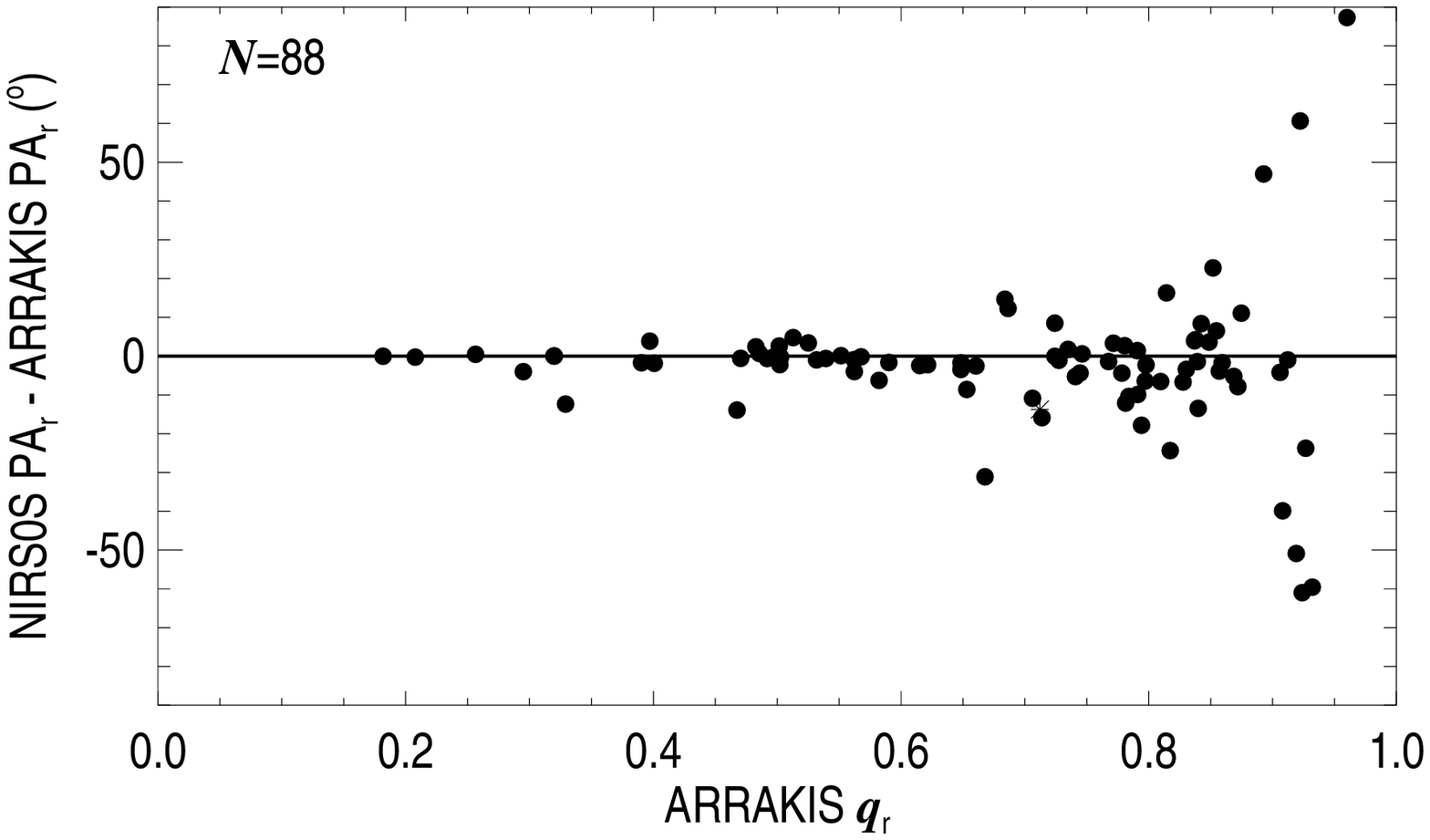}\\
  \end{tabular}
  \caption{\label{fNIRS0S} As in Figure~\ref{fCSRG}, but with data from NIRS0S.}
\end{figure*}

NIRS0S is a ground-based survey of 206 nearby early-type disc galaxies with a sample dominated by S0s. The survey was produced using the $K_{\rm s}$ band. NIRS0S does not reach as deep as the S$^4$G but has a better angular resolution (typical pixel resolution of $0\farcs3$ and typical seeing around $1\arcsec$). NIRS0S and the S$^4$G have           93
\unskip galaxies in common.

The ring sizes measured in NIRS0S agree very well with ours (Figure~\ref{fNIRS0S}). When comparing the ring diameters from ARRAKIS and NIRS0S (in both cases we use the internal error function from Equation~\ref{e1} because in both cases the rings are measured on digital images), we obtain the following excellent linear fit
\begin{eqnarray}
 d_{\rm r},D_{\rm r}({\rm ARRAKIS})= \nonumber\\
 0\farcm006\pm0\farcm009+(0.993\pm0.005)(d_{\rm r},D_{\rm r}) (\rm NIRS0S)
\unskip,
\end{eqnarray}
and the correlation factor is $r=0.988$
\unskip.

Figure~\ref{fNIRS0S} also shows that the agreement in the measured position angle of the ring major axis is very good and much better than when comparing this paper with the CSRG. One possible reason is that NIRS0S is a digital survey where ${\rm PA_{r}}$ can be measured more easily. The second reason is that in the CSRG the position angle difference between the ring and the bar major axis was measured. This measurement includes two errors: that related to the orientation of the ring, and that related to the orientation of the bar, and thus causes larger discrepancies than those found when comparing NIRS0S with this paper. As we found when we compared ARRAKIS with the CSRG and from our internal errors, the uncertainty in the orientation measurement increases for smaller and/or rounder rings.

\section{Results}

\label{sresults}

We recall that in this paper rings refers to the set including both open pseudorings and closed features. Here, the set including only closed features is referred to as closed rings.

\subsection{Fraction of rings as a function of the Hubble stage}

\label{sstage}

\begin{figure*}
  \includegraphics[width=0.9\textwidth]{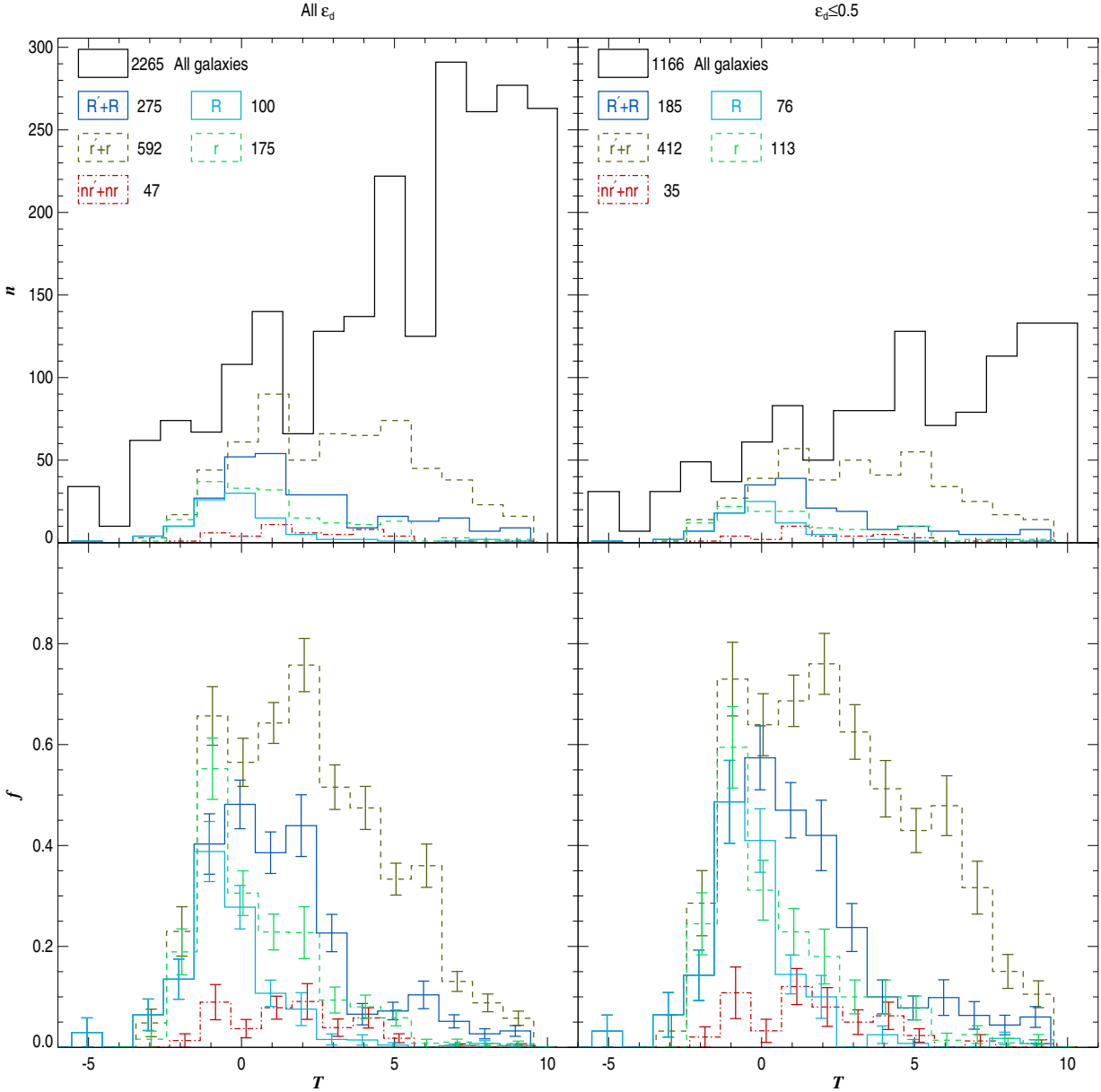}
  \caption{\label{fmorph} Top panels: stage distribution of the S$^4$G sample (black line), of galaxies hosting outer rings (continuous darker blue line), inner rings (dashed darker green line), and nuclear rings (dash-point red line). Lighter blue and green lines indicate histograms considering only galaxies hosting closed outer and closed inner rings, respectively. Bottom panels: fraction of galaxies with resonant features for a given stage colour-coded as in the two top panels. Left panels are for the whole sample and right panels for galaxies with a disc ellipticity $\epsilon_{\rm d}\leq0.5$. The numbers in the top panels indicate the number of galaxies included in the histograms that corresponds to the colour and line pattern of the adjacent box.}
\end{figure*}

Stage classification is commonly described by a numerical code ranging from -5 to 10 corresponding to the sequence going from E to I galaxies. However, in some cases, a more detailed classification can be made and is then denoted with intermediate stages such as Sb${\underline{\rm c}}$ or S${\underline{\rm c}}$d, which in turn are codified as $T=4.5$ and $T=5.5$. In the few particular cases for which such a classification is provided, we rounded up the stage so that, for example, Sb${\underline{\rm c}}$ becomes Sc ($T=5$) and S${\underline{\rm c}}$d becomes Scd ($T=6$).

Some galaxies in Buta et al.~(in preparation) have not been classified into categories that fit in the traditional Hubble fork. Indeed, based on the \citet{BERGH76} hypothesis that S0 galaxies form a sequence from early to late galaxies just like normal disc galaxies do, some galaxies have been classified as S0$_{\rm b}$, S0$_{\rm bc}$, S0$_{\rm c}$, S0$_{\rm cd}$, S0$_{\rm d}$, and S0$_{\rm m}$. This is true for           36
\unskip S$^4$G galaxies, three
\unskip of which host rings. These galaxies were included in the bin corresponding to their counterpart in the traditional sequence (e.g., S0$_{\rm b}$ galaxies were grouped together with Sb galaxies).

A few galaxies (          12
\unskip) have a double stage \citep[for details on double-stage galaxies see][]{BU10}, five
\unskip of which are ringed. These galaxies were not included in the stage statistics. We also excluded the dwarf elliptical galaxies from the results in this subsection (          34
\unskip galaxies, none of which is ringed). Finally, we also excluded           12
\unskip S$^4$G galaxies that are so difficult to classify that there were assigned no stage. None of these galaxies is ringed

The distribution of resonant features varies greatly with the stage of the galaxies, as seen in Figure~\ref{fmorph}. The bottom-row panels in that figure have $1-\sigma$ (68\%) confidence levels calculated using binomial distribution statistics,
\begin{equation}
\label{epoisson}
 \Delta(p)=\sqrt{\frac{p\left(1-p\right)}{n}},
\end{equation}
where $p$ is the probability of a galaxy in a given stage bin to have a specific type of ring and $n$ the total number of galaxies in that stage bin. All the uncertainties given in this section were calculated using this expression. When the data in this section are quantitatively compared with the data in other papers, we verified that the uncertainties in those papers were computed in the same way as we did.

Inclination has to be taken into account before discussing in detail the stage distributions statistics. Indeed, it is more likely for inclined galaxies to have unidentified or misinterpreted rings. We found that of the whole sample of galaxies, 31\%
\unskip have some resonant feature, but that this fraction is increased to 41\%
\unskip for the subsample with $\epsilon_{\rm d}\leq0.5$. These numbers indicate that as much as $\sim50\%$ of the resonant features are missed in inclined galaxies. This effect is especially pronounced for inner features, most likely because they are found in regions of galaxies that are hard to interpret because of bars, spiral arms, and/or intense star formation regions.

\subsubsection{Outer rings}

The stage distribution of outer rings based on the S$^4$G subsample with $\epsilon_{\rm d}\leq0.5$ indicates the following (see also continuous blue lines in Figure~\ref{fmorph}):
\begin{itemize}
 \item They are found in $16\pm1\%$
\unskip of S$^4$G galaxies.
 \item They are mostly found at $-1\leq T \leq 2$ where they appear in over 40\% of galaxies. This number is similar to the $53\pm5\%$ fraction of outer features found in NIRS0S for stages $-1\leq T\leq1$ \citep{LAU11}.
 \item For $T\geq4$ they are quite rare, and their frequency drops below 10\%.
 \item Outer closed rings (continuous light blue lines) are most frequent for $T\leq0$, with a peak frequency of $\sim50\%$ at $T=-1$. Above $T=0$ their frequency drops fast with increasing stage. This behaviour is also seen in NIRS0S.
 \item Outer closed rings and closed ring-lenses account for 100\% of the outer features for $T\leq-1$. This is a difference with NIRS0S: the fraction of pseudorings for $T\leq1$ NIRS0S galaxies is non-zero, though still rather small, $19\pm7\%$. Of the five NIRS0S galaxies with $T\leq-1$ and with an outer pseudoring, two are included in the S$^4$G sample. These two galaxies have been classified with the same stage in NIRS0S and Buta et al.~(in preparation), which means that the difference between NIRS0S and ARRAKIS probably caused by the different appearance of outer features at different wavelengths and not by problems in the stage classification.
\end{itemize}

\citet{BU96} found qualitatively similar results using RC3 data. Using data from VB80, \citet{EL92} also found that earlier-type galaxies are more likely to have outer rings.

\subsubsection{Inner rings}

The stage distribution of inner rings based on the S$^4$G subsample with $\epsilon_{\rm d}\leq0.5$ indicates the following (see also dashed green lines in Figure~\ref{fmorph}):
\begin{itemize}
 \item They are found in $35\pm1\%$
\unskip of S$^4$G galaxies.
 \item They have a frequency of over 40\% for stages $-1\leq T\leq6$. The distribution is not very peaked and has its maximum  for $-1\leq T\leq3$ (more than 60\% of these galaxies have inner features).
 \item Their frequency drops below 20\% for galaxies with $T\leq-3$ and $T\geq8$. They are therefore very frequent in most disc galaxy stages.
 \item The peak in the inner closed ring frequency is found at $T=-1$ (light green lines) as also found in NIRS0S. The inner closed ring frequency distribution is shifted towards earlier types than that of the inner ring distribution.
 \item Inner closed rings are more frequent than inner pseudorings for $T\leq1$. This agrees very well with the results in NIRS0S. There the fraction of inner features that are inner closed rings in galaxies with $-3\leq T\leq-1$ is $75\pm7\%$, we found $81\pm6\%$
\unskip. For galaxies with types $0\leq T\leq1$ NIRS0S reports a fraction of inner closed rings among inner features of $38\pm8\%$, we found $40\pm5\%$
\unskip.
\end{itemize}

These results differ from those presented in \citet{BU96}, which were based on RC3 data. These authors found the inner closed ring fraction to be roughly constant for all types of disc galaxies with $T\leq5$ (here the distribution is peaked at $T=-1$) and also that they are more frequent than pseudorings for $T\leq3$ (here, this only occurs for $T\leq-1$). These differences are too large to be a consequence of the one stage shift caused by classifying galaxies in $3.6\mu$m instead of the $B$ band.

The stage distribution of outer and inner rings is quite different.  While the outer ring distribution drops between types $T=3$ and $T=4$, the inner rings distribution does so  between types $T=7$ and $T=8$.

\subsubsection{Nuclear rings}

Nuclear rings (dash-point red lines in Figure~\ref{fmorph}) are found in galaxies with $-1\leq T\leq6$, which is similar to that obtained in AINUR ($-3\leq T\leq7$). However, the peak in the distribution ($T=2$) does not coincide with those in AINUR ($T=-1$ and $T=4$). These differences are probably not significant, because the number of nuclear rings here is a factor of three smaller than in AINUR.

In AINUR  nuclear rings were found in $\sim20\%$ of galaxies with stages $-3\leq T\leq7$. A similar fraction was given by \citet{KNA05}. However, the fraction of low inclination ($\epsilon_{\rm d}\leq0.5$) S$^4$G galaxies that host nuclear rings is $5\pm1\%$
\unskip in the same range of stages. The reason for the discrepancy is that, because of angular resolution problems, many of the smaller rings are undetected or are classified as nuclear lenses in the S$^4$G.

\subsection{Fraction of rings as a function of the galaxy family}

\label{sfamily}

\begin{figure*}
  \includegraphics[width=0.9\textwidth]{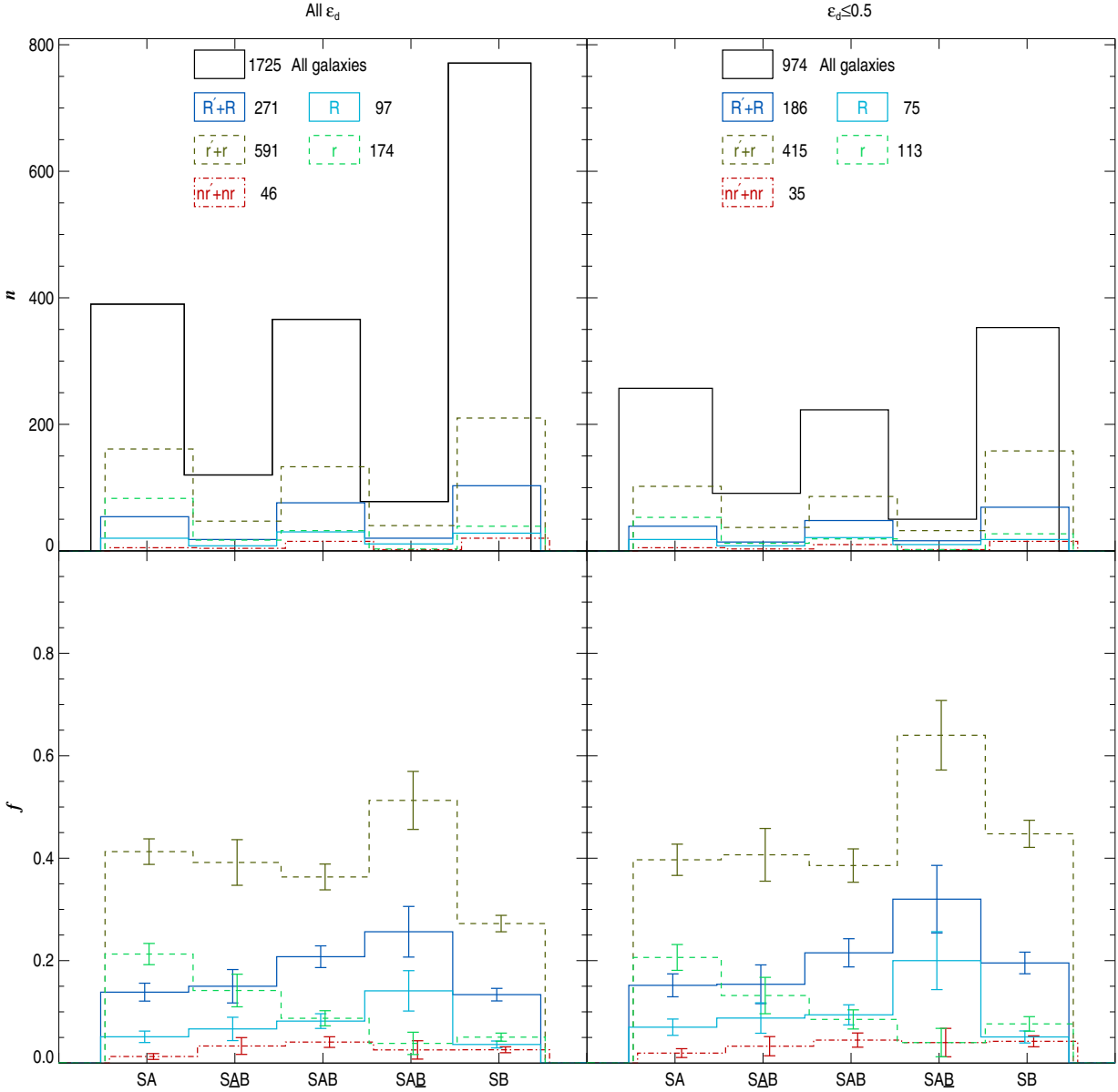}
  \caption{\label{fbar} Top panels: family distribution of the S$^4$G sample and ringed galaxies colour- and line-coded as in Figure~\ref{fmorph}. Bottom panels: fraction of galaxies with rings for a given family colour- and line-coded as in the two top panels. Left panels are for the whole sample and right panels for galaxies with a disc ellipticity $\epsilon_{\rm d}\leq0.5$. The numbers in the top panels indicate the number of galaxies included in the histogram that corresponds to the colour- and line pattern of the adjacent box.}
\end{figure*}

Of the         2348
\unskip S$^4$G galaxies that have been classified in Buta et al.~(in preparation),         1725
\unskip are disc galaxies ($-3\leq T\leq9$) to which a family has been assigned (SA, S${\underline{\rm A}}$B, SAB, SA${\underline{\rm B}}$, SB). Some disc galaxies (         233
\unskip) have no assigned family and are mostly close to edge-on galaxies without obvious signatures of edge-on bars such as boxy inner isophotes. The ring distribution with family is shown in Figure~\ref{fbar}.

\subsubsection{Outer rings}

The fraction of galaxies with $\epsilon_{\rm d}\leq0.5$ that host outer rings (blue continuous lines in Figure~\ref{fbar}) increases steadily from SA ($15\pm2\%$
\unskip) to SA${\underline{\rm B}}$ ($32\pm7\%$
\unskip). For stronger bars (SB), it drops to $20\pm2\%$
\unskip.

The distribution of outer closed rings (light blue continuous lines) is qualitatively similar: the fraction of galaxies that host closed rings rises from $7\pm2\%$
\unskip to $20\pm6\%$
\unskip between SA and SA${\underline{\rm B}}$ and then drops down to $5\pm1\%$
\unskip for the SB family.

\subsubsection{Inner rings}

The inner ring family distribution is quite different from that for the outer rings (green dashed lines in Figure~\ref{fbar}). The fraction of galaxies that host inner rings remains roughly constant from SA to SAB at a level of $\sim40\%$ and then suddenly increases to $64\pm7\%$
\unskip for SA${\underline{\rm B}}$ galaxies. The fraction of SB galaxies that host inner rings is similar to that in SA to SAB galaxies.

We also find that galaxies with no bars or weak bars (SA and S${\rm{\underline A}}$B) have a higher fraction of inner closed rings (light green dashed lines) than galaxies with stronger bars.

\subsubsection{Nuclear rings}

The number of nuclear rings in ARRAKIS is small and thus the validity of the statistics is uncertain. However, we can see that the nuclear feature fraction increases from families SA to SAB and then remains roughly constant (dash-point red lines in Figure~\ref{fbar}).

\subsection{Ring shapes}

\label{sshape}

\subsubsection{Projected ring shapes}

\begin{figure*}
  \includegraphics[width=0.9\textwidth]{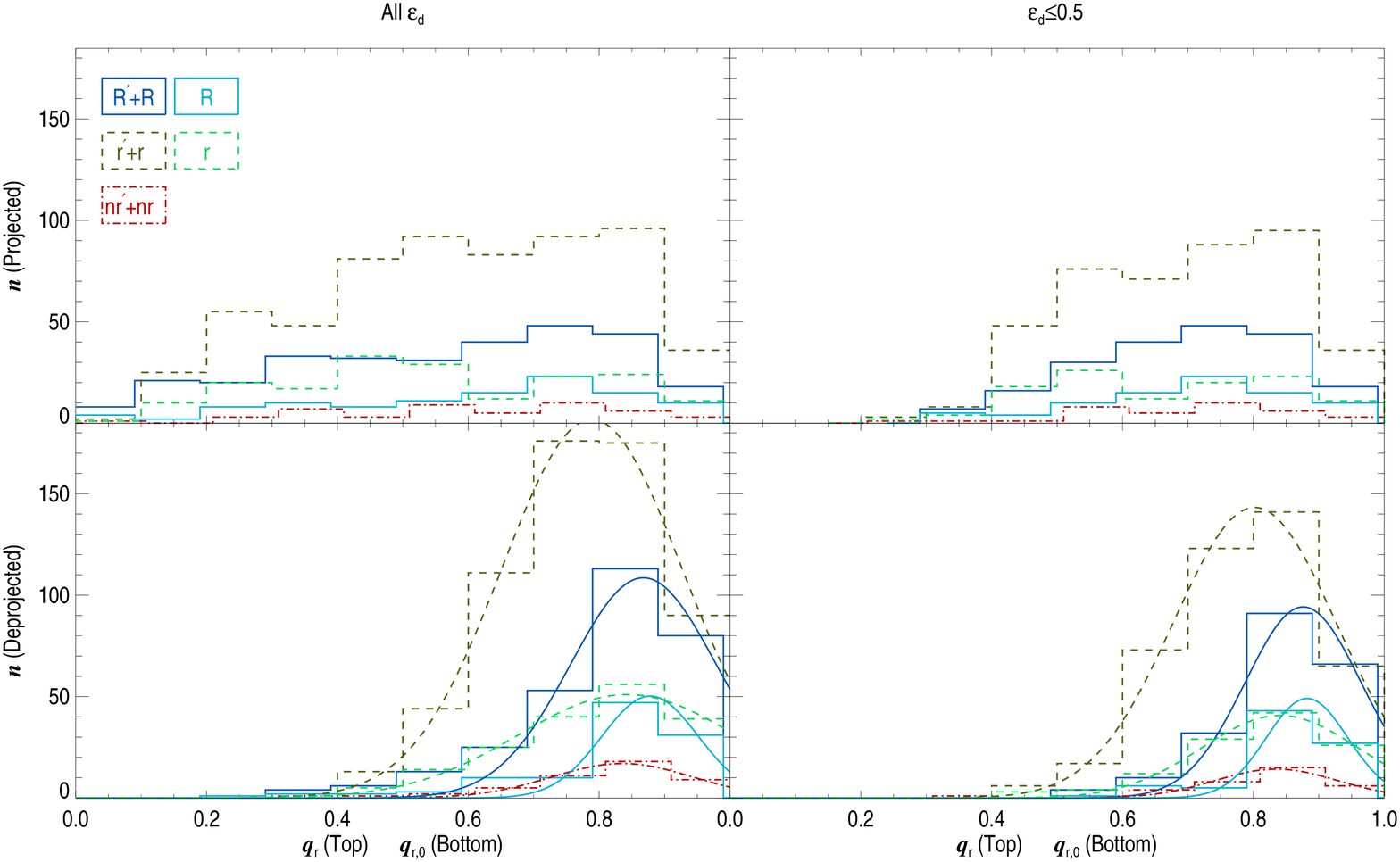}
  \caption{\label{fshape} Distribution of the resonance ring axis ratios colour- and line-coded as in Figure~\ref{fmorph}. The left-hand plots are for the whole ARRAKIS sample and the right-hand plots correspond to rings in host galaxies with $\epsilon_{\rm d}\leq0.5$. The top-row plots show the distribution of projected axis ratios, the bottom row shows the distribution of deprojected ones.}
\end{figure*}

The top-left panel in Figure~\ref{fshape} shows the projected axis ratio of resonance rings, under the assumption that they can be described by an ellipse coplanar with the disc of the galaxy. It is qualitatively similar to that found in the CSRG (top panels of its Figure~7): the distribution grows gently from zero and has a maximum for quite round rings ($0.8<q_{\rm r}\leq0.9$ for inner and outer rings here and $0.7<q_{\rm r}\leq0.8$ in the CSRG). It then  drops abruptly for even rounder rings. 

The statistical expectation is that if rings were intrinsically circular and sitting in perfectly circular discs, with their rotation axis with random orientations, the projected ring axis ratio probability density function would be uniform. As explained in the CSRG, the reason to justify the lack of rings with a low apparent axis ratio is that they become increasingly harder to identify in inclined galaxies. It is interesting to see that just like in the CSRG, for the axis ratios close to $q_{\rm r}=0$, outer rings show a behaviour closer to the statistical expectation than inner rings. This is because they are found in outer parts of the galaxy that are less messy and easier to interpret than inner regions when seen close to edge-on. We do not observe the distribution to be uniform near $q_{\rm r}=1$ because there is a dip in the observed distribution of rings. The reason why we do not see more rings that appear nearly round in projection is that they are not intrinsically circular, as described in the next subsubsection.

\subsubsection{Deprojected ring shapes}

\begin{figure*}
 \includegraphics[width=0.90\textwidth]{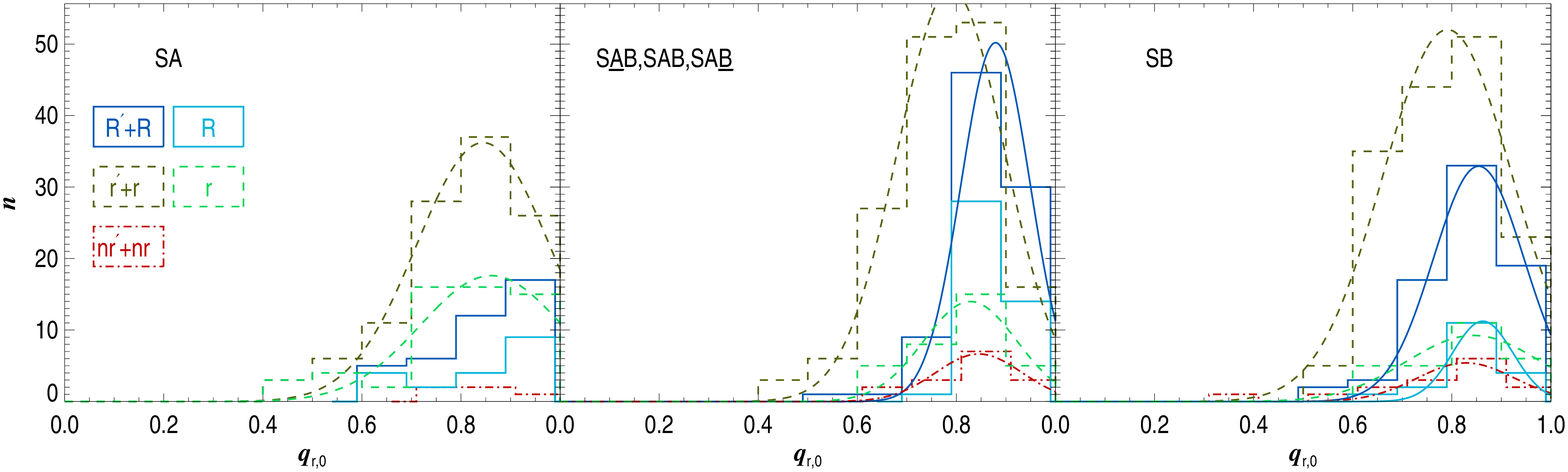}
  \caption{\label{fshapebar} Distribution of the resonance ring deprojected axis ratios colour- and line-coded as in Figure~\ref{fmorph} for different galaxy families and for $\epsilon_{\rm d}\leq0.5$. The left panel shows SA galaxies, the middle one S${\underline{\rm A}}$B to SA${\underline{\rm B}}$ galaxies, the right panel SB galaxies.}
\end{figure*}

\begin{figure*}
 \includegraphics[width=0.90\textwidth]{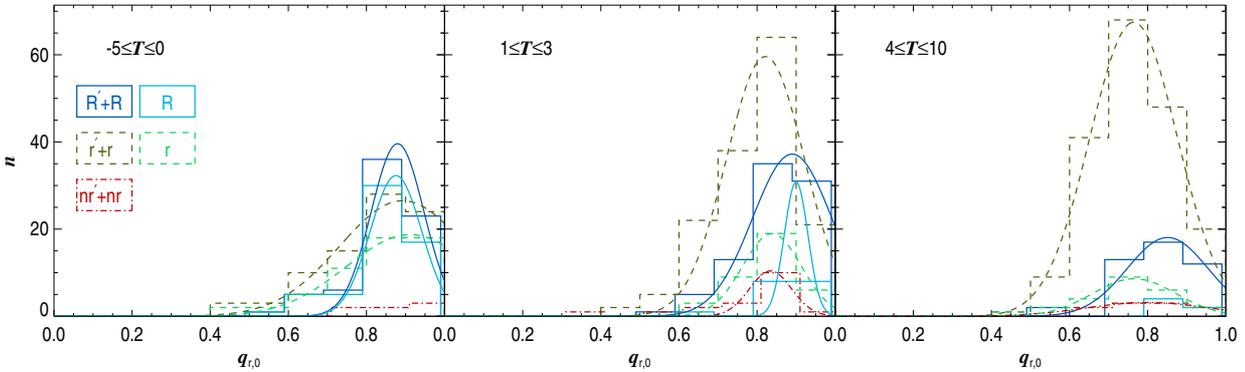}
  \caption{\label{fshapemorph} Distribution of the resonance ring deprojected axis ratios colour- and line-coded as in Figure~\ref{fmorph} for different galaxy stage ranges and for $\epsilon_{\rm d}\leq0.5$. The left panel represents galaxies with $-5\leq T\leq0$, the middle panel galaxies with $1\leq T\leq3$, the right panel galaxies with $4\leq T\leq10$ .}
\end{figure*}

\begin{table*}
\caption{Number of galaxies in the histograms in Figures~\ref{fshape}, \ref{fshapebar}, and \ref{fshapemorph} and parameters of the Gaussian fits to the $q_{\rm r,0}$ distributions}
\label{tellip}
\centering
\begin{tabular}{c c | c c c}
\hline\hline
Galaxy subsample&Feature type&$n$&$\left<q_{\rm r,0}\right>$&$\sigma$\\
\hline
\multirow{5}{*}{All galaxies} & Outer rings&295&0.87 (0.01)&0.11 (0.01)\\
&Inner rings&610&0.79 (0.00)&0.14 (0.00)\\
&Nuclear rings&47&0.84 (0.01)&0.11 (0.01)\\
&Outer closed rings&106&0.88 (0.01)&0.07 (0.01)\\
&Inner closed rings&180&0.84 (0.02)&0.17 (0.02)\\
\hline
\multirow{5}{*}{$\epsilon_{\rm d}\leq0.5$} & Outer rings&203&0.88 (0.01)&0.09 (0.01)\\
&Inner rings&425&0.80 (0.01)&0.12 (0.01)\\
&Nuclear rings&35&0.84 (0.01)&0.09 (0.01)\\
&Outer closed rings&82&0.88 (0.01)&0.06 (0.01)\\
&Inner closed rings&116&0.84 (0.01)&0.12 (0.01)\\
\hline
\multirow{2}{*}{$\epsilon_{\rm d}\leq0.5$} & Outer rings&40&$-$&$-$\\
&Inner rings&111&0.84 (0.01)&0.13 (0.01)\\
&Nuclear rings&5&$-$&$-$\\
\multirow{2}{*}{SA} &Outer closed rings&19&$-$&$-$\\
&Inner closed rings&56&0.86 (0.04)&0.14 (0.04)\\
\hline
\multirow{2}{*}{$\epsilon_{\rm d}\leq0.5$} & Outer rings&87&0.88 (0.00)&0.07 (0.00)\\
&Inner rings&156&0.79 (0.01)&0.11 (0.01)\\
&Nuclear rings&15&0.85 (0.03)&0.09 (0.03)\\
\multirow{2}{*}{S${\underline{\rm A}}$B,SAB,SA${\underline{\rm B}}$} &Outer closed rings&43&$-$&$-$\\
&Inner closed rings&33&0.83 (0.03)&0.09 (0.03)\\
\hline
\multirow{2}{*}{$\epsilon_{\rm d}\leq0.5$} & Outer rings&74&0.85 (0.00)&0.09 (0.00)\\
&Inner rings&158&0.79 (0.02)&0.13 (0.02)\\
&Nuclear rings&15&0.83 (0.02)&0.10 (0.02)\\
\multirow{2}{*}{SB} &Outer closed rings&18&0.86 (0.01)&0.06 (0.01)\\
&Inner closed rings&27&0.84 (0.06)&0.14 (0.07)\\
\hline
\multirow{2}{*}{$\epsilon_{\rm d}\leq0.5$} & Outer rings&71&0.88 (0.01)&0.07 (0.01)\\
&Inner rings&83&0.89 (0.03)&0.16 (0.03)\\
&Nuclear rings&7&$-$&$-$\\
\multirow{2}{*}{$-5\leq T\leq0$} &Outer closed rings&58&0.88 (0.01)&0.07 (0.01)\\
&Inner closed rings&56&0.91 (0.02)&0.16 (0.02)\\
\hline
\multirow{2}{*}{$\epsilon_{\rm d}\leq0.5$} & Outer rings&85&0.89 (0.01)&0.10 (0.01)\\
&Inner rings&152&0.82 (0.02)&0.10 (0.02)\\
&Nuclear rings&18&0.83 (0.01)&0.05 (0.01)\\
\multirow{2}{*}{$1\leq T\leq3$} &Outer closed rings&17&0.90 (0.00)&0.03 (2.97)\\
&Inner closed rings&37&0.84 (0.01)&0.08 (0.01)\\
\hline
\multirow{2}{*}{$\epsilon_{\rm d}\leq0.5$} & Outer rings&44&0.85 (0.02)&0.10 (0.02)\\
&Inner rings&187&0.76 (0.00)&0.11 (0.00)\\
&Nuclear rings&10&0.80 (0.00)&0.16 (0.00)\\
\multirow{2}{*}{$4\leq T\leq10$} &Outer closed rings&6&$-$&$-$\\
&Inner closed rings&24&0.77 (0.01)&0.11 (0.01)\\
\hline
\end{tabular}
\tablefoot{$\left<q_{\rm r,0}\right>$ stands for the fitted centre of the Gaussian distribution and $\sigma$ for its width. The numbers in brackets indicate 1-sigma errors of the fits.}
\end{table*}

One of the improvements of this study over the largest systematic study of rings conducted so far, the CSRG, is that the S$^4$G P4 provides us with reliable orientations of the galaxies. This allows us to obtained intrinsic ring shapes and orientations, as explained in Section~\ref{sdeproj}.

In the CSRG, the deprojected ring axis ratio distribution was studied under several assumptions and using a statistical treatment. One of the assumptions was that the actual shape of the deprojected axis ratio distribution is Gaussian, with a peak at $\left<q_{\rm r,0}\right>$ and a dispersion $\sigma$. With our deprojected data we do not need this assumption.

The deprojected ring distribution is shown in the bottom panels of Figure~\ref{fshape}. We overlay Gaussian fits to show how well the Gaussian distribution hypothesis works. The parameters of the Gaussian fits in this section can be found in Table~\ref{tellip}. Although the fitted Gaussians are nearly the same when fitting rings in all galaxies and when fitting them only for galaxies with $\epsilon_{\rm d}\leq0.5$, the distributions are slightly different in the low-$q_{\rm r}$ wing; indeed, when considering all galaxies, the left wing for outer and inner rings extends farther than when considering only galaxies with $\epsilon_{\rm d}\leq0.5$. This is because for close to edge-on galaxies deprojections based on the shape of outer isophotes of discs become unreliable due to the non-zero thickness of discs (see Section~\ref{sreliability2}) and/or the presence of extended bulges and haloes. Thus, for some galaxies, the obtained deprojection parameters can be considered to be tentative at best and may cause the calculated deprojected ring shapes to be unrealistically stretched. A notorious case is NGC~4594 (the Sombrero galaxy), where outer isophotes have an ellipticity of about $\epsilon_{\rm d}=0.58$ even though the real figure is probably closer to $\epsilon_{\rm d}=0.83$ \citep[see, e.g.,][]{JAR11}. All but one ring with deprojected axis ratio $q_{\rm r,0}\leq0.4$ are found in galaxies with fitted disc ellipticity $\epsilon_{\rm d}>0.5$ (IC~5264, NGC~4594, NGC~5078, NGC~5746, NGC~5777, and NGC~5900). The exception is NGC~986, a galaxy with $\epsilon_{\rm d}=0.1$ and a peculiar elongated nuclear ring. Because this effect may also occur (although in a less obvious way) in other inclined galaxies, we continue our analysis of intrinsic ring shapes based only on galaxies with $\epsilon_{\rm d}\leq0.5$.

Based on the fitted $\left<q_{\rm r,0}\right>$, we find that outer rings are generally rounder than the inner ones. More specifically, for the ARRAKIS sample we find $\left<q_{\rm r,0}({\rm O})\right>=0.88
\unskip$ and $\left<q_{\rm r,0}({\rm I})\right>=0.80
\unskip$, respectively. Outer closed rings have the same average $\left<q_{\rm r,0}({\rm O})\right>$ as outer features in general, but inner closed rings are on average slightly rounder than the average for inner features.

Rings become systematically more elliptical when moving along the family sequence from SA to SB (Figure~\ref{fshapebar}). However, the dispersion in each of the Gaussian distributions is always wide enough to allow very round rings to exist, even for SB galaxies. Fits for outer rings in SA galaxies are not shown because they yield that the centre of the Gaussian would be located at $\left<q_{\rm r,0}({\rm O})\right>>1$.

Inner rings tend to become more elliptical with increasing galaxy stage. This trend is not found for the outer features, although this might be explained by the low number of them found at stages $T\geq4$.

\subsection{Ring orientations}

\label{sorientation}

\begin{figure*}
  \includegraphics[width=0.9\textwidth]{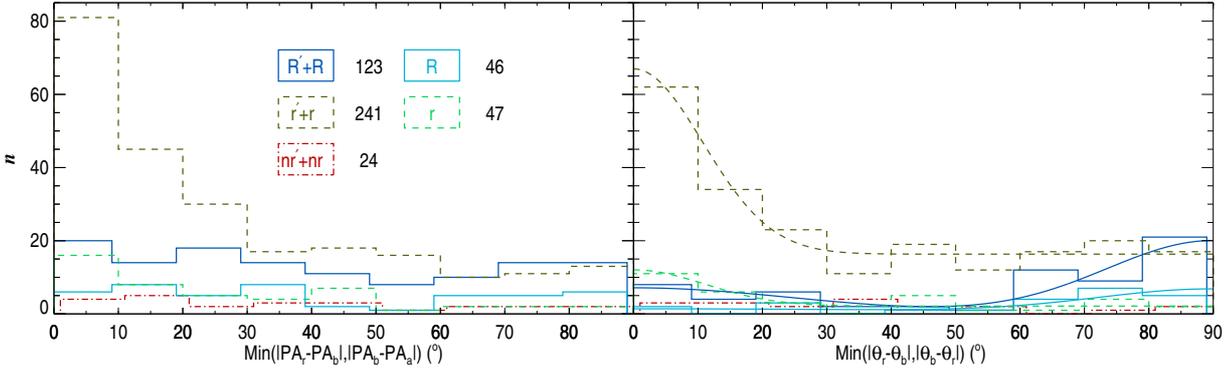}
  \caption{\label{fangle} Distribution of the projected resonance ring PA differences with respect to the bar (left panel) and ring orientation with respect to the bar in the deprojected galaxies (right panel). The histograms are colour- and line-coded as in Figure~\ref{fmorph}. The plots only include rings with $q_{\rm r}\leq0.85$ (left panel) and $q_{\rm r,0}\leq0.85$ (right panel) and in host galaxies with $\epsilon_{\rm d}\leq0.5$. The numbers in the top-left panel indicate the number of galaxies included in the histogram corresponding to the colour of the adjacent box.}
\end{figure*}

\begin{figure*}
 \includegraphics[width=0.90\textwidth]{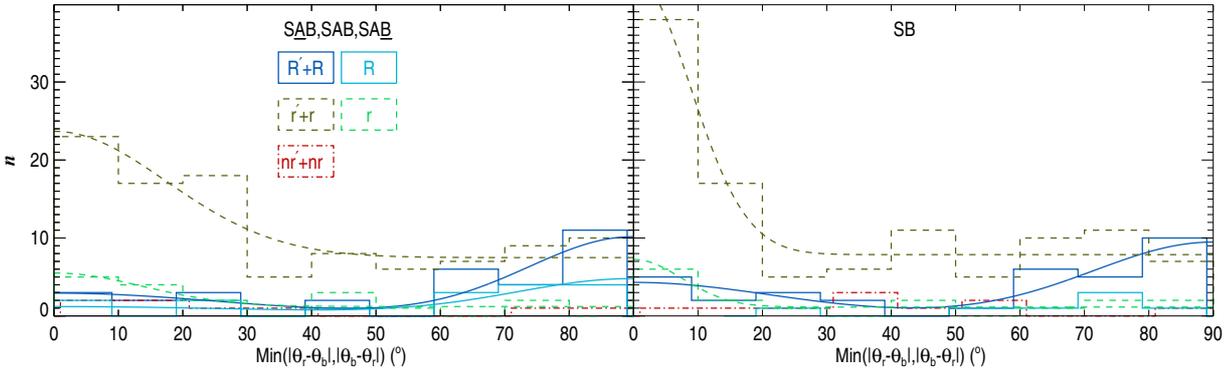}
  \caption{\label{fanglebar} Distribution of the resonance ring deprojected major axis angle difference with the bar colour- and line-coded as in Figure~\ref{fmorph} for different galaxy families and for $\epsilon_{\rm d}\leq0.5$. The left panel is for S${\underline{\rm A}}$B to SA${\underline{\rm B}}$ galaxies, the right panel for SB galaxies. The plots only include rings with $q_{\rm r,0}\leq0.85$.}
\end{figure*}

\begin{figure*}
 \includegraphics[width=0.90\textwidth]{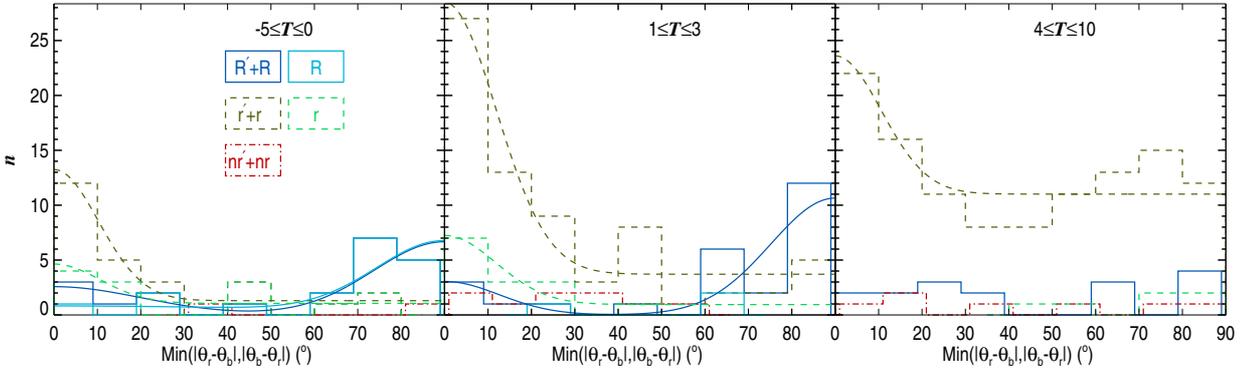}
  \caption{\label{fanglemorph} Distribution of the resonance ring deprojected major axis angle difference with the bar colour- and line-coded as in Figure~\ref{fmorph} for different galaxy stage ranges and for $\epsilon_{\rm d}\leq0.5$. The left panel represents galaxies with $-5\leq T\leq0$, the middle panel galaxies with $1\leq T\leq3$, the right panel galaxies with $4\leq T\leq10$. The plots only include rings with $q_{\rm r,0}\leq0.85$.}
\end{figure*}

\begin{table*}
\caption{Number of galaxies in the histograms in Figures~\ref{fangle}, \ref{fanglebar}, and \ref{fanglemorph} and parameters of the Gaussian fits to the $q_{\rm r,0}$ distributions}
\label{tangle}
\centering
\begin{tabular}{c c | c c c c c c}
\hline\hline
Galaxy subsample&Feature type&$n$&$\sigma_{0^{\rm o}}$ $(^{\rm o})$&$w_{0^{\rm o}}$&$\sigma_{90^{\rm o}}$ $(^{\rm o})$&$w_{90^{\rm o}}$&$w_{\rm c}$\\
\hline
\multirow{2}{*}{$\epsilon_{\rm d}\leq0.5$} & Outer rings&65&25 (3)&0.35 (0.18)&17 (2)&0.65 (0.39)&$-$\\
&Inner rings&215&11 (1)&0.48 (0.12)&$-$&$-$&0.52 (0.10)\\
&Nuclear rings&18&$-$&$-$&$-$&$-$&$-$\\
\multirow{2}{*}{Barred} &Outer closed rings&23&53 (4)&0.39 (0.11)&18 (1)&0.61 (0.20)&$-$\\
&Inner closed rings&33&11 (1)&0.59 (0.18)&$-$&$-$&0.41 (0.16)\\
\hline
\multirow{2}{*}{$\epsilon_{\rm d}\leq0.5$} & Outer rings&31&27 (2)&0.32 (0.08)&16 (1)&0.68 (0.19)&$-$\\
&Inner rings&103&17 (1)&0.51 (0.11)&$-$&$-$&0.49 (0.09)\\
&Nuclear rings&8&$-$&$-$&$-$&$-$&$-$\\
\multirow{2}{*}{S${\underline{\rm A}}$B,SAB,SA${\underline{\rm B}}$} &Outer closed rings&16&37 (8)&0.34 (0.27)&18 (1)&0.66 (0.59)&$-$\\
&Inner closed rings&17&14 (1)&0.58 (0.27)&$-$&$-$&0.42 (0.30)\\
\hline
\multirow{2}{*}{$\epsilon_{\rm d}\leq0.5$} & Outer rings&34&22 (2)&0.35 (0.17)&19 (1)&0.65 (0.44)&$-$\\
&Inner rings&110&9 (1)&0.53 (0.18)&$-$&$-$&0.47 (0.15)\\
&Nuclear rings&10&$-$&$-$&$-$&$-$&$-$\\
\multirow{2}{*}{SB} &Outer closed rings&7&$-$&$-$&$-$&$-$&$-$\\
&Inner closed rings&16&7 (1)&0.53 (0.17)&$-$&$-$&0.47 (0.22)\\
\hline
\multirow{2}{*}{$\epsilon_{\rm d}\leq0.5$} & Outer rings&21&20 (1)&0.32 (0.11)&16 (0)&0.68 (0.36)&$-$\\
&Inner rings&26&10 (1)&0.72 (0.41)&$-$&$-$&0.28 (0.28)\\
&Nuclear rings&2&$-$&$-$&$-$&$-$&$-$\\
\multirow{2}{*}{$-5\leq T\leq0$} &Outer closed rings&18&60 (8)&0.33 (0.15)&16 (1)&0.67 (0.31)&$-$\\
&Inner closed rings&13&11 (2)&0.50 (0.35)&$-$&$-$&0.50 (0.50)\\
\hline
\multirow{2}{*}{$\epsilon_{\rm d}\leq0.5$} & Outer rings&26&13 (0)&0.19 (0.03)&15 (0)&0.81 (0.21)&$-$\\
&Inner rings&70&12 (2)&0.69 (0.33)&$-$&$-$&0.31 (0.20)\\
&Nuclear rings&9&$-$&$-$&$-$&$-$&$-$\\
\multirow{2}{*}{$1\leq T\leq3$} &Outer closed rings&3&$-$&$-$&$-$&$-$&$-$\\
&Inner closed rings&14&12 (1)&0.69 (0.18)&$-$&$-$&0.31 (0.16)\\
\hline
\multirow{2}{*}{$\epsilon_{\rm d}\leq0.5$} & Outer rings&16&$-$&$-$&$-$&$-$&$-$\\
&Inner rings&116&11 (1)&0.25 (0.06)&$-$&$-$&0.75 (0.10)\\
&Nuclear rings&7&$-$&$-$&$-$&$-$&$-$\\
\multirow{2}{*}{$4\leq T\leq10$} &Outer closed rings&0&$-$&$-$&$-$&$-$&$-$\\
&Inner closed rings&6&$-$&$-$&$-$&$-$&$-$\\
\hline
\end{tabular}
\tablefoot{$\sigma_{0^{\rm o}}$ and $\sigma_{90^{\rm o}}$ stand for the fitted widths of the Gaussian components set to be centred at $\theta_{\rm r}-\theta_{\rm b}=0^{\rm o}$ and $\theta_{\rm r}-\theta_{\rm b}=90^{\rm o}$. $w_{0^{\rm o}}$, $w_{90^{\rm o}}$, and $w_{\rm c}$ stand for the fitted fraction of rings found to be in one of the two Gaussian components and that in the component of rings randomly orientated with respect to the bar. The numbers in brackets indicate 1-sigma errors of the fit.}
\end{table*}

In this subsection we describe the orientations of the rings with respect to bars. We focus only on barred galaxies for which we are able to accurately measure their orientation (that is, most of those in galaxies with $\epsilon_{\rm d}\leq0.5$). We compared the PA of the major axis of the ring, ${\rm PA_{r}}$, and the bar, ${\rm PA_{b}}$. We denote the difference between those major axis as ${\rm PA_{r}}-{\rm PA_{b}}$. However, because we define it to be in the first quadrant, it should be formally written as ${\rm minimum}\left(\left|{\rm PA_r}-{\rm PA_b}\right|,\left|{\rm PA_b}-{\rm PA_r}\right|\right)$. Since ${\rm PA_{r}}$ is ill-defined for rings that are almost round in projection, we only studied features with a projected axis ratio $q_{\rm r}\leq0.85$.

Because for deprojected rings and bars the concept of PA is no longer applicable, we defined the deprojected orientation of a ring or a bar to be the counter-clockwise difference in angle between the orientation of its major axis and the line of nodes, $\theta_{\rm r}$ and $\theta_{\rm b}$. Thus, for deprojected galaxies we denote the orientation difference between a ring and the bar as $\theta_{\rm r}-\theta_{\rm b}$ [although it is defined as ${\rm minimum}\left(\left|\theta_{\rm r}-\theta_{\rm b}\right|,\left|\theta_{\rm b}-\theta_{\rm r}\right|\right)$]. Because $\theta_{r}$ is ill-defined for rings that are almost round in deprojection, we only studied features with a deprojected axis ratio $q_{\rm r,0}\leq0.85$. In Figure~\ref{fangle} we present the distributions of ${\rm PA_{r}}-{\rm PA_{b}}$ and $\theta_{\rm r}-\theta_{\rm b}$.

For the outer rings ${\rm PA_{r}(O)}-{\rm PA_{b}}$ there is some hint that the distribution has two mild peaks at ${\rm PA_{r}(O)}-{\rm PA_{b}}=0^{\rm o}$ and ${\rm PA_{r}(O)}-{\rm PA_{b}}=90^{\rm o}$ divided by a shallow valley. This behaviour is similar to that found in the CSRG, although there the peak at ${\rm PA_{r}(O)}-{\rm PA_{b}}=0^{\rm o}$ has more contrast; this may be due to our relatively low number of outer features (our plot has five times fewer objects than Figure~7g in the CSRG).

The distribution for inner rings has a large peak at ${\rm PA_{r}(I)}-{\rm PA_{b}}=0^{\rm o}$ and the the distribution drops until it stabilises between ${\rm PA_{r}(I)}-{\rm PA_{b}}=50^{\rm o}$ and ${\rm PA_{r}(I)}-{\rm PA_{b}}=90^{\rm o}$.

In deprojected galaxies, the outer rings have two preferred orientations, namely  parallel and perpendicular to bars [$\theta_{\rm r}({\rm O})-\theta_{\rm b}=0^{\rm o}$ and $\theta_{\rm r}({\rm O})-\theta_{\rm b}=90^{\rm o}$], which can be associated with the $R_2$ and $R_1$ ring types in the simulations of \citet{SCH81, SCH84} as well as with the predictions from non-deprojected data in the CSRG.

The intrinsic inner ring orientation distribution has a clear peak parallel to the bar [$\theta_{\rm r}({\rm I})-\theta_{\rm b}=0^{\rm o}$], as was previously hinted at in other works \citep[e.g., the CSRG,][]{VAU64, SCH81, SCH84, BU86b, ATH09A}. However, we also found that a significant fraction of rings have random orientations with respect to the bar, as indicated by the flat distribution for $\theta_{\rm r}({\rm I})-\theta_{\rm b}>30^{\rm o}$. Such a randomly oriented inner ring fraction, although sometimes suspected on the basis of a few individual cases of obvious bar/ring misalignment [ESO~565-11 \citep{BU95B}, NGC~309 \citep{BU07}, NGC~4319 \citep{SUL87}, and NGC~6300 \citep{BU87}] has never been described before.

We fitted the outer ring $\theta_{\rm r}({\rm O})-\theta_{\rm b}$ distribution with the superposition of two Gaussian curves centred at $\theta_{\rm r}({\rm O})-\theta_{\rm b}=0^{\rm o}$ and $\theta_{\rm r}({\rm O})-\theta_{\rm b}=90^{\rm o}$, as assumed in the CSRG. For the inner rings, we fitted the superposition of a Gaussian centred at $\theta_{\rm r}({\rm I})-\theta_{\rm b}=0^{\rm o}$, describing the population of inner features parallel to the bar and a constant distribution representing the population of rings oriented at random with respect to the bar. The fitted parameters and the number of galaxies in each histogram are tabulated in Table~\ref{tangle}. The fit also provides the fraction of rings that belong to the components parallel and anti-parallel to the bar for outer rings ($w_{0^{\rm o}}$ and $w_{90^{\rm o}}$) and parallel and oriented at random with respect to the bar for inner rings ($w_{0^{\rm o}}$ and $w_{\rm c}$). We found that the fraction of outer rings aligned parallel to the bar is $35\pm18\%$
\unskip and that of features oriented perpendicular to the bar is $65\pm39\%$
\unskip. For inner rings, we found that some $48\pm12\%$
\unskip are aligned parallel to the bar and the other $52\pm10\%$
\unskip have random orientations.

The distribution of the intrinsic orientations of outer features does not change much when looking separately at weakly and strongly barred galaxies (Figure~\ref{fanglebar}). The inner ring distribution changes more, since the width of the distribution of galaxies parallel to the bar decreases from $17\pm1^{\rm o}$
\unskip for galaxies with families S${\underline{\rm A}}$B to SA${\underline{\rm B}}$ to only $9\pm1^{\rm o}$
\unskip for SB galaxies.

The outer ring orientation distribution is also independent of the galaxy stage (Figure~\ref{fanglemorph}). But again, this is not the case for the inner ring distribution. Indeed, the later the stage, the larger the fraction of inner rings oriented at random with respect to the bar. This fraction increases from $28\pm28\%$
\unskip ($-5\leq T\leq0$) and $31\pm20\%$
\unskip ($1\leq T\leq3$) to $75\pm10\%$
\unskip for galaxies with $4\leq T\leq10$. For inner rings in late-type stages there is even some hint of a preferred orientation perpendicular to the bar (right panel in Figure~\ref{fanglemorph}).

\subsection{Absolute ring sizes}

\label{sringdiam}

\begin{figure*}
  \includegraphics[width=0.9\textwidth]{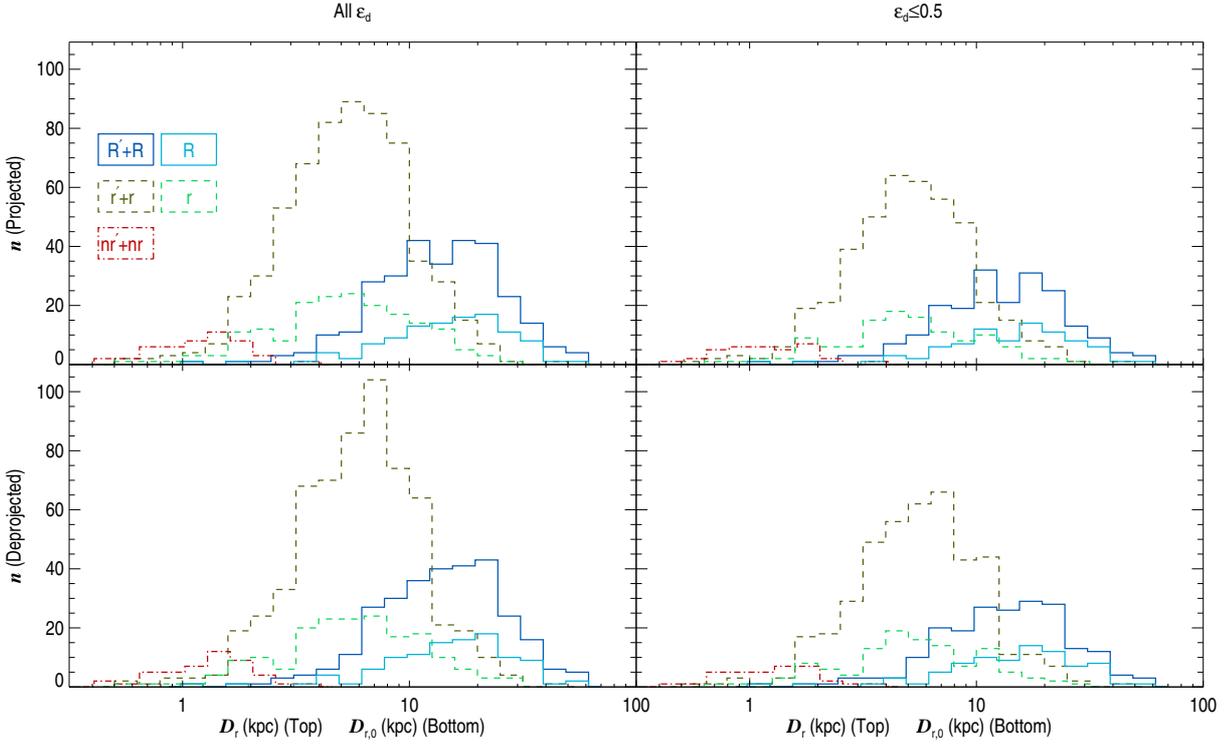}
  \caption{\label{fabsdiam} Distribution of the resonance ring absolute diameters. The top row shows the projected diameters and the bottom row the deprojected ones. The left column shows all galaxies, the right column galaxies with $\epsilon_{\rm d}\leq0.5$. The histograms are colour- and line-coded as in Figure~\ref{fmorph}.}
\end{figure*}

\begin{figure*}
 \includegraphics[width=0.9\textwidth]{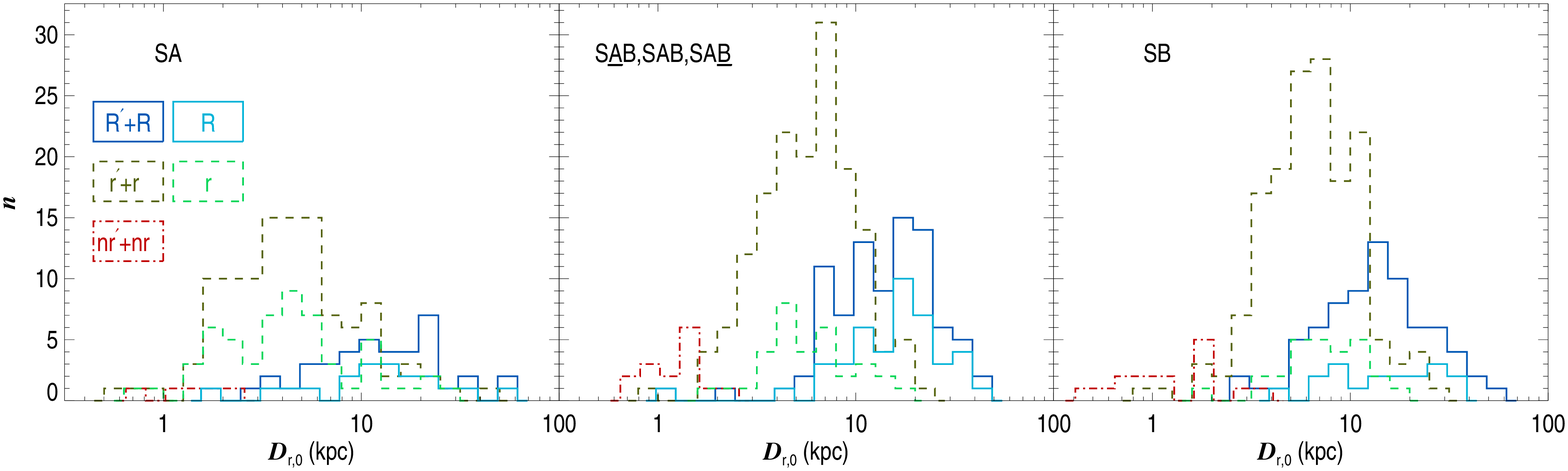}
  \caption{\label{fabsdiambar} Distribution of the resonance ring deprojected major axis absolute diameter colour- and line-coded as in Figure~\ref{fmorph} for different galaxy families and for $\epsilon_{\rm d}\leq0.5$. The left panel shows SA galaxies, the middle one S${\underline{\rm A}}$B to SA${\underline{\rm B}}$ galaxies, the right panel SB galaxies.}
\end{figure*}

\begin{figure*}
 \includegraphics[width=0.9\textwidth]{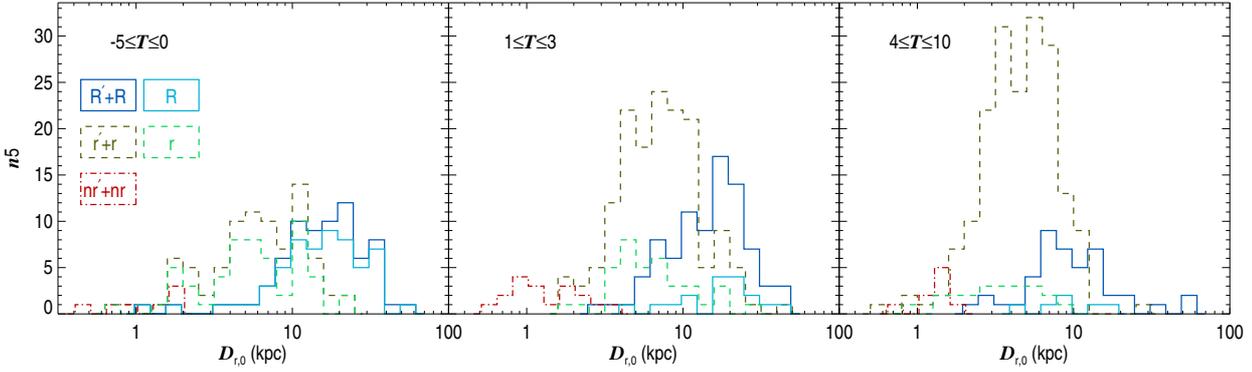}
  \caption{\label{fabsdiammorph} Distribution of the resonance ring deprojected major axis absolute diameter colour- and line-coded as in Figure~\ref{fmorph} for different galaxy stage ranges and for $\epsilon_{\rm d}\leq0.5$. The left panel represents galaxies with $-5\leq T\leq0$ galaxies, the middle panel galaxies with $1\leq T\leq3$, the right panel galaxies with $4\leq T\leq10$. }
\end{figure*}

\begin{table*}
\caption{Number of galaxies in the histograms in Figures~\ref{fabsdiam}, \ref{fabsdiambar}, and \ref{fabsdiammorph} and average and standard deviation of the absolute ring deprojected diameters for several ARRAKIS subsamples}
\label{tabsdiam}
\centering
\begin{tabular}{c c |c c c}
\hline\hline
Galaxy subsample&Feature type&$n$&$\left<D_{\rm r,0}\right>$ (kpc)&$\sigma\left(D_{\rm r,0}\right)$ (kpc)\\
\hline
\multirow{5}{*}{All galaxies} & Outer rings&295&17.0&10.3\\
&Inner rings&610&7.1&4.7\\
&Nuclear rings&47&1.4&0.6\\
&Outer closed rings&106&18.1&10.2\\
&Inner closed rings&180&7.2&5.3\\
\hline
\multirow{5}{*}{$\epsilon_{\rm d}\leq0.5$} & Outer rings&203&16.5&10.3\\
&Inner rings&425&6.7&4.6\\
&Nuclear rings&35&1.4&0.6\\
&Outer closed rings&82&17.8&10.3\\
&Inner closed rings&116&6.7&4.9\\
\hline
\multirow{2}{*}{$\epsilon_{\rm d}\leq0.5$} & Outer rings&40&16.0&12.1\\
&Inner rings&111&6.0&5.9\\
&Nuclear rings&5&1.4&0.6\\
\multirow{2}{*}{SA}&Outer closed rings&19&16.5&12.7\\
&Inner closed rings&56&5.9&5.8\\
\hline
\multirow{2}{*}{$\epsilon_{\rm d}\leq0.5$} & Outer rings&87&16.4&9.1\\
&Inner rings&156&6.7&3.7\\
&Nuclear rings&15&1.3&0.4\\
\multirow{2}{*}{S${\underline{\rm A}}$B,SAB,SA${\underline{\rm B}}$}&Outer closed rings&43&18.0&9.4\\
&Inner closed rings&33&6.7&3.8\\
\hline
\multirow{2}{*}{$\epsilon_{\rm d}\leq0.5$} & Outer rings&74&16.5&10.8\\
&Inner rings&158&7.4&4.3\\
&Nuclear rings&15&1.4&0.8\\
\multirow{2}{*}{SB}&Outer closed rings&18&17.8&9.9\\
&Inner closed rings&27&8.2&3.7\\
\hline
\multirow{2}{*}{$\epsilon_{\rm d}\leq0.5$} & Outer rings&71&18.5&10.4\\
&Inner rings&83&7.3&4.6\\
&Nuclear rings&7&1.2&0.5\\
\multirow{2}{*}{$-5\leq T\leq0$}&Outer closed rings&58&17.9&10.5\\
&Inner closed rings&55&7.1&4.7\\
\hline
\multirow{2}{*}{$\epsilon_{\rm d}\leq0.5$} & Outer rings&85&17.0&9.4\\
&Inner rings&152&8.4&5.6\\
&Nuclear rings&18&1.4&0.8\\
\multirow{2}{*}{$1\leq T\leq3$}&Outer closed rings&17&20.2&10.6\\
&Inner closed rings&37&7.7&5.7\\
\hline
\multirow{2}{*}{$\epsilon_{\rm d}\leq0.5$}& Outer rings&44&12.5&11.2\\
&Inner rings&184&5.1&2.9\\
&Nuclear rings&9&1.3&0.3\\
\multirow{2}{*}{$4\leq T\leq10$}&Outer closed rings&6&10.7&5.2\\
&Inner closed rings&22&3.9&2.2\\
\hline
\end{tabular}
\end{table*}

Absolute ring diameters were obtained using the average of the redshift-independent distances provided by NED as the host galaxy distance. If such a distance measurement was not available, we used the NED Hubble-Lema\^itre flow distance corrected for the effects of the Virgo cluster, the Great Atractor, and the Shapley cluster (Hubble-Lema\^itre constant $H_{0}=73\,{\rm km\,s^{-1}\,Mpc^{-1}}$). Because the S$^4$G sample selection was based on uncorrected flow distances, this means that some of the galaxy distances listed in Appendix~A exceed 41\,Mpc. Of these galaxies with rings, only six are farther away than 60\,Mpc (ESO~576-1, NGC~4722, NGC~5600, PGC~47721, and UGC~5814). Such high distances might indicate either inaccurate redshift-based or redshift-independent distance determinations. However, since these galaxies are only a few, we did not exclude them from the study of absolute ring sizes.

Here, we define the size of a ring to be its major axis diameter. That is, $D_{\rm r}$ in the projected case and $D_{\rm r,0}$ in the deprojected case. Because rings are in general almost circular, their absolute size distribution does not vary much when comparing the projected and the deprojected sizes (Figure~\ref{fabsdiam}). Moreover, because the ring size is easier to measure than its orientation, the size distribution does not vary much when studying our whole sample or a subsample restricted by disc ellipticity. The averages and the standard deviations of each deprojected distribution in the plots are listed in Table~\ref{tabsdiam}.

The average absolute size for each of the three ring flavours is nearly independent of galaxy family (Figure~\ref{fabsdiambar} and Table~\ref{tabsdiam}). It is noteworthy that the dispersion in the size of inner rings in SA galaxies is higher than that in barred galaxies. This high dispersion can also be seen by examining the smallest and largest rings in the top panel in Figure~\ref{fabsdiambar}; the smallest resonance feature in SA galaxies is an inner pseudoring. This is also the case for the third-largest feature in this panel. Such a behaviour is not seen in barred galaxies, where the small- and large-size tails are dominated by nuclear and outer rings, respectively. A possible explanation of this high dispersion in the size of inner features found in SA galaxies is that a fraction of outer and nuclear resonant features have been misclassified as inner rings.

Outer and inner rings in galaxies with stages $4\leq T\leq10$ are on average much smaller than those found in earlier stages (Figure~\ref{fabsdiammorph}). This effect is not seen for nuclear features, but this may well be because some of them are too small to be resolved in the S$^4$G.

\subsection{Ring sizes relative to the size of the galaxy}

\label{sreldiam}

\begin{figure*}
  \includegraphics[width=0.9\textwidth]{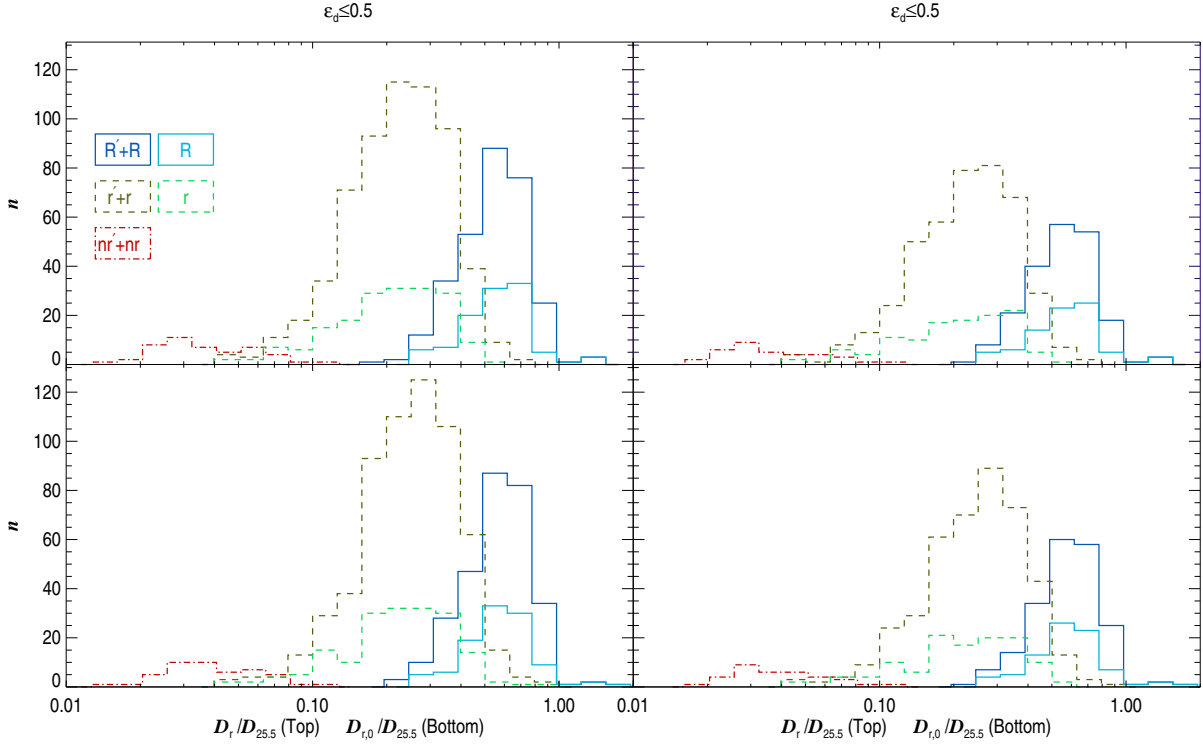}
  \caption{\label{freldiam} Distribution of the resonance ring deprojected major axis diameters relative to $D_{25.5}$. The left column shows all galaxies, the right column galaxies with $\epsilon_{\rm d}\leq0.5$. The histograms are colour and line-coded as in Figure~\ref{fmorph}.}
\end{figure*}

\begin{figure*}
 \includegraphics[width=0.9\textwidth]{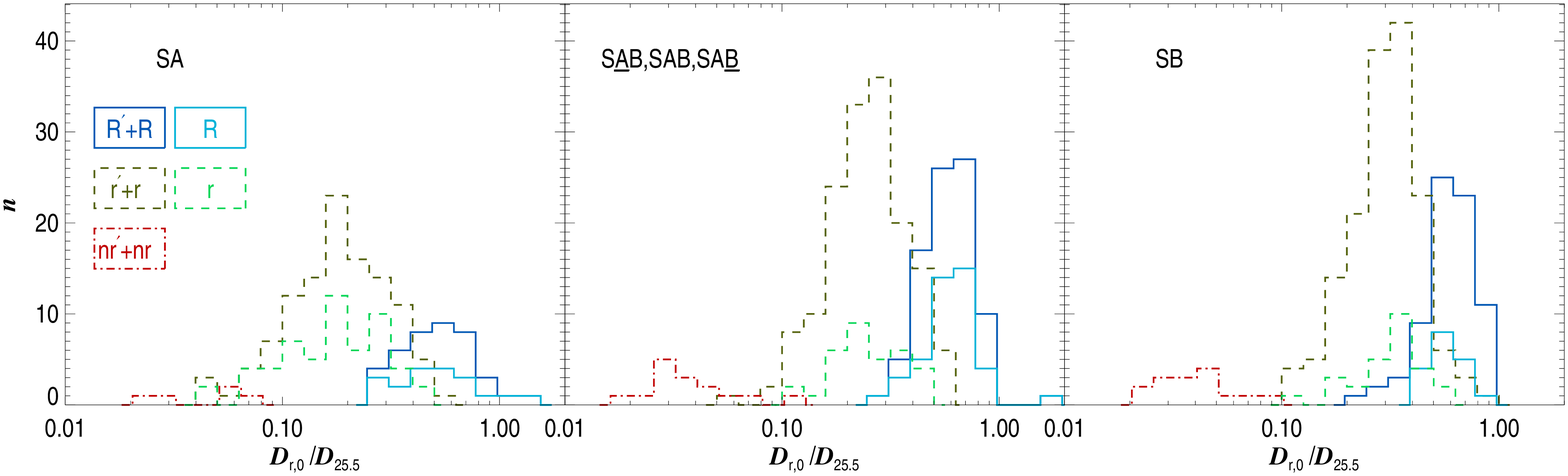}
  \caption{\label{freldiambar} Distribution of the resonance ring deprojected major axis relative to $D_{25.5}$ colour- and line-coded as in Figure~\ref{fmorph} for different galaxy families and for $\epsilon_{\rm d}\leq0.5$. The left panel shows SA galaxies, the middle one S${\underline{\rm A}}$B to SA${\underline{\rm B}}$ galaxies, the right panel SB galaxies.}
\end{figure*}

\begin{figure*}
 \includegraphics[width=0.9\textwidth]{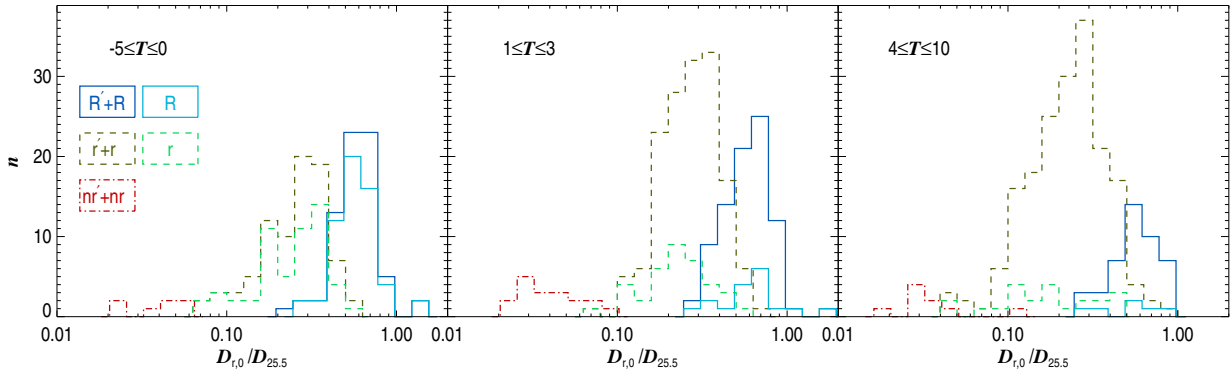}
  \caption{\label{freldiammorph} Distribution of the resonance ring deprojected major axis diameter relative to $D_{25.5}$ colour and line-coded as in Figure~\ref{fmorph} for different galaxy stage ranges and for $\epsilon_{\rm d}\leq0.5$. The left panel is for $-5\leq T\leq0$ galaxies, the middle panel is for $1\leq T\leq3$ galaxies, and the right panel is for $4\leq T\leq10$ galaxies.}
\end{figure*}

\begin{table*}
\caption{Number of galaxies in the histograms in Figures~\ref{freldiam}, \ref{freldiambar}, and \ref{freldiammorph} and average and standard deviation of the ring deprojected diameters relative to $D_{25.5}$ for several ARRAKIS subsamples.}
\label{treldiam}
\centering
\begin{tabular}{c c |c c c}
\hline\hline
Galaxy subsample&Feature type&$n$&$\left<D_{\rm r,0}/D_{25.5}\right>$&$\sigma\left(D_{\rm r,0}/D_{25.5}\right)$\\
\hline
\multirow{5}{*}{All galaxies} & Outer rings&295&0.60&0.19\\
&Inner rings&608&0.27&0.12\\
&Nuclear rings&47&0.04&0.02\\
&Outer closed rings&106&0.63&0.22\\
&Inner closed rings&180&0.25&0.12\\
\hline
\multirow{5}{*}{$\epsilon_{\rm d}\leq0.5$} & Outer rings&203&0.62&0.20\\
&Inner rings&424&0.27&0.12\\
&Nuclear rings&35&0.04&0.02\\
&Outer closed rings&82&0.64&0.24\\
&Inner closed rings&116&0.24&0.11\\
\hline
\multirow{2}{*}{$\epsilon_{\rm d}\leq0.5$} & Outer rings&40&0.58&0.24\\
&Inner rings&111&0.20&0.10\\
&Nuclear rings&5&0.05&0.02\\
\multirow{2}{*}{SA}&Outer closed rings&19&0.60&0.30\\
&Inner closed rings&56&0.19&0.10\\
\hline
\multirow{2}{*}{$\epsilon_{\rm d}\leq0.5$} & Outer rings&87&0.63&0.19\\
&Inner rings&155&0.27&0.11\\
&Nuclear rings&15&0.04&0.02\\
\multirow{2}{*}{S${\underline{\rm A}}$B,SAB,SA${\underline{\rm B}}$}&Outer closed rings&43&0.65&0.22\\
&Inner closed rings&33&0.26&0.09\\
\hline
\multirow{2}{*}{$\epsilon_{\rm d}\leq0.5$} & Outer rings&74&0.62&0.16\\
&Inner rings&158&0.32&0.12\\
&Nuclear rings&15&0.04&0.02\\
\multirow{2}{*}{SB}&Outer closed rings&18&0.60&0.14\\
&Inner closed rings&27&0.33&0.10\\
\hline
\multirow{2}{*}{$\epsilon_{\rm d}\leq0.5$} & Outer rings&71&0.62&0.20\\
&Inner rings&83&0.27&0.11\\
&Nuclear rings&7&0.04&0.01\\
\multirow{2}{*}{$-5\leq T\leq0$}&Outer closed rings&58&0.62&0.20\\
&Inner closed rings&55&0.26&0.11\\
\hline
\multirow{2}{*}{$\epsilon_{\rm d}\leq0.5$} & Outer rings&85&0.63&0.21\\
&Inner rings&152&0.29&0.11\\
&Nuclear rings&18&0.04&0.02\\
\multirow{2}{*}{$1\leq T\leq3$}&Outer closed rings&17&0.70&0.33\\
&Inner closed rings&37&0.25&0.10\\
\hline
\multirow{2}{*}{$\epsilon_{\rm d}\leq0.5$} & Outer rings&44&0.60&0.18\\
&Inner rings&183&0.25&0.13\\
&Nuclear rings&9&0.04&0.02\\
\multirow{2}{*}{$4\leq T\leq10$}&Outer closed rings&6&0.59&0.25\\
&Inner closed rings&22&0.20&0.13\\
\hline
\end{tabular}
\end{table*}

One of the outputs of the S$^4$G pipeline~3 (Mu\~noz-Mateos et al., in preparation) is the diameter at the $\mu_{\rm 3.6\,\mu m}=25.5\,{\rm mag\,arcsec^{-2}}$ level, $D_{25.5}$. This diameter can be used as an indicator of the size of the disc, although in a few cases it may be overestimating its true size because of the effect of bright extended spheroidal components (an obvious case probably is the Sombrero galaxy). Since $D_{25.5}$ is located in the outskirts of the galaxy, in the region where its orientation is typically measured, it does not need to be deprojected. $D_{25.5}$ is available for all sample galaxies but two, NGC~253 and NGC~4647. For NGC~253 this is because the galaxy is tightly fitted in the S$^4$G frame and the $D_{25.5}$ diameter is close to the limits of the frame or even beyond. For NGC~4647 this is because the galaxy isophotes are distorted because of a close galaxy in the same frame. 

The distribution of projected and deprojected relative radii of the rings, $D_{\rm r}/D_{25.5}$ and $D_{\rm r,0}/D_{25.5}$, (Figure~\ref{freldiam}) shows that these magnitudes are much more useful than $D_{\rm r}$ and $D_{\rm r,0}$ for separating the three flavours of resonance features. The overlap in radius of the different types of features is much smaller than that found when looking at absolute radii because distributions are much more peaked. Indeed, comparing Table~\ref{tabsdiam} and Table~\ref{treldiam} shows that for outer rings in galaxies with $\epsilon_{\rm d}\leq0.5$ the ratio between the peak value of the distribution and its dispersion is $\left[D_{\rm r,0}({\rm O})/D_{25.5}\right]/\sigma\left[D_{\rm r,0}({\rm O})/D_{25.5}\right]=3.2
\unskip$, which is almost twice higher than $D_{\rm r,0}({\rm O})/\sigma\left[D_{\rm r,0}({\rm O})\right]=1.6
\unskip$. For inner rings the values are $\left[D_{\rm r,0}({\rm I})/D_{25.5}\right]/\sigma\left[D_{\rm r,0}({\rm I})/D_{25.5}\right]=2.2
\unskip$ and $D_{\rm r,0}({\rm I})/\sigma\left[D_{\rm r,0}({\rm I})\right]=1.5
\unskip$.

This effect is not seen for nuclear rings; a reason may be that outer and inner rings are associated to a single resonance or manifold structure that is found at a given radius relative to the bar (with small variations due to the bar pattern speed and the galaxy rotation curve). Instead, nuclear rings are associated to the range of radii between the two ILRs, or to the range of radii between the centre and the ILR, in case of galaxies with only one such resonance. A complementary explanation is that we are missing all nuclear features below a given angular size. If enough galaxies of a possible $D_{\rm r,0}({\rm N})/D_{25.5}$ peak were smaller than that threshold, the scatter of the nuclear ring absolute size distribution would not be reduced by much when using the ring diameters normalised with $D_{25.5}$.

When segregating galaxies according to their family (Figure~\ref{freldiambar}), we found a large overlap between the range of relative radii occupied by the inner and nuclear rings in SA galaxies. Such a large overlap is not seen in barred galaxies and may indicate, as commented before when looking at the absolute radii of rings, that some nuclear rings may have been classified as inner features in unbarred galaxies. We also found that the peak in the inner ring relative size distribution increases systematically with increasing bar strength. This is probably because of a slight tendency of stronger bars (SB) to be longer on average relative to galaxy size than weaker bars (S${\underline{\rm A}}$B, SAB, SA${\underline{\rm B}}$), at least for late-type galaxies. This was originally shown by \citet{ER05}, based on the data of \citet{MAR95}. \citet{ER05} used the RC3 $R_{25}$ measurement as the galaxy size.

Finally, we find that the relative ring size distributions are not sensitive to galaxy stage (Figure~\ref{freldiammorph}).

\subsection{Outer ring sizes relative to those of inner rings}

\label{sdouble}

\begin{figure*}
  \includegraphics[width=0.9\textwidth]{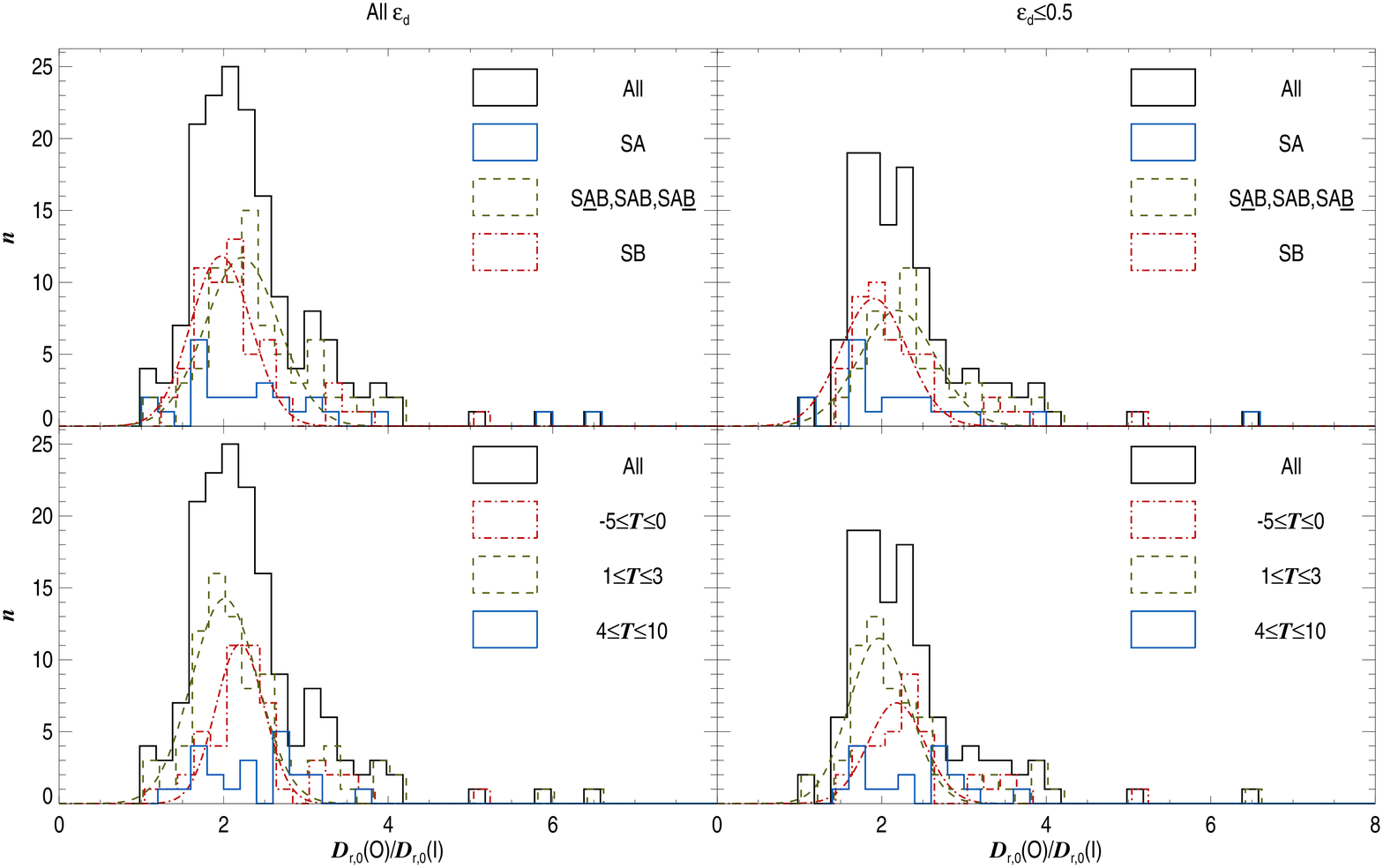}
  \caption{\label{fdouble} Distribution of the ratios of the diameters of outer and inner rings. The left column shows all galaxies, the right column galaxies with $\epsilon_{\rm d}\leq0.5$. The of histograms are colour- and line-coded according to the host galaxy family (top panels) and stage (bottom panels). One galaxy, NGC~5033, has $D_{\rm r,0}({\rm O})/D_{\rm r,0}({\rm I})=17.5$ and is beyond the horizontal limits of the plots.}
\end{figure*}

\begin{table*}
\caption{Number of galaxies in the histograms in Figure~\ref{fdouble} and parameters of the Gaussian fits to the $D_{\rm r,0}({\rm O})/D_{\rm r,0}({\rm I})$ distributions}
\label{tdouble}
\centering
\begin{tabular}{c c | c c c}
\hline\hline
Galaxy&Subsample&$n$&$\left<D_{\rm r,0}({\rm O})/D_{\rm r,0}({\rm I})\right>$&$\sigma$\\
inclinations&&&&\\
\hline
\multirow{7}{*}{All $\epsilon_{\rm d}$}&SA&29&$-$&$-$\\
&S${\underline{\rm A}}$B,SAB,SA${\underline{\rm B}}$&75&2.22 (0.06)&0.45 (0.06)\\
&SB&59&1.97 (0.04)&0.37 (0.04)\\
&&&&\\
&$-5\leq T\leq0$&52&2.20 (0.03)&0.29 (0.03)\\
&$1\leq T\leq3$&87&2.01 (0.03)&0.40 (0.03)\\
&$4\leq T\leq10$&23&$-$&$-$\\
\hline
\multirow{7}{*}{$\epsilon_{\rm d}\leq0.5$}&SA&22&$-$&$-$\\
&S${\underline{\rm A}}$B,SAB,SA${\underline{\rm B}}$&51&2.19 (0.08)&0.46 (0.08)\\
&SB&45&1.91 (0.04)&0.40 (0.05)\\
&&&&\\
&$-5\leq T\leq0$&36&2.18 (0.05)&0.34 (0.06)\\
&$1\leq T\leq3$&63&1.97 (0.03)&0.36 (0.03)\\
&$4\leq T\leq10$&18&$-$&$-$\\
\hline
\end{tabular}
\tablefoot{$\left<D_{\rm r,0}({\rm O})/D_{\rm r,0}({\rm I})\right>$ stands for the fitted centre of the Gaussian distribution and $\sigma$ for its width. The numbers in brackets indicate 1-sigma errors of the fits.}
\end{table*}

Our sample contains          163
\unskip galaxies with both outer and inner features. We calculated the ratio of the major axis of these features, $D_{\rm r,0}({\rm O})/D_{\rm r,0}({\rm I})$. Our sample has           19
\unskip galaxies with outer and inner features that also have more than one outer ring and/or more than one inner ring. When that was the case, the average of the major axis diameters of the outer (inner) features was used as the diameter of the outer (inner) feature.

The resulting diameter ratio distribution is represented in Figure~\ref{fdouble}, colour- and line-coded according to different host galaxy families and stages. One galaxy, NGC~5033, has $D_{\rm r,0}({\rm O})/D_{\rm r,0}({\rm I})=17.5$ and is beyond the horizontal limits of the plots. This high ratio may indicate that the innermost resonance feature in this unbarred galaxy is a nuclear and not an inner ring. In fact, this innermost feature in NGC~5033 is classified as a nuclear ring in AINUR.

In Figure~\ref{fdouble}, the histograms for barred galaxies and those for stages $-5\leq T\leq3$ have a shape that can be roughly approximated by a Gaussian, therefore we fitted them with that function. Other histograms (SA galaxies and galaxies with stages $4\leq T\leq10$) have too few galaxies to be fitted. The histogram for SA galaxies is very flat and has no distinct peak. The results of the fits and the number of galaxies included in each histogram, are listed in Table~\ref{tdouble}.

We found that $D_{\rm r,0}({\rm O})/D_{\rm r,0}({\rm I})$ is larger for weaker bars and also for earlier-type galaxies.

The peaks of the fitted distributions in Figure~\ref{fdouble} are always about $D_{\rm r,0}({\rm O})/D_{\rm r,0}({\rm I})=2$, which is consistent with the CSRG, \citet{KOR79}, \citep{ATH82}, and \citep{BU86a}. 

\section{Discussion}

\label{sdiscuss}

\subsection{Why is the stage distribution of inner and outer resonance features different?}
\label{sbarfreq1}

\begin{table*}
 \caption{Simplified behaviour of the probability of finding outer and inner rings for different galaxy stage ranges}
 \label{tschema}
 \centering
 \begin{tabular}{c c |c c c c}
 \hline\hline
 & &$-5\leq T\leq-2$&$-1\leq T\leq3$&$4\leq T\leq7$&$8\leq T\leq10$\\
 \hline
 \multirow{2}{*}{Can one frequently find\ldots?}& Outer rings & NO & YES & NO &  NO\\
				   & Inner rings & NO & YES & YES & NO\\
 \hline
 \end{tabular}
\end{table*}

\begin{figure}
  \includegraphics[width=0.45\textwidth]{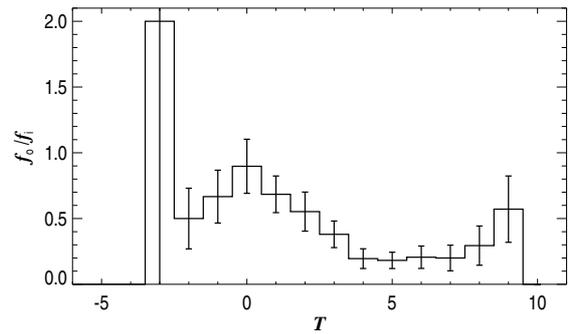}\\
  \caption{\label{funder} Fraction of galaxies with $\epsilon_{\rm d}\leq0.5$ with outer rings divided by that with inner rings, as a function of the galaxy stage.}
\end{figure}

In Section~\ref{sstage} we found that the distribution of stages (morphological type) for inner and outer rings is significantly different. Indeed, the outer ring fraction is high (above $20\%$) for $-1\leq T\leq3$ before dropping at later stages, but the inner ring distribution only drops at $T=6$ or $T=7$ (see a schematic view of this in Table~\ref{tschema} where the stages with a frequency of rings one fourth that of the peak for that ring flavour are considered to rarely host rings). For stages below $T=3$ outer rings are at least 40\% as abundant as inner rings, but this ratio decreases for later stages. Thus, there is a significant under-abundance of outer rings for late-type galaxies (Figure~\ref{funder}).

A possible justification for this effect can be found in the simulations by \citet{COM93}, who compared models designed to mimic both early- and late-type spirals distinguished by steeply and slowly rising inner rotation curves, respectively. However, to what extent their galaxy models are realistic is a matter of debate, since their $t=0$ models were sharply truncated at a radius of only a few disc scale-lengths (as little as 1.5 scale-lengths in some cases). Therefore, any explanation based on these simulations has to be taken with some caution. \citet{COM93} found that for early-type galaxies the CR radius is well inside the optical disc of the galaxy, but for later-type galaxies the CR is moved farther out. As a consequence, for a late-type galaxy, the OLR radius is typically found in a low-density region, and very little material is rearranged into a ring shape. Eventually, an outer feature forms when particles near the CR obtain angular momentum and are moved outwards, but this is a secular process that is even slower in late-type galaxies because regions near the CR have a lower particle density than that in earlier-type galaxies. Our results show that for this particular problem, the transition between an early- and a late-type disc galaxy is at $T=3$. The connection discussed here between simulations and observational data was previously made by \citet{COM93}.

\subsection{Why is the family distribution of inner and outer resonance features different?}
\label{sbarfreq2}

In Section~\ref{sfamily} we showed that the outer and inner ring distributions differ in the range of families from SA to SAB. In that range, the fraction of galaxies that host inner rings is constant within the error bars, but that of galaxies that host outer features increases steadily ($15\pm2\%$
\unskip for SA galaxies, $15\pm4\%$
\unskip for S${\underline{\rm A}}$B galaxies, and $22\pm3\%$
\unskip for SAB galaxies). A straightforward explanation for this effect is that outer rings do not form easily in unbarred galaxies. Indeed, weaker bars are likely to be less effective at redistributing material and angular momentum. It is therefore reasonable to assume that some have not yet had time to bring enough material from the CR region to the OR and, as a consequence, there is not enough material there to build an outer ring.

\subsection{Rings in unbarred galaxies}
\label{sunbarred}

Based on the information in Figure~\ref{fbar} we found that $15\pm2\%$
\unskip of SA galaxies have an outer ring. This fraction increases to $21\pm2\%$
\unskip for galaxies with bars (S${\underline{\rm A}}$B, SAB, SA${\underline{\rm B}}$, SB). For inner rings the values are $40\pm3\%$
\unskip and $44\pm2\%$
\unskip, respectively.

In the range $-1\leq T\leq3$, which is where outer features are frequently found according to Table~\ref{tschema}, outer rings are found in $29\pm5\%$
\unskip of the SA galaxies and in $49\pm3\%$
\unskip of the barred galaxies. In the range $-1\leq T\leq7$, where most inner features are found, the fraction of SA galaxies with inner rings is $46\pm4\%$
\unskip and that of barred ones hosting them is $59\pm2\%$
\unskip.

These numbers indicate that outer rings are 1.7 times more frequent in the S${\underline{\rm A}}$B to SB families than in the SA family. For inner rings the fraction in barred galaxies is only larger by a factor of about 1.3 than that in SA galaxies.

Our results are again mostly similar to those that can be obtained from the NIRS0S galaxy classifications \citep{LAU11}, which include disc galaxies in the range of stages $-3\leq T\leq1$. For outer features, we (they) found that the fraction of SA galaxies hosting them is $27\pm5\%$
\unskip ($15\pm5\%$) and that the fraction of barred galaxies hosting them is $49\pm4\%$
\unskip ($47\pm5\%$). For inner features in that range of stages, we (they) found that the fraction of SA galaxies hosting them is $51\pm5\%$
\unskip ($24\pm6\%$) and that the fraction of barred galaxies hosting them is $60\pm4\%$
\unskip ($56\pm5\%$). The NIRS0S statistics were calculated here excluding the galaxies labelled as spindle in \citet{LAU11} and ours were calculated with galaxies with $\epsilon_{\rm d}\leq0.5$.

The fact that outer features are more sensitive to the presence of a bar than inner features fits naturally with a scenario in which the rings in SA galaxies have formed due to broad oval distortions or long-lasting spiral modes \citep{RAU00} or with a model in which bars have been destroyed or dissolved after forming the rings. In the first case, simulations show that outer resonance features take longer to form than their inner counterparts. In the second case, the deficit of outer rings in unbarred galaxies can be explained because they can be destroyed easily by the interaction with companion galaxies \citep{EL92}. Alternatively, this effect can be attributed to the fact that a significant fraction of outer rings may have been misclassified as inner rings in unbarred galaxies (Section~\ref{sringdiam}).

Quantifying the link between resonance features and ovals or spiral patterns requires studying parameters such as the non-axisymmetric torque, $Q$. This was defined by \citet{COM81} and indicates the strength of the non-axisymmetries of the galactic potential. This parameter can be local ($Q_{\rm r}$, the local non-axisymmetric torque) or global ($Q_{\rm g}$, the maximum of all $Q_{\rm r}$). In a study that contained 16 unbarred and 31 barred galaxies, \citet{GROU10} pointed out a correlation between the ring axis ratio and the $Q_{\rm r}$ at its radius independently of whether the host galaxy is barred or not. This shows that, indeed, not only bars are responsible for shaping the rings, but also other non-axisymmetries in the galactic potential such as those caused by ovals and spiral arms. Further investigation will require analysing larger samples, which may be drawn from the S$^4$G.

\subsection{Why is the inner ring orientation dependent on the host galaxy stage?}

\label{sdiscorientation}

\begin{figure}
  \includegraphics[width=0.45\textwidth]{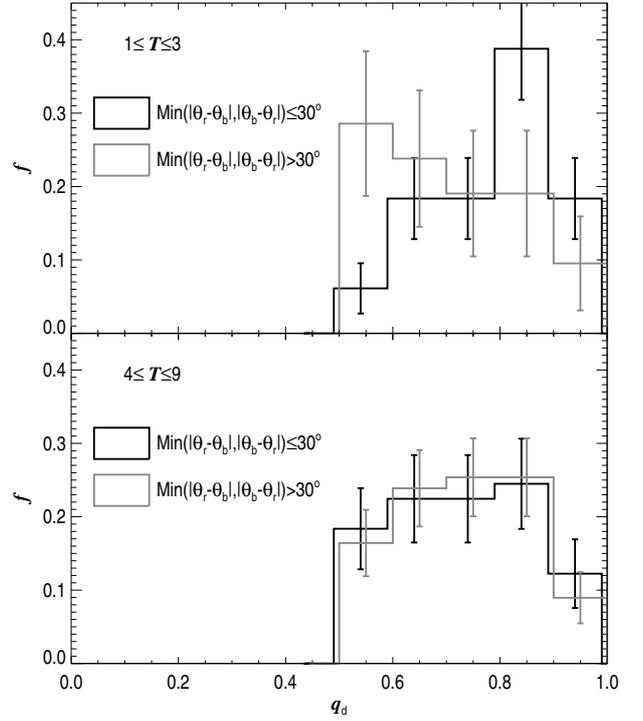}\\
  \caption{\label{fmisincl} Fraction of galaxies with inner rings aligned with the bar ($\theta_{\rm r}-\theta_{\rm b}\leq30^{\rm o}$; in black) and inner rings misaligned with the bar ($\theta_{\rm r}-\theta_{\rm b}>30^{\rm o}$; in grey) that have a given disc axis ratio. The top panel shows galaxies with stages $1\leq T\leq3$, the bottom panel galaxies with stages $4\leq T\leq 10$. Only galaxies with deprojected ring axis ratios $q_{\rm r,0}({\rm I})<0.85$ and disc axis ratios $q_{\rm d}\leq0.5$ are included. The error bars are calculated using binomial statistics.}
\end{figure}

\begin{figure}
  \includegraphics[width=0.45\textwidth]{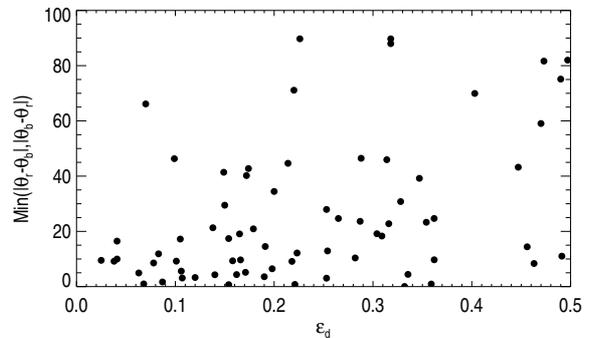}\\
  \caption{\label{fmisinclb} Ring deprojected major axis angle difference with the bar as a function of the galaxy disc ellipticity for $\epsilon\leq0.5$.}
\end{figure}

\begin{figure*}
\setlength{\tabcolsep}{2.5pt}
  \begin{tabular}{c c c c}
  \includegraphics[width=0.24\textwidth]{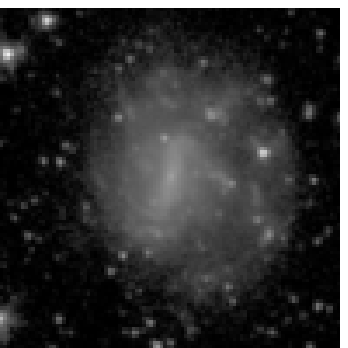}&
  \includegraphics[width=0.24\textwidth]{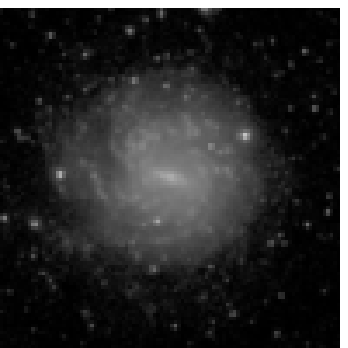}&
  \includegraphics[width=0.24\textwidth]{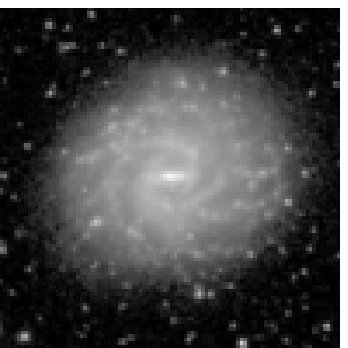}&
  \includegraphics[width=0.24\textwidth]{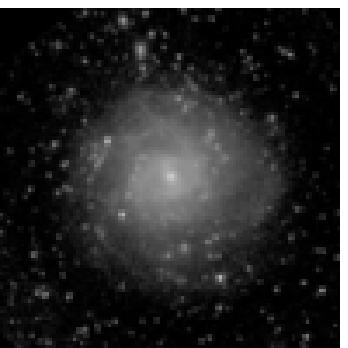}\\
  IC~4536\hspace{3mm}SB(r${\underline{\rm s}}$)cd&
  NGC~1493\hspace{3mm}SB(rs)c&
  NGC~3346\hspace{3mm}SB(rs)cd&
  NGC~4136\hspace{3mm}S${\underline{\rm A}}$B(rs)bc\\
  $|\theta_{\rm r}-\theta_{\rm b}|=71^{\rm o}$&
  $|\theta_{\rm r}-\theta_{\rm b}|=38^{\rm o}$&
  $|\theta_{\rm r}-\theta_{\rm b}|=74^{\rm o}$&
  $|\theta_{\rm r}-\theta_{\rm b}|=84^{\rm o}$\\
  \end{tabular}
  \caption{\label{fmisal} Selection of close to face-on S$^4$G galaxies with stages $4\leq T\leq10$ that host inner pseudorings misaligned with the bar. The images have the same properties as those in Figure~\ref{for}. The name of the galaxies, their morphological classification, and the bar-pseudoring intrinsic misalignment can be found below each image.}
\end{figure*}

In Section~\ref{sorientation} we found that inner rings prefer to be aligned in parallel with the bar major axis. However, we also found that a significant fraction of inner rings ($\sim50\%$) is oriented at random with respect to the bar. The fraction of rings with random orientations is low ($20-30\%$) for early-type galaxies, but increases to $\sim70\%$ for galaxies with $T\geq4$.

We investigated whether this effect might be related to a high galaxy inclination, which would prevent the accurate measurement of bar and ring orientations. To do this, we divided the galaxies that host inner rings into two groups: those whose inner rings are roughly aligned with the bar ($\theta_{\rm r}-\theta_{\rm b}\leq30^{\rm o}$), and those that are misaligned with it ($\theta_{\rm r}-\theta_{\rm b}>30^{\rm o}$). For each of these groups we plotted the fraction of galaxies that can be found at a given disc axis ratio, restricting ourselves to $\epsilon\leq0.5$. The expectation for circular discs with their rotation axis pointing at random is that the distribution is uniform. We calculated this for galaxies with $1\leq T\leq3$ and $4\leq T\leq10$ (top and bottom panels in Figure~\ref{fmisincl} respectively). Earlier-type galaxies have nearly no misaligned inner rings.

The top panel in Figure~\ref{fmisincl} suggests that misaligned rings tend to be found more frequently in the bins with more highly inclined galaxies, and the reversed trend is seen for galaxies oriented parallel with the ring (if we except the bin centred at $q_{\rm d}=0.95$). To verify whether this trend is statistically significant, we plotted in Figure~\ref{fmisinclb} the intrinsic ring/bar misalignments as a function of the disc ellipticity for all galaxies in the top panel of Figure~\ref{fmisincl}. We can see that misaligned rings have a preference for inclined galaxies. We calculated the Spearman coefficient of the data in Figure~\ref{fmisinclb} and found that $\rho=0.37$ and that the probability of the variables in the two axes to be uncorrelated is $p=0.002$. It is therefore very likely that at least some of the misalignments are caused by incorrect deprojections. We see no such effect for galaxies with $4\leq T\leq10$ (bottom panel in Figure~\ref{fmisincl}), which suggests that these deprojection problems might be related to the thickness of the discs. Indeed, as discussed in Section~\ref{sreliability2}, earlier-type galaxies have thicker discs.

For late-type discs, the distributions in Figure~\ref{fmisincl} differ from the uniform expectation in the sense that the bin for rounder discs is less populated than the others. This effect is well known for disc galaxies in general and has been attributed to genuine disc deviations from circularity of the order of $b/a=0.9$ or to the presence of substructure in the discs \citep{LAM92}. We examined whether these deviations from perfectly circular discs might be responsible for rings misaligned with their bars if we were wrongly assuming purely circular discs. We did this by projecting several thousand elliptical discs with random orientations. These discs had a bar also oriented at random and rings with an orientation differing from that of the bar following a normal distribution with a dispersion of the order of $10^{\rm o}$. Then, we deprojected the galaxies, but this time we assumed that they were circular. We found that for moderate disc intrinsic ellipticities ($b/a\gtrsim0.8$), this did not cause a fraction of $\sim70\%$ rings apparently oriented at random with respect to the bar, as we observe for the late spirals. However, the fraction of rings oriented at random with respect to the bar for galaxies with $1\leq T\leq3$ can be explained in this way. This is natural since, as we already argued, at least some ring/bar misalignments in early spirals might be caused by errors in the deprojections.

Therefore our results indicate that at least for late-type galaxies misalignments are not caused by uncertainties in bar and ring properties measured in highly inclined discs or because discs might not exactly be circular, as assumed when calculating the deprojections. Therefore, a significant fraction of inner rings are intrinsically misaligned, especially those in late-type galaxies. A selection of very low-inclination galaxies with such misaligned inner rings is shown in Figure~\ref{fmisal}.

The existence of misaligned inner features was also shown in simulations by \citet{RAU00}. In their study, this misalignment appears because a spiral mode with a pattern speed lower than that of the bar dominates the Fourier amplitude spectrum at the radius of the inner ring. This means that our results would be consistent with spiral modes rotating with a pattern speed different from that of the bar and with a significant contribution at the I4R radius for late-type galaxies. Alternatively, this is caused by a rotation of the Lagrangian points from their usual position, in the direction of the bar major axis, by an angle that increases with increasing spiral amplitude \citep[e.g.,][]{ATH10}. A way to explore these possibilities would be to compare the amplitude of the bar and spiral arm torque at the ring radius in our sample galaxies as reported, for instance, by \citet{BLOCK04} and \citet{SA10}. The latter authors described that the spiral density amplitude correlates well with the local bar torque as long as it is measured within 1.5 times the bar length. Since bars in late-type galaxies can significantly shorter than their CR radius \citep[see, e.g.,][]{COM93}, a correlation between the bar and the spiral torque at the inner ring radius may not be found.

An alternative possibility is that the apparent random orientation between bars and rings is caused by errors in measuring the inner feature position angle. This is because the later a galaxy is, the more ill-defined its rings become. Also, inner pseudorings in late-type galaxies tend to be more spiral-like than those in early-type galaxies, which partly invalidates the approximation made when fitting them with ellipses.

\section{Summary and conclusions}

\label{sconclusions}

\begin{figure}
  \includegraphics[width=0.45\textwidth]{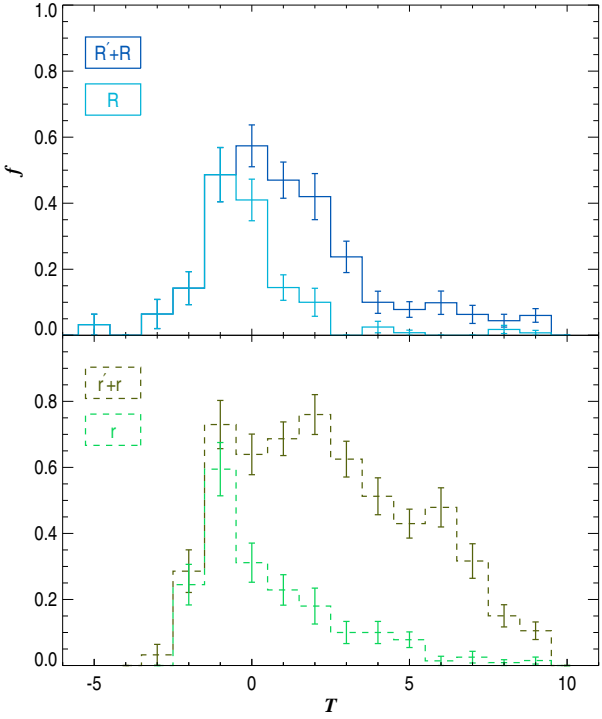}\\
  \caption{\label{fmdetail} Fraction of galaxies with outer rings (top panel) and inner rings (bottom panel) for a given stage. The plots only include galaxies with a disc ellipticity $\epsilon_{\rm d}\leq0.5$. In each of the panels, both rings as a whole and closed rings are indicated.}
\end{figure}

\begin{figure}
  \includegraphics[width=0.45\textwidth]{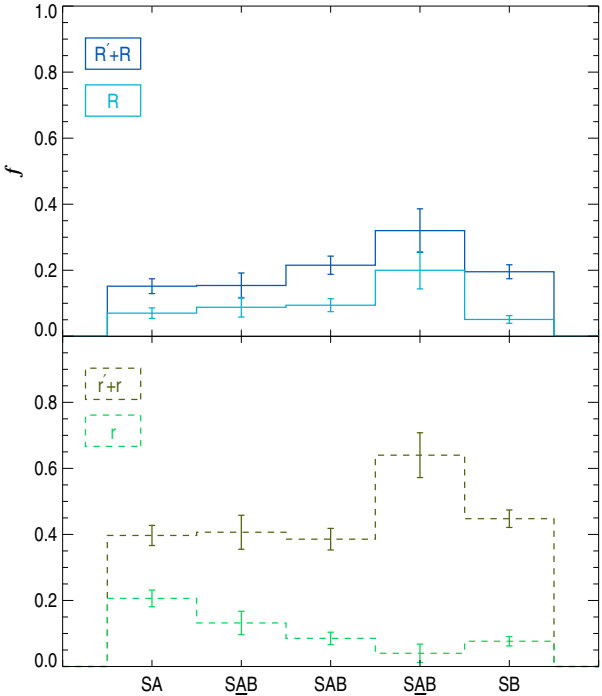}\\
  \caption{\label{fbdetail} As in Figure~\ref{fmdetail}, but now with the bar family instead of stage.}
\end{figure}

We presented ARRAKIS, the atlas of resonant rings as known in the S$^4$G, and a catalogue of the ring and pseudoring properties. The preliminary analysis of the data contained in the catalogue (Appendix~A) and the atlas (Appendix~B) was also presented. In this paper rings and pseudorings have been collectively termed rings.

The most common rings are resonance rings. These rings are thought to be due to gathering of gas, which can later be transformed into stars, at or close to the radii of the resonances caused by non-axisymmetries such as bars, ovals, and strong spiral patterns \citep[e.g.,][]{SCH81, SCH84, SELL93, BYRD94, RAU00}. They come in three flavours (outer, inner, and nuclear), depending on their radius relative to that of the bar. Outer features are thought to be related to the OLR or occasionally to the O4R, inner features to the I4R, and nuclear features to the ILRs. Inner and outer rings might also be based on flux tube manifolds triggered by the bar or oval distortion \citep{ROM06, ROM07, ATH09A, ATH09B, ATH10, ATH12B}. Either way, rings are tracers of the underlying galactic potential, which makes them important clues for the understanding of the secular evolution of disc galaxies.

The S$^4$G is a survey of 2352 galaxies representative of the local Universe observed at 3.6 and $4.5\,\mu{\rm m}$ wavelengths. The galaxies in the S$^4$G were classified by R.~J.~Buta, who also identified the rings (Buta et al.~in preparation). We studied all the rings in the S$^4$G frames, but our focus was on resonance rings in the S$^4$G galaxy sample. Because dust obscuration is weak in the mid-infrared, the chances that rings are hidden by dust are much lower than at optical wavelengths. This makes the S$^4$G a very good tool for studying rings.

We described the ring shape and orientation by marking their contours and fitting ellipses to them. To do this, we used model-subtracted images of the galaxies produced by the P4 of the S$^4$G (Salo et al.~in preparation). We obtained intrinsic ring shapes and orientations by deprojecting the fitted ellipses. The deprojection parameters were obtained from the P4 ellipse fits to the outer parts of S$^4$G galaxies. Bar orientations and ellipticities (a rough indicator of bar strength) were measured by looking at the ellipse fit corresponding to the radius where the highest ellipticity was found within the bar. They were deprojected in the same way as for rings.

We summarise our findings as follows:
\begin{itemize}
 \item {\bf Section~\ref{sstage} and Figures~\ref{fmorph} and \ref{fmdetail}}: Outer rings are found in $16\pm1\%$
\unskip of the S$^4$G galaxies and are more frequent for stages $-1\leq T\leq2$ (over 40\% frequency). Outer closed rings account for 100\% of outer features with $T\leq-1$. Inner rings are the most frequent resonance feature in the local Universe ($35\pm1\%$
\unskip frequency in the S$^4$G sample). They are typically found in the range $-1\leq T\leq6$ (over $40\%$ frequency) and their frequency peaks at stages $-1\leq T\leq3$ (over $60\%$ frequency). The inner closed ring distribution is shifted to earlier stages than that of inner features.
 \item {\bf Sections~\ref{sfamily} and \ref{sbarfreq2} and Figures~\ref{fbar} and \ref{fbdetail}}: The outer ring frequency increases from $15\pm2\%$
\unskip to $32\pm7\%$
\unskip along the family sequence from SA to SA${\underline{\rm B}}$, and decreases again to $20\pm2\%$
\unskip for SB galaxies. The inner ring frequency is qualitatively different, since it is roughly constant at $\sim40\%$ for all families except for a peak at $64\pm7\%$
\unskip frequency for SA${\underline{\rm B}}$ galaxies. The reason might be that outer rings take a longer time than inner rings to be built in galaxies with weaker non-axisymmetries. We also found that inner closed rings have a preference for unbarred and weakly barred galaxies.
 \item {\bf Section~\ref{sunbarred}}: Barred galaxies have outer rings 1.7 times more often than unbarred ones. Barred galaxies have $1.3$ more inner rings than their unbarred counterparts. This indicates that although they are stimulated by bars, rings do not need them to exist. Simulations suggest \citep[e.g.,][]{SA99, RAU00} that rings in unbarred galaxies may be related to weak ovals and/or long-lived spiral modes. Alternatively, they may have formed because of bars that have since been destroyed or have dissolved.
 \item {\bf Sections~\ref{sstage} and \ref{sunbarred}}: We confirmed the results of the NIRS0S survey regarding the frequencies of outer and inner rings as a function of the stage and the family for galaxies with stages $-3\leq T\leq1$.
 \item {\bf Section~\ref{sshape} and Figures~\ref{fshape}, \ref{fshapebar}, and \ref{fshapemorph}}: The axis ratio distribution of each of the three ring flavours - outer, inner, and nuclear - can be fitted with a Gaussian curve. Their intrinsic axis ratios typically range from $q_{\rm r,0}=0.6$ to $q_{\rm r,0}=1.0$. Inner rings are in general more elliptical than their outer counterparts. We found that outer and inner rings become on average more elliptical when the bar strength increases (when the galaxy family changes from SA to SB) and that inner rings are more elliptical in late-stage galaxies than in earlier-stage galaxies.
 \item {\bf Section~\ref{sorientation} and Figure~\ref{fangle}}: We confirmed that outer rings have two preferred orientations, namely parallel and perpendicular to the bar. Of the outer rings in barred galaxies $35\pm18\%$
\unskip are parallel to the bar and $65\pm39\%$
\unskip are oriented perpendicular to it.
 \item {\bf Sections~\ref{sorientation} and \ref{sdiscorientation} and Figures~\ref{fangle} and \ref{fanglemorph}}: We found that many inner rings have their major axes oriented parallel to the bar, as predicted in simulations and as previously reported from observations. However, we also found that maybe as much as $50\%$ of inner rings have random orientations with respect to the bar. These misaligned inner rings are mostly found in late-type spirals ($T\geq4$). We speculate that this may be because the Fourier amplitude spectrum at the radius of the I4R is dominated by spiral modes with a pattern speed different from that of the bar late-type galaxies. Alternatively, this can be due to the increased difficulty of measuring inner ring properties when they become ill-defined at late stages.
 \item {\bf Sections~\ref{sringdiam} and \ref{sreldiam} and Figures~\ref{fabsdiam}, \ref{fabsdiambar}, \ref{freldiam}, and \ref{freldiambar}}: The size of a ring relative to that of the galaxy is a much better indicator of its flavour (outer, inner, or nuclear) than its absolute size. The wings of both the absolute and relative size distributions of inner rings are more extended in unbarred galaxies than for those with bars. This may indicate that several outer and nuclear rings have been misclassified as inner features in unbarred galaxies.
 \item {\bf Section~\ref{sdouble}}: Several of our sample galaxies have both an outer and an inner ring. The distribution of their diameter ratio peaks at $D_{\rm r,0}({\rm O})/D_{\rm r,0}({\rm I})\sim2$, which agrees with the literature.

\end{itemize}

\subsection{Open questions}

We discussed two very interesting questions on rings which still have no clear solution. Why are some rings found in apparently unbarred galaxies? And why do inner rings in late-type galaxies have a higher tendency to be oriented at random with respect to the bar than those in earlier-type galaxies?

As developed in the discussion in Section~\ref{sdiscuss}, galaxies with these properties have occasionally been found in simulations. However, to test whether the mechanism creating rings in unbarred galaxies and the one shaping inner rings that are misaligned with the bar is the same in nature and in simulations, one has to study each galaxy at a deeper level of detail than we did here. Indeed, visual identification of the bar and a rough estimate of its orientation may be insufficient and one may need to perform a Fourier analysis to unveil the effects of non-axisymmetries induced by spiral arms or broad weak ovals. Again, the S$^4$G is the right tool for this type of research because the mid-infrared is an excellent tracer of the stellar mass and thus of the baryonic matter contribution to the galactic potential.

\begin{acknowledgements}

We thank our anonymous referee, who very carefully read this paper. The authors wish to thank the entire S$^4$G team for their efforts in this project. We thank Pertti Rautiainen for useful discussions on the ring formation in his simulations, Sim\'on D\'iaz-Garc\'ia for helping with fundamental statistical concepts, and Glenn van de Ven for his useful comments.

We acknowledge financial support to the DAGAL network from the People Programme (Marie Curie Actions) of the European Union's Seventh Framework Programme FP7/2007-2013/ under REA grant agreement number PITN-GA-2011-289313. EA and AB acknowledge financial support from the CNES (Centre National d'\'Etudes Spatiales - France). KS, J-CM-M, TK, and TM acknowledge support from the National Radio Observatory, which is a facility of the National Radio Astronomy Observatory operated under cooperative agreement by Associated Universities, Inc. SC, HS, EL, MH-H, and JL acknowledge support from the Academy of Finland. 

This work is based on observations and archival data made with the Spitzer Space Telescope, which is operated by the Jet Propulsion Laboratory, California Institute of Technology under a contract with NASA. We are grateful to the dedicated staff at the Spitzer Science Center for their help and support in planning and execution of this Exploration Science program. We also gratefully acknowledge support from NASA JPL/Spitzer grant RSA 1374189 provided for the S$^4$G project.

This research has made use of SAOImage DS9, developed by Smithsonian Astrophysical Observatory.

This research has made use of the NASA/IPAC Extragalactic Database (NED) which is operated by the Jet Propulsion Laboratory, California Institute of Technology, under contract with the National Aeronautics and Space Administration.

\end{acknowledgements}

\bibliographystyle{aa}
\bibliography{arrakis.bib}

\Online
\appendix
\onecolumn
\section{Catalogue}
The catalogue contains two tables. One for galaxies included in the S$^4$G sample and a second one for galaxies that appear in S$^4$G frames, but are not included in the original sample. The information in the tables in the catalogue is organized as follows:
\begin{table}[!h]
\centering
\begin{tabular}{r l}
Column~1 & Galaxy ID\\
Column~2 & NED redshift-independent distance if available, NED redshift distance corrected for Virgo Cluster, Great Attractor\\
 &  and Shapley Cluster effects with $H_{0}=73\,{\rm km\,s^{-1}\,Mpc^{-1}}$ otherwise\\
Column~3 & Type of feature (ring, pseudoring or bar) described in that row\\
Column~4 & Feature major projected diameter (maximum ellipticity major diameter for bars)\\
Column~5 & Feature minor projected diameter (maximum ellipticity minor diameter for bars)\\
Column~6 & Feature major axis PA\\
Column~7 & Feature major intrinsic diameter (maximum ellipticity major diameter for bars)\\
Column~8 & Feature minor intrinsic diameter (maximum ellipticity minor diameter for bars)\\
Column~9 & Counter-clockwise angular distance between the line of nodes of the galaxy deprojection and the major axis of the feature\\
\end{tabular}
\end{table}
 
Deprojected ring parameters are not given for polar rings.
\clearpage
\subsection{(Pseudo)Rings in S$^4$G sample galaxies}
\begin{longtable}{l c c c c c c c c}
\caption{Catalogue of the properties of ringed galaxies included in the S$^4$G sample}\\
\hline\hline
ID &  $D$ &  Feature & $D_{\rm r,b}$ & $d_{\rm r,b}$ & ${\rm PA_{r,b}}$ & $D_{\rm r,b,0}$ & $d_{\rm r,b,0}$ & $\theta_{\rm r,b}$\\
 & (Mpc) & &(\arcmin) & (\arcmin) & ($^{\rm o}$) & (\arcmin) & (\arcmin) & ($^{\rm o}$)\\
 (1) & (2) & (3) & (4) & (5) & (6) & (7) & (8) & (9)\\
\hline
\endfirsthead
\caption{continued.}\\
\hline\hline
ID &  $D$ &  Feature & $D_{\rm r,b}$ & $d_{\rm r,b}$ & ${\rm PA_{r,b}}$ & $D_{\rm r,b,0}$ & $d_{\rm r,b,0}$ & $\theta_{\rm r,b}$\\
 & (Mpc)& & (\arcmin) & (\arcmin) & ($^{\rm o}$) & (\arcmin) & (\arcmin) & ($^{\rm o}$)\\
 (1) & (2) & (3) & (4) & (5) & (6) & (7) & (8) & (9)\\
\hline
\endhead
\multirow{2}{*}{ESO~13-16} & \multirow{2}{*}{23.0} & rs & 0.78 & 0.66 & 161.7 & 1.01 & 0.78 & 92.5\\
\nopagebreak
 & & bar & 0.30 & 0.08 & 166.0 & 0.30 & 0.12 & 0.5\\
\nopagebreak
\hline
\nopagebreak
\multirow{2}{*}{ESO~26-1} & \multirow{2}{*}{19.2} & R$_{2}^{\prime}$ & 1.25 & 0.92 & 37.4 & 1.26 & 0.97 & 22.3\\
\nopagebreak
 & & bar & 0.23 & 0.07 & 66.9 & 0.24 & 0.08 & 49.8\\
\nopagebreak
\hline
\nopagebreak
\multirow{1}{*}{ESO~54-21} & \multirow{1}{*}{16.5} & R$^{\prime}$ & 3.15 & 1.55 & 101.6 & 3.78 & 2.65 & 53.9\\
\nopagebreak
\hline
\nopagebreak
\multirow{1}{*}{ESO~119-16} & \multirow{1}{*}{26.6} & R$^{\prime}$ & 2.24 & 0.56 & 30.3 & 2.33 & 1.50 & 22.4\\
\nopagebreak
\hline
\nopagebreak
\multirow{2}{*}{ESO~202-41} & \multirow{2}{*}{17.0} & rs & 0.85 & 0.75 & 174.4 & 1.13 & 0.83 & 82.9\\
\nopagebreak
 & & bar & 0.73 & 0.22 & 166.9 & 0.75 & 0.32 & 19.9\\
\end{longtable}
\clearpage
\subsection{(Pseudo)Rings not in S$^4$G sample galaxies}
\begin{longtable}{l c c c c c c c c}
\caption{Catalogue of the properties of ringed galaxies that appear in S$^4$G frames, but are not included in the S$^4$G sample}\\
\hline\hline
ID &  $D$ &  Feature & $D_{\rm r,b}$ & $d_{\rm r,b}$ & ${\rm PA_{r,b}}$ & $D_{\rm r,b,0}$ & $d_{\rm r,b,0}$ & $\theta_{\rm r,b}$\\
 & (Mpc) & &(\arcmin) & (\arcmin) & ($^{\rm o}$) & (\arcmin) & (\arcmin) & ($^{\rm o}$)\\
 (1) & (2) & (3) & (4) & (5) & (6) & (7) & (8) & (9)\\
\hline
\endfirsthead
\caption{continued.}\\
\hline\hline
ID &  $D$ &  Feature & $D_{\rm r,b}$ & $d_{\rm r,b}$ & ${\rm PA_{r,b}}$ & $D_{\rm r,b,0}$ & $d_{\rm r,b,0}$ & $\theta_{\rm r,b}$\\
 & (Mpc)& & (\arcmin) & (\arcmin) & ($^{\rm o}$) & (\arcmin) & (\arcmin) & ($^{\rm o}$)\\
 (1) & (2) & (3) & (4) & (5) & (6) & (7) & (8) & (9)\\
\hline
\endhead
\multirow{1}{*}{IC~3062} & \multirow{1}{*}{115.7} & rs & 0.27 & 0.21 & 149.9 & 0.30 & 0.23 & 119.1\\
\nopagebreak
\hline
\nopagebreak
\multirow{1}{*}{NGC~2856} & \multirow{1}{*}{40.9} & r & 0.37 & 0.16 & 144.6 & 0.43 & 0.28 & 47.1\\
\nopagebreak
\hline
\nopagebreak
\multirow{3}{*}{NGC~2950} & \multirow{3}{*}{19.1} & RL & 1.99 & 1.50 & 115.9 & 2.25 & 1.96 & 104.8\\
\nopagebreak
 & & r${\underline{\rm l}}$ & 0.97 & 0.68 & 122.2 & 1.00 & 0.97 & 93.9\\
\nopagebreak
 & & bar & 0.80 & 0.47 & 155.7 & 0.98 & 0.57 & 56.5\\
\nopagebreak
\hline
\nopagebreak
\multirow{2}{*}{NGC~4175} & \multirow{2}{*}{42.2} & r & 0.41 & 0.11 & 130.4 & 0.41 & 0.32 & 176.0\\
\nopagebreak
 & & PRG & 0.29 & 0.03 & 38.7 & $-$ & $-$ & $-$\\
\nopagebreak
\hline
\nopagebreak
\multirow{2}{*}{NGC~4277} & \multirow{2}{*}{13.7} & rs & 0.57 & 0.43 & 117.5 & 0.59 & 0.57 & 87.9\\
\end{longtable}

\end{document}